\newcommand{\blind}{1}

\documentclass[authoryear,11pt]{article}
\usepackage[a4paper, total={6.5in, 10in}]{geometry}
\usepackage{setspace}

\usepackage[english]{babel}
\usepackage[utf8]{inputenc}
\usepackage[T1]{fontenc}

\usepackage{amsthm,amsmath,amssymb,mathrsfs,dsfont,mathtools}

\usepackage{float}
\usepackage{graphicx} % Required for inserting images
\usepackage{xcolor}
\usepackage{algorithm}
\usepackage{algpseudocode}
\algblock{ParFor}{EndParFor}
\algnewcommand\algorithmicparfor{\textbf{for}}
\algnewcommand\algorithmicpardo{\textbf{do Parallel}}
\algnewcommand\algorithmicendparfor{\textbf{end\ Parallel for}}
\algrenewtext{ParFor}[1]{\algorithmicparfor\ #1\ \algorithmicpardo}
\algrenewtext{EndParFor}{\algorithmicendparfor}
\usepackage{dsfont}
\usepackage{mathrsfs}
\usepackage{bm}
\usepackage[colorinlistoftodos,textsize=tiny,backgroundcolor=red!10,bordercolor=red]{todonotes}
\usepackage[colorlinks=true, allcolors=blue,hypertexnames=false]{hyperref}
\usepackage{booktabs,subcaption}
\usepackage[font=scriptsize,labelfont=bf]{caption}
\usepackage{hyperref}
\usepackage[all]{nowidow}
\usepackage{tabulary}
\usepackage{tabularx}
\usepackage{colortbl}

\usepackage{setspace}
\usepackage{tikz} % For the plot

\usepackage[sf]{titlesec}
\usepackage{sectsty}
\usepackage[round]{natbib}

\theoremstyle{definition}

\usepackage{multirow}
\definecolor{rowgray}{gray}{0.93}

\makeatletter
\def\env@cases{%
  \let\@ifnextchar\new@ifnextchar
  \left\lbrace
  \def\arraystretch{0.8}%
  \array{@{}l@{\quad}l@{}}}
\makeatother

\algblockdefx[MyFunc]{MyFunction}{EndMyFunction}%
    [2]{\textbf{function:} \textsc{#1} #2}% #1 = name, #2 = parameters
    [0]{\textbf{end function}}% End block text, no parameters

\newcommand \indep {\perp \!\!\!\! \perp}

\def\T{{ \mathrm{\scriptscriptstyle T} }}

\newcommand \dd  { \,\textup d}

\newcommand \MN {\operatorname{MN}}
\newcommand \IW {\operatorname{IW}}
\newcommand \mT {\operatorname{T}}

\usepackage{xr}
\usepackage{xr-hyper}
\makeatletter
\newcommand*{\addFileDependency}[1]{% argument=file name and extension
  \typeout{(#1)}
  \@addtofilelist{#1}
  \IfFileExists{#1}{}{\typeout{No file #1.}}
}
\makeatother

\newcommand*{\myexternaldocument}[1]{%
    \externaldocument{#1}%
    \addFileDependency{#1.tex}%
    \addFileDependency{#1.aux}%
}
\myexternaldocument{supplement_jcgs}

\usepackage{authblk}

\begin{document}
\def\spacingset#1{\renewcommand{\baselinestretch}%
        {#1}\small\normalsize} \spacingset{1}

\if1\blind{
\title{\huge \bf Dynamic Bayesian Predictive Stacking via Markovian Spatiotemporal Propagation}
\date{}
\author[1]{Luca Presicce}
\author[2]{Sudipto Banerjee}
%\author[1]{}
\affil[1]{Department of Economics, Management and Statistics, Università degli studi Milano-Bicocca, Milano, Italy}
\affil[2]{Department of Biostatistics, University
of California, Los Angeles, California}
\maketitle
}\fi

\if0\blind 
{
    % \title{\huge \bf Adaptive Markovian Spatiotemporal Transfer Learning in Multivariate Bayesian Modeling}
    \title{\huge \bf Dynamic Bayesian Predictive Stacking via Markovian Spatiotemporal Propagation}
    \author[]{}
    \maketitle
} \fi

\begin{abstract}
This manuscript develops computationally efficient online learning for multivariate spatiotemporal models. The proposed framework relies on matrix-variate Gaussian distributions, dynamic linear models, and Bayesian predictive stacking to efficiently share information across temporal data shards. The model facilitates effective information propagation over time while seamlessly integrating spatial components within a dynamic framework, building a Markovian dependence structure between datasets at successive time instants. This structure supports flexible, high-dimensional modeling of complex dependence patterns, as commonly found in spatiotemporal phenomena, where computational challenges arise rapidly with increasing dimensions. The proposed approach further manages exact inference through predictive stacking, enhancing robustness and interoperability. Combining sequential and parallel processing of temporal shards, each unit passes assimilated information forward and then back-smooths it to improve posterior estimation, incorporating all available information. This framework advances the scalability and adaptability of spatiotemporal modeling, making it suitable for dynamic, multivariate, and data-rich environments. Simulation experiments and an extracted data analysis from the Copernicus Data Space Ecosystem (CDSE) help evaluate and illustrate the framework.
\end{abstract}

\spacingset{1.9}
% \textbf{Keywords:} \textit{Bayesian inference; dynamic linear models, Bayesian predictive stacking; Spatiotemporal models; Markovian dependence.}

% ################################################
\section{Introduction}\label{sec:intro}

Spatiotemporal process models are a key statistical tool for analyzing data referenced by spatial locations and time points at which variables of interest have been measured \citep[see, e.g.,][]{gelfand_handbook_2010, cressie_statistics_2011, banerjeeEtAl2025book}. In particular, dynamic spatiotemporal models or \textsc{dstm}s \citep[][]{stroud_dynamic_2001, gelfandBanerjeeGamerman2005envcs, wikle_general_2010}, which build on a venerable literature on dynamic linear models \citep[see the authoritative text by][]{west_bayesian_1997}, offer probabilistic inference for the underlying stochastic processes \citep[][]{ali_analysis_1979, cressie_statistics_2011, schmidt_dynamic_2019}. These models can be recast as first-order affine discrete dynamical systems \citep{sandefur_discrete_1990, abdallaEtAl2020technometrics}, which provide an inferential framework for modeling Markovian dependence structures. As state-space models, they encompass a variety of modeling frameworks adapted to a wide range of applications \citep[e.g.][]{nobre_spatio-temporal_2005, gamerman_spatial_2008, hefley_dynamic_2017, pherwani_spatiotemporal_2024, idjigberou_spatio-temporal_2025}.

% Recent developments have increasingly focused on integrating ideas from statistical modeling, machine learning, physics, signal processing, and engineering. Probabilistic machine learning methods for physical systems, in particular, have relied on dynamic spatiotemporal modeling \citep[see][for examples in population dynamics and ecology]{czaran_spatiotemporal_1992, chen_spatio-temporal_2011}. 
Developments in spatiotemporal analysis have attempted to integrate probabilistic machine learning with statistical inference based on Bayesian dynamic linear models (\textsc{dlm}s) \citep[e.g.][]{ivanovic_trajectron_2019, zammit-mangion_deep_2020, banerjee_dynamic_2025}. 

We focus on multivariate dynamic linear models \citep[see, e.g.,][for diverse modeling and inferential perspectives]{west_bayesian_1997, gelfandBanerjeeGamerman2005envcs, prado_time_2021} and develop an inferential framework grounded in the closed-form distribution theory of the matrix-normal-inverse-Wishart (\textsc{mniw}) family of conjugate \textsc{dlm}s \citep[][]{quintana_analysis_1987, west_bayesian_1997, landim_dynamic_2000, jimenez_assessing_2021, elkhouly_dynamic_2021, banerjee_dynamic_2025}. 
The core contribution is a dynamic adaptation of Bayesian predictive stacking that preserves this conjugacy while scaling exact, simulation-free inference to massive spatiotemporal datasets. 
Specifically, the approach targets settings in which large data volumes arise from long streams of $T$ temporal snapshots, each constituting a spatial field of moderate size $n$.
The framework is general; the Earth-observation data from the \href{https://www.copernicus.eu}{Copernicus Data Space Ecosystem} analyzed in Section~\ref{sec:dataappl} provide a large-scale application that motivates the demand for such scalable inference.

Fully Bayesian inference for such models is typically carried out with iterative algorithms such as Markov chain Monte Carlo (\textsc{mcmc}) or Integrated Nested Laplace Approximation (\textsc{inla}), which are prohibitively expensive, motivating inferential procedures that return exact posterior summaries in closed form.

Rather than using iterative algorithms, we exploit the analytical closed-form distribution theory available from the \textsc{mniw} family. Here, we average over exact posterior distributions conditional on weakly identified hyperparameters that are, in any case, not consistently estimable \citep{zhang_inconsistent_2004, tang_identifiability_2021}. This approach is referred to as Bayesian predictive stacking  \citep[(\textsc{bps})][]{wolpert_stacked_1992,yao_using_2018} and has recently been demonstrated by \cite{zhangEtAl2025jasa} as an effective inferential tool for Bayesian geostatistics. 

We develop a dynamic adaptation of Bayesian predictive stacking for spatiotemporal dynamic linear models. The proposed dynamic Bayesian predictive stacking (\textsc{dynbps}) leverages the analytical tractability of the \textsc{dlm} and alleviates the computational cost of \textsc{bps}. Data are processed online, in a sequential stream of $T$ time points, each corresponding to a temporal snapshot of a spatial random field.
At each temporal data shard, we infer in parallel from closed-form posteriors for $J$ models, each corresponding to fixed values of spatial kernel hyperparameters, and propagate inference to the next time shard using forward filtering. For each time shard, we assimilate inference across the $J$ models using \textsc{bps} and proceed with backward sampling from $t=T, T-1,\ldots,0$ to complete the inference. A flow chart is supplied in Figure~\ref{fig:dataset_dependence} comprising three steps within each temporal ``cell'': (i) parallel computation of posterior distributions for each model specification; (ii) assimilate spatiotemporal inferences by stacking conjugate posteriors in closed form; and (iii) propagate information to the next step to deliver posterior-to-prior update. 

The manuscript is organized as follows. Section~\ref{sec:methodology} presents adaptive dynamic modeling and collects some of the necessary results on distribution theory that we will use. Section~\ref{sec:computational} provides details on the computational features of our inferential framework. Section~\ref{sec:simulations} provides results from simulation studies and Section~\ref{sec:dataappl} analyzes the extracted multivariate data from the \href{https://www.copernicus.eu}{Copernicus Data Space Ecosystem (CDSE)}. The paper concludes with remarks and implications for future research in Section~\ref{sec:discuss}. Supplements to Sections~\ref{sec:methodology},~\ref{sec:computational}~and~\ref{sec:simulations} are provided in Sections~\ref{sec:appendixA_theory},~\ref{sec:appendixB_algs}~and~\ref{sec:appendixC_sim}, respectively. Section~\ref{sec:appendix_eda} provides an overview of the specific dataset we analyze. Finally, Section~\ref{sec:appendixD_graphics} collects additional figures that complement the core exposition. 

% Figure 1
\usetikzlibrary{arrows.meta, positioning, calc, decorations.pathmorphing, backgrounds, fit, petri, decorations.footprints}
\begin{figure}[t!]
    \centering
% \hspace{-1cm}
\scalebox{1}[1]{
    \begin{tikzpicture}

    % envelope detail into rectangular
    \draw[very thick,blue,fill=cyan!10,rounded corners=1cm] (-6,-5.5) rectangle (-1,-0.75) node[black, fill=black!5, circle, draw, minimum size=1cm, scale=1,above left,yshift=0.25cm,xshift=-5cm] (Dt1) {$\mathscr{D}_{1}$};
    \draw[very thick,blue,fill=cyan!10,rounded corners=1cm] (4,-5.5) rectangle (9,-0.75) node[black, fill=black!5, circle, draw, minimum size=1cm, scale=1,above left,yshift=0.25cm,xshift=-5cm] (Dt) {$\mathscr{D}_{T}$};

    % Parallel Computing Models Text for each dataset
    \node (parallel1) [black!60!green,fill=black!60!green!5,shape=rectangle,rounded corners,draw, scale=1.2] at (-3.5,-2.5) {\tiny$\quad\begin{aligned} p(\cdot\mid&\mathscr{D}_{1},\mathscr{M}_{1}) \\[-1ex] &\cdots \\[-1ex] p(\cdot\mid&\mathscr{D}_{1},\mathscr{M}_{j}) \\[-1ex] &\cdots \\[-1ex] p(\cdot\mid&\mathscr{D}_{1},\mathscr{M}_{J})\end{aligned}\quad$};
    \node [above of= parallel1, yshift=0.35cm, black!60!green] {\tiny PARALLEL COMPUTING};
    
    \draw[line width=2pt,color=black,rounded corners=0.5cm] (Dt1.south) |- ([xshift=-1cm]parallel1.west);
    \draw[black!60!green,line width=1.25pt,arrows = {-Stealth[length=10pt,width=5pt,round]}] ([xshift=-1cm]parallel1.west) -- ([yshift=0.9cm]parallel1.west);
    \draw[black!60!green, line width=1.25pt,arrows = {-Stealth[length=10pt,width=5pt,round]}] ([xshift=-1cm]parallel1.west) -- ([yshift=0.45cm]parallel1.west);
    \draw[black!60!green, line width=1.25pt,arrows = {-Stealth[length=10pt,width=5pt,round]}] ([xshift=-1cm]parallel1.west) -- ([yshift=0cm]parallel1.west);
    \draw[black!60!green, line width=1.25pt,arrows = {-Stealth[length=10pt,width=5pt,round]}] ([xshift=-1cm]parallel1.west) -- ([yshift=-0.45cm]parallel1.west);
    \draw[black!60!green, line width=1.25pt,arrows = {-Stealth[length=10pt,width=5pt,round]}] ([xshift=-1cm]parallel1.west) -- ([yshift=-0.9cm]parallel1.west);
    
    \draw[black!60!green,line width=1.25pt,arrows = {-Stealth[length=10pt,width=5pt,round]}] ([yshift=0.9cm]parallel1.east) -- ([xshift=1cm]parallel1.east);
    \draw[black!60!green,line width=1.25pt,arrows = {-Stealth[length=10pt,width=5pt,round]}] ([yshift=0.45cm]parallel1.east) -- ([xshift=1cm]parallel1.east);
    \draw[black!60!green,line width=1.25pt,arrows = {-Stealth[length=10pt,width=5pt,round]}] ([yshift=0cm]parallel1.east) -- ([xshift=1cm]parallel1.east);
    \draw[black!60!green,line width=1.25pt,arrows = {-Stealth[length=10pt,width=5pt,round]}] ([yshift=-0.45cm]parallel1.east) -- ([xshift=1cm]parallel1.east);
    \draw[black!60!green,line width=1.25pt,arrows = {-Stealth[length=10pt,width=5pt,round]}] ([yshift=-0.9cm]parallel1.east) -- ([xshift=1cm]parallel1.east);

    \node (parallel2) [black!60!green,fill=black!60!green!5,shape=rectangle,rounded corners,draw, scale=1.2] at (6.5,-2.5) {\tiny$\quad\begin{aligned} p(\cdot\mid&\mathscr{D}_{T},\mathscr{M}_{1}) \\[-1ex] &\cdots \\[-1ex] p(\cdot\mid&\mathscr{D}_{T},\mathscr{M}_{j}) \\[-1ex] &\cdots \\[-1ex] p(\cdot\mid&\mathscr{D}_{T},\mathscr{M}_{J})\end{aligned}\quad$};
    \node [black!60!green,above of= parallel2, yshift=0.35cm] {\tiny PARALLEL COMPUTING};
        
    \draw[line width=2pt,color=black,rounded corners=0.5cm] (Dt.south) |- ([xshift=-1cm]parallel2.west);
    \draw[black!60!green,line width=1.25pt,arrows = {-Stealth[length=10pt,width=5pt,round]}] ([xshift=-1cm]parallel2.west) -- ([yshift=0.9cm]parallel2.west);
    \draw[black!60!green,line width=1.25pt,arrows = {-Stealth[length=10pt,width=5pt,round]}] ([xshift=-1cm]parallel2.west) -- ([yshift=0.45cm]parallel2.west);
    \draw[black!60!green,line width=1.25pt,arrows = {-Stealth[length=10pt,width=5pt,round]}] ([xshift=-1cm]parallel2.west) -- ([yshift=0cm]parallel2.west);
    \draw[black!60!green,line width=1.25pt,arrows = {-Stealth[length=10pt,width=5pt,round]}] ([xshift=-1cm]parallel2.west) -- ([yshift=-0.45cm]parallel2.west);
    \draw[black!60!green,line width=1.25pt,arrows = {-Stealth[length=10pt,width=5pt,round]}] ([xshift=-1cm]parallel2.west) -- ([yshift=-0.9cm]parallel2.west);
    
    \draw[black!60!green,line width=1.25pt,arrows = {-Stealth[length=10pt,width=5pt,round]}] ([yshift=0.9cm]parallel2.east) -- ([xshift=1cm]parallel2.east);
    \draw[black!60!green,line width=1.25pt,arrows = {-Stealth[length=10pt,width=5pt,round]}] ([yshift=0.45cm]parallel2.east) -- ([xshift=1cm]parallel2.east);
    \draw[black!60!green,line width=1.25pt,arrows = {-Stealth[length=10pt,width=5pt,round]}] ([yshift=0cm]parallel2.east) -- ([xshift=1cm]parallel2.east);
    \draw[black!60!green,line width=1.25pt,arrows = {-Stealth[length=10pt,width=5pt,round]}] ([yshift=-0.45cm]parallel2.east) -- ([xshift=1cm]parallel2.east);
    \draw[black!60!green,line width=1.25pt,arrows = {-Stealth[length=10pt,width=5pt,round]}] ([yshift=-0.9cm]parallel2.east) -- ([xshift=1cm]parallel2.east);

    % Stacked filtered posterior
    \node (stack1) [black!30!red,fill=red!5,shape=rectangle,rounded corners,draw,below=of parallel1, yshift=0cm, scale=0.8] {\color{black}$\sum_{j=1}^{J}\hat{w}_{j}^{(1)}\; p(\cdot\mid \mathscr{D}_{1},\mathscr{M}_{j})$};
    \node[very thick, black, fill=black!5, circle, draw, minimum size=1cm, scale=0.75,above left,yshift=-10cm,xshift=-7.25cm] (post1)
    {\scriptsize$\{\hat{Y},\hat{\Theta}\}_{1}$};
    \draw[line width=2pt,arrows = {-Stealth[length=15pt,width=10pt,round]},color=black,rounded corners=0.5cm] (stack1.west) -| (post1.north);
    \node (stack2) [black!30!red,fill=red!5,shape=rectangle,rounded corners,draw,below=of parallel2, yshift=0cm, scale=0.8] {\color{black}$\sum_{j=1}^{J}\hat{w}_{j}^{(T)}\; p(\cdot\mid \mathscr{D}_{T},\mathscr{M}_{j})$};
    \node[very thick, black, fill=black!5, circle, draw, minimum size=1cm, scale=0.75,above left,yshift=-10cm,xshift=6cm] (post2)
    {\scriptsize$\{\hat{Y},\hat{\Theta}\}_{T}$};
    \draw[line width=2pt,arrows = {-Stealth[length=15pt,width=10pt,round]},color=black,rounded corners=0.5cm] (stack2.west) -| (post2.north);

    % Draw arrows from the Bayesian Linear Model definitions to stacked filtered posterior
    \draw[line width=1.25pt,arrows = {-Stealth[length=10pt,width=5pt,round]},black!30!red] ([xshift=-0.5cm]parallel1.south) -- ([xshift=-0.5cm]stack1.north) node [right,pos=0.75]  {\tiny STACKING};
    \draw[line width=1.25pt,arrows = {-Stealth[length=10pt,width=5pt,round]},black!30!red] ([xshift=-0.5cm]parallel2.south) -- ([xshift=-0.5cm]stack2.north) node [right,pos=0.75] {\tiny STACKING};
    
    % Draw line for the borrowing of approximate information
    \node (ff) [black,right=of parallel1, xshift=2cm, yshift=0cm] {$\cdots$};
    \node [black,above of= ff, yshift=-0.5cm] {\tiny FORWARD FILTERING};
    \node [black,below of= ff, yshift=0.5cm] {\tiny for $\{\mathscr{D}_t\}_{t=1,\dots,T-1}$};
    \draw[line width=2pt,arrows = {-Stealth[length=15pt,width=10pt,round]},color=black,rounded corners=1cm] ([xshift=1cm]parallel1.east) -- (ff.west);
    \draw[line width=2pt,color=black,rounded corners=1cm] (ff.east) -- ([xshift=-1cm]parallel2.west);

    % Draw arrows for backward sampling
    \node (bs) [black,left=of post2, xshift=-2.5cm, yshift=0cm] {$\cdots$};
    \node [black,above of= bs, yshift=-0.5cm] {\tiny BACKWARD SAMPLING};
    \node [black,below of= bs, yshift=0.5cm] {\tiny for $\{\mathscr{D}_t\}_{t=T-1,\dots,0}$};
    \draw[line width=2pt,arrows = {-Stealth[length=15pt,width=10pt,round]},color=black,rounded corners=1cm] ([xshift=1cm]parallel2.east) -| ([xshift=5cm]post2.east) -- (post2.east);
    \draw[line width=2pt,arrows = {-Stealth[length=15pt,width=10pt,round]},color=black,rounded corners=1cm] (post2.west) -- (bs.east);
    \draw[line width=2pt,arrows = {-Stealth[length=15pt,width=10pt,round]},color=black,rounded corners=1cm] (bs.west) -- (post1.east);
    
\end{tikzpicture}
}
    \caption{Data shards dynamics dependencies representation}
    \label{fig:dataset_dependence}
\end{figure}

% ################################################
\section{Dynamic Spatiotemporal Multivariate Modeling}\label{sec:methodology}
Spatiotemporal modeling focuses on sequentially updating inference across time to efficiently learn from multivariate spatiotemporal processes.
In this setting, our aim is to infer latent spatiotemporal fields, which vary continuously in space and discretely in time. We automate inference propagation %while addressing the complexity of multivariate spatiotemporal dependencies by leveraging the conjugacy of matrix-variate dynamic spatiotemporal models and implementing 
using a dynamic Bayesian predictive stacking framework. 

% ################################################
\subsection{Matrix-variate Dynamic linear models}\label{sec:dlm}

Let $\left\{Y_t, t \in \mathcal{T}\right\}$ be a matrix-valued time series indexed by a set $\mathcal{T} = \{0, \pm 1, \pm 2, \pm 3, \cdots\}$ of discrete time points, where each $Y_t = \left\{Y_{i,j}^{(t)}\right\}$ is $n\times q$ with the $(i,j)$-th element $Y_{i,j}^{(t)}$ representing the recorded measurement of the variable $j\in\{1,\dots,q\}$ at spatial location $i\in\{1,\dots,n\}$. We let $\mathscr{D}_t$ denote the data available to us up to time $t$, including the current observation $Y_{t}$, any covariates observed in $t$ and all relevant past information $\mathscr{D}_{t-1}$
 %.
% Moreover, the set of observed temporal information $\mathscr{D}=\{\mathscr{D}_{1},\dots,\mathscr{D}_{T}\}$ can be thought as a filtration, such that $\mathscr{D}_{1}\subset\mathscr{D}_{2}\subset\cdots\subset\mathscr{D}_{T-1}\subset\mathscr{D}_{T}\equiv\mathscr{D}$. For the sake of complete notation, we introduce $\mathscr{D}_{0}\subset\mathscr{D}_{1}$ as the prior information, while $\mathscr{D}_{1}$ represent the first observed dataset.
with $\mathscr{D}_{0}$ denoting prior information without data%, while $\mathscr{D}_{1}$ represents the first observed dataset
.

A dynamic linear model decomposes each observation matrix $Y_{t}$ into two time-dependent elements: the linear trend modeled as $F_{t} \Theta_{t}$ and the error $\Upsilon_{t}$, which follows a matrix-variate normal distribution with row covariance $V_{t}$ and column covariance $\Sigma$. The (latent) $(p\times q)$ matrix $\Theta_{t}$, known as the ``state'' matrix at $t$, evolves smoothly over time, only depending on the previous state matrix $\Theta_{t-1}$. Specifically, we formulate a matrix-variate \textsc{dlm} as  
\begin{equation}\label{eq:DLM_matrix}
\begin{aligned}
    Y_{t} &= F_{t} \Theta_{t} + \Upsilon_{t},\quad &\Upsilon_{t} \sim\MN(0, V_{t}, \Sigma);\quad &\Theta_{t} = G_{t} \Theta_{t-1} + \Xi_{t} , \quad &\Xi_{t} \sim\MN(0, W_{t},\Sigma),
\end{aligned}
\end{equation}
where $\Theta_{0}\mid\Sigma \sim \MN(m_{0}, C_{0}, \Sigma)$ represents the initial information on the state matrix, with $m_{0}$ and $C_{0}$ being the known $(p \times q)$ mean and $(p \times p)$ row covariance matrices, respectively. The prior information for the column covariance matrix is captured by $\Sigma\sim\IW(\nu_{0},\Psi_{0})$ with the scalar parameter $\nu_{0}$ and the scale matrix $\Psi_{0}$. Here, $F_{t}$ is $n \times p$ with columns corresponding to explanatory predictors, $G_{t}$ is $p \times p$ modeling the evolution of the state matrix, and $W_{t}$ is $p \times p$ defining the dependence structure across the rows of $\Theta_{t}$. Equations~\eqref{eq:DLM_matrix} are commonly called the ``observation'' and ``state'' equations. 
All error terms are conditionally independent, given the column covariance matrix $\Sigma$. In particular, the observation and state disturbances $\{\Upsilon_{t}\}_{t\in\mathcal{T}}$ and $\{\Xi_{t}\}_{t\in\mathcal{T}}$ are each independent across time and mutually independent given $\Sigma$, so that $\Upsilon_{t}\indep\Xi_{t'}\mid\Sigma$ for all $t,t'\in\mathcal{T}$, and are independent of the initial state $\Theta_{0}$ given $\Sigma$. Together with the state recursion, these assumptions yield the Markovian structure of the model and ensure that $Y_{t}$ is conditionally independent of the past given $\Theta_{t}$ and $\Sigma$.
The column covariance $\Sigma$ is shared between the observation and state equations to ensure conjugacy, %which is analogous to sharing the error variance in Bayesian linear regression, and it encodes the assumption that the cross-variable (column) dependence is common across the levels of the hierarchy. The 
while %
row covariances $V_{t}$ and $W_{t}$, which govern the dependence among statistical units, remain free to differ across levels. The assumption can be relaxed at the cost of losing conjugacy and the closed-form \textsc{ffbs} %and 
leading to increased costs in %
stacking computations% on which our approach relies
.

%A \textsc{dlm} is fully characterized by the quadruple $\left\{F_t, G_t, V_t, W_t\right\}$. If we assume that $V_{t} = V$ and $W_{t} = W$ for all $t \in \mathcal{T}$, the model is referred to as a constant model. This type includes the local level and local trend models, the trigonometric seasonal model, and the time-varying parameter auto-regressive model. Often, the matrices in the quadruple can be decomposed into block-diagonal forms to represent conditional independence assumptions. Dynamic linear models rely on the Markov property that the current state $\Theta_{t}$ depends solely on the previous state $\Theta_{t-1}$. 
The Markovian structure in \eqref{eq:DLM_matrix} facilitates efficient inference and prediction using the Kalman filter. Consequently, \textsc{dlm}s are particularly well-suited for online learning applications, enabling real-time updates to estimates as new data is observed. Sequential updating, forecasting, and retrospective smoothing naturally follow from familiar distribution theory %for univariate and multivariate models 
\citep[see, e.g.,][for necessary derivations]{west_bayesian_1997,schmidt_dynamic_2019}.

The forward filtering backward sampling (\textsc{ffbs}) algorithm updates information using  
\begin{equation}
    \Theta_{t}\mid \mathscr{D}_{t-1},\Sigma \sim\MN(A_{t}, R_{t}, \Sigma)\;, \label{eq:Theta_FFprior}
\end{equation}
which constitutes forward filtering, and the one-step-ahead predictive distribution specified as
\begin{equation}
    Y_{t}\mid \mathscr{D}_{t-1},\Sigma \sim\MN(q_{t}, Q_{t}, \Sigma) \label{eq:Y_FFpredictive},
\end{equation} 
where $A_{t}=G_{t}m_{t-1}$, $R_{t}=G_{t}C_{t-1}G_{t}^{\T} +W_{t}$, $q_{t}=F_{t}A_{t}$, and $Q_{t}=F_{t}R_{t}F_{t}^{\T} +V_{t}$. The filtered posterior distribution for $\{\Theta_t, \Sigma\}$ is specified by the pair of distributions,
\begin{align}
    \Theta_{t}\mid \mathscr{D}_{t},\Sigma &\sim\MN(m_{t}, C_{t}, \Sigma) \label{eq:Theta_FFposterior}\\
    \Sigma\mid \mathscr{D}_{t} &\sim\IW(\nu_{t},\Psi_{t}) \label{eq:Sigma_FFposterior}.
\end{align}
% \begin{equation}\label{eq:ThetaSigma_FFposterior}
%     \Theta_{t}\mid \mathscr{D}_{t},\Sigma \sim\MN(m_{t}, C_{t}, \Sigma), \quad
%     \Sigma\mid \mathscr{D}_{t} \sim\IW(\nu_{t},\Psi_{t}),
% \end{equation}
with $m_{t} = C_{t}\left[R_{t}^{-1}A_{t}+F_{t}^{\T}V_{t}^{-1}Y_{t}\right]$, $C_{t}=\left[R_{t}^{-1}+F_{t}^{\T}V_{t}^{-1}F_{t}\right]^{-1}$, $\nu_{t}=\nu_{t-1} + n$ and $\Psi_{t} = \Psi_{t-1}+ (Y_{t}-q_{t})^{\T}Q_{t}^{-1}(Y_{t}-q_{t})$. 
Once the filtered posterior distribution $p(\Theta_{t}\mid \mathscr{D}_{t},\Sigma)$ is recovered for all $t=1,\dots,T$, the ``backward sampling'' algorithm is applied to smooth and refine state matrix estimates using both past and future knowledge. The smoothed posterior is obtained by sampling $\Theta_{T}\mid\mathscr{D}_{T},\Sigma$ from the filtered posterior and then, for each $t=T-1,\ldots,0$, from the distribution
% sampling backward for each $t=T, T-1,\ldots,0$ from the distribution 
\begin{equation}\label{eq:Theta_BSposterior}
    \Theta_{t}\mid\Theta_{t+1},\mathscr{D}_{T},\Sigma\sim\MN(h_{t}, H_{t}, \Sigma)\;,
\end{equation}
where $h_{t}=H_{t}[C_{t}^{-1}m_{t}+G_{t+1}^{\T}W_{t+1}^{-1}\Theta_{t+1}]$ and $H_{t}=[C_{t}^{-1}+G_{t+1}^{\T}W_{t+1}^{-1}G_{t+1}]^{-1}$.
% Posterior inference for \textsc{dlm}s just involves two steps: forward filtering information using Equations from~\eqref{eq:Theta_FFprior} to Equation~\eqref{eq:Sigma_FFposterior}, and smoothed future information backward by \eqref{eq:Theta_BSposterior}.

Bayesian predictive forecasts are also straightforward. Given a prediction horizon $k$, the joint predictive distribution for the state and observation matrices $\Theta_{T+k},Y_{T+k}$ is
\begin{equation}\label{eq:Joint_forecast}
    \begin{bmatrix} \Theta_{T+k} \\ Y_{T+k} \end{bmatrix}\mid \mathscr{D}_{T},\Sigma\sim\MN\left(\underbrace{\begin{bmatrix} A_{T}(k) \\ q_{T}(k) \end{bmatrix}}_{A^{\star}(k)},\underbrace{\begin{bmatrix} R_{T}(k) & R_{T}(k)F_{T+k}^{\T} \\ F_{T+k}R_{T}(k) & Q_{T}(k) \end{bmatrix}}_{R^{\star}(k)},\Sigma\right)\;,
\end{equation}
where $A_{T}(k)=G_{T+k}A_{T}(k-1)$, $R_{T}(k)=G_{T+k}R_{T}(k-1)G_{T+k}^{\T}+W_{T+k}$, $q_{T}(k)=F_{T+k}A_{T}(k)$, and $Q_{T}(k)=F_{T+k}R_{T}(k)F_{T+k}^{\T}+V_{T+k}$, with $A_{T}(0)=m_{T}$, $R_{T}(0)=C_{T}$. Parameters are updated recursively and calculated for previous $k-1$ steps to obtain $ k$-step-ahead forecasts.

We introduce spatial dependence in \eqref{eq:DLM_matrix} using an augmented linear model \citep[see, e.g.,][for a univariate spatial example]{banerjee_modeling_2020} that absorbs the spatial effects. We write $F_{t}=[X_{t}:\mathbb{I}_{n}]$ $(n\times (p+n))$ and $\Theta_{t}=[B_{t}^{\T}:\Omega_{t}^{\T}]^{\T}$ $((p+n)\times q)$, where $p$ is the number of available predictors in $X_t$, $B_{t}$ is the matrix ($p\times q$) of dynamic regression coefficients, and $\Omega_{t}$ is the latent matrix ($n\times q$) of spatial process realizations in $n$ locations and $q$ variables. 
We further assume $G_{t}=\begin{bsmallmatrix}\mathbb{I}_{p} & 0_{p\times n}\\0_{n\times p} &\mathbb{I}_{n}\end{bsmallmatrix}$, so that the regression coefficients $B_{t}$ and the spatial process $\Omega_{t}$ evolve separately over time.
% We further assume $B_{t}\indep\Omega_{t}\;\forall t\in\mathcal{T}$, leading to $G_{t}=\begin{bsmallmatrix}\mathbb{I}_{p} & 0_{p\times n}\\0_{n\times p} &\mathbb{I}_{n}\end{bsmallmatrix}$. 
This yields
% \begin{equation}\label{eq:DLM_spatiotemporal}
% \begin{aligned}
%     Y_{t} &= \underbrace{\begin{bmatrix} X_{t} & \mathbb{I}_{n} \end{bmatrix}}_{F_{t}} \underbrace{\begin{bmatrix} B_{t} \\ \Omega_{t} \end{bmatrix}}_{\Theta_{t}} + \Upsilon_{t}\;,\;\; \Upsilon_{t} \sim\MN(0, V_{t}(\alpha), \Sigma)\;; \\ 
%     \underbrace{\begin{bmatrix} B_{t} \\ \Omega_{t} \end{bmatrix}}_{\Theta_{t}} &= \underbrace{\begin{bmatrix} \mathbb{I}_{p} & 0_{p\times n} \\ 0_{n\times p} & \mathbb{I}_{n} \end{bmatrix}}_{G_{t}} \underbrace{\begin{bmatrix} B_{t-1} \\ \Omega_{t-1} \end{bmatrix}}_{\Theta_{t-1}} + \Xi_{t}\;,\;\; \Xi_{t} \sim\MN(0, W_{t}(\phi),\Sigma),
% \end{aligned}
% \end{equation}
\begin{equation}\label{eq:DLM_spatiotemporal}
    Y_{t} = \underbrace{\begin{bmatrix} X_{t} & \mathbb{I}_{n} \end{bmatrix}}_{F_{t}} \underbrace{\begin{bmatrix} B_{t} \\ \Omega_{t} \end{bmatrix}}_{\Theta_{t}} + \Upsilon_{t}\;,\qquad 
    \underbrace{\begin{bmatrix} B_{t} \\ \Omega_{t} \end{bmatrix}}_{\Theta_{t}} = \underbrace{\begin{bmatrix} \mathbb{I}_{p} & 0_{p\times n} \\ 0_{n\times p} & \mathbb{I}_{n} \end{bmatrix}}_{G_{t}} \underbrace{\begin{bmatrix} B_{t-1} \\ \Omega_{t-1} \end{bmatrix}}_{\Theta_{t-1}} + \Xi_{t}\;,
\end{equation}
where $\Upsilon_{t} \sim\MN(0, V_{t}(\alpha), \Sigma)$, $\Xi_{t} \sim\MN(0, W_{t}(\phi),\Sigma)$, and $V_{t}(\alpha)=(\frac{1-\alpha}{\alpha})\mathbb{I}_{n}$, $W_{t}(\phi)=\begin{bsmallmatrix} W^{B}_{t} & 0_{p\times n} \\ 0_{n\times p} & \mathcal{K}_{t}(\mathcal{S},\mathcal{S};\phi) \end{bsmallmatrix}$. $W^{B}_{t}$ is a generic $p\times p$ row-covariance matrix for dynamic coefficients and $\mathcal{K}_{t}(\mathcal{S},\mathcal{S};\phi)$ is an $n\times n$ spatial correlation matrix using locations enumerated in the set $\mathcal{S}=\{s_{1},\dots,s_{n}\}$. The $(i,j)$-th element of $\mathcal{K}_{t}(\mathcal{S},\mathcal{S};\phi)$ is the value of a spatial correlation function (e.g., an exponential or Mat\'ern with parameters $\phi$) evaluated at locations $s_i$ and $s_j$. 
% where $V_{t}(\alpha)=(\frac{1-\alpha}{\alpha})\mathbb{I}_{n}$, $W_{t}(\phi)=\begin{bsmallmatrix} W^{B}_{t} & 0_{p\times n} \\ 0_{n\times p} & \mathcal{K}_{t}(\mathcal{S},\mathcal{S};\phi) \end{bsmallmatrix}$, $W^{B}_{t}$ is a generic $p\times p$ row-covariance matrix for dynamic coefficients and $\mathcal{K}_{t}(\mathcal{S},\mathcal{S};\phi)$ is an $n\times n$ spatial correlation matrix using locations enumerated in the set $\mathcal{S}=\{s_{1},\dots,s_{n}\}$. The $(i,j)$-th element of $\mathcal{K}_{t}(\mathcal{S},\mathcal{S};\phi)$ is the value of a spatial correlation function (e.g., an exponential or Mat\'ern with parameters $\phi$) evaluated at locations $s_i$ and $s_j$. 
Together with the block-diagonal $G_{t}$, the block-diagonal structure of $W_{t}(\phi)$ makes $B_{t}\indep\Omega_{t}$ a priori. 
%A posteriori, they are dependent through $F_{t}$, as reflected by the block $C_{t,B\Omega}$ in Section~\ref{sec:complexity}.
The block $W^{B}_{t}$ can be any $p\times p$ covariance matrix; in what follows we set $W^{B}_{t}=\mathbb{I}_{p}$, which imposes no structure on the joint evolution of the regression coefficients.
The specification for $V_{t}(\alpha)$ introduces a discontinuity accounting for micro-scale measurement error. In doing so, we introduce a parameter $\alpha$ that captures the proportion of total variability expressed by the spatial component. The model in \eqref{eq:DLM_spatiotemporal} %produces the same structure as
is, again, %
a \textsc{dlm}. Hence, the conditional posterior distribution of $\{\Theta_t,\Sigma\}$, given fixed values of $\{\alpha,\phi\}$ is available in closed form from \eqref{eq:Theta_FFposterior}~and~\eqref{eq:Sigma_FFposterior}.

A remark on the evolution matrix is in order. Setting $G_{t}=\mathbb{I}_{p+n}$ specifies a random walk, $\Theta_{t}=\Theta_{t-1}+\Xi_{t}$, so that $\Theta_{t}$ retains full memory of its past and its variance accumulates over time.
Under the random walk specification, $\Theta_{t}=\Theta_{0}+\sum_{s=1}^{t}\Xi_{s}$, with $\Xi_{s}=\left[(\Xi^{B}_{s})^{\T}:(\Xi^{\Omega}_{s})^{\T}\right]^{\T}$. For a given $\phi$, the spatial correlation $\mathcal{K}_{t}(\mathcal{S},\mathcal{S};\phi)$ does not change with $t$. 
%Then, the row-covariance of $\Omega_{t}$ is a multiple of $\mathcal{K}(\mathcal{S},\mathcal{S};\phi)$ that grows linearly with $t$. 
Then, the accumulated spatial innovations $\sum_{s=1}^{t}\Xi^{\Omega}_{s}$ have row covariance $t\,\mathcal{K}_{t}(\mathcal{S},\mathcal{S};\phi)$. 
This is the origin of the factor $t$ in $\Tilde{W}_{t}(\phi)$ below in~\eqref{eq:Predictive_posterior}, while the kriging weights are unaffected.
A white-noise evolution would require $G_{t}=0_{p+n}$, giving $\Theta_{t}=\Xi_{t}$. More generally, $G_{t}$ governs the temporal dynamics of the state and can encode stationary behavior, for instance $G_{t}=\rho_{t}\mathbb{I}_{p+n}$ with $|\rho_{t}|<1$, akin to an Ornstein-Uhlenbeck mechanism. We adopt the block-diagonal identity because it preserves both the conjugacy and the block structure on which the scalability of the forward filter rests (Section~\ref{sec:complexity}).

In addition to temporal forecasting, we consider spatial interpolation. The target involves interpolating the latent spatial process for unobserved points. Here, we consider space as continuous and time as discrete. Thus, we are looking for spatial interpolation at each time point. To be precise, for $t\in\mathcal{T}$, we let $\Tilde{Y}_{t}$ denote the matrix of unknown values of the outcome on a set of $m$ unobserved spatial locations enumerated as $\mathcal{U}=\{u_{1},\dots,u_{m}\}$. 
Here, $\Tilde{F}_{t}=\left[\Tilde{X}_{t}:\mathbb{I}_{m}\right]$ denotes the observation matrix associated with the unobserved locations $\mathcal{U}$, while $\Tilde{\Omega}_{t}$ collects the realizations of the latent spatial process at those locations.
Thus, we seek the posterior predictive distribution defined as
\begin{equation}\label{eq:Predictive_integral}
    p(\Tilde{Y}_{t},\Tilde{\Omega}_{t}\mid \mathscr{D}_{t}) = \int\left[\int p(\Tilde{Y}_{t},\Tilde{\Omega}_{t}\mid \mathscr{D}_{t},\Theta_{t},\Sigma) p(\Theta_{t}\mid \mathscr{D}_{t},\Sigma) \dd\Theta_{t}\right] p(\Sigma\mid \mathscr{D}_{t})\dd\Sigma.
\end{equation}
For the model in Equation~\eqref{eq:DLM_spatiotemporal}, the integral in~\eqref{eq:Predictive_integral} can be computed analytically \citep[using similar arguments used in][Supplement 1.1]{presicce_bayesian_2024}. Full predictive inference on outcomes and spatial latent processes follows from the posterior predictive distribution,
\begin{equation}\label{eq:Predictive_posterior}
    p(\Tilde{Y}_{t},\Tilde{\Omega}_{t}\mid \mathscr{D}_{t}) = \mT_{2m,q}(\nu_{t}, \mu_{t}, E_{t}, \Psi_{t}),
\end{equation}
where $\mT_{2m,q}$ is the matrix-variate $\mT$ distribution with dimension $2m \times q$ \citep{gupta_matrix_2000, banerjee_dynamic_2025, presicce_bayesian_2024}, $\mu_{t}=\chi_{t}m_{t}$, $E_{t}=\chi_{t}C_{t}\chi_{t}^{\T}+N_{t}$, with $\chi_{t}=\begin{bsmallmatrix} \Tilde{X}_{t} & \Tilde{M}_{t} \\ 0 & \Tilde{M}_{t} \end{bsmallmatrix}$, $N_{t}=\begin{bsmallmatrix} \Tilde{V}_{t}(\alpha)+\Tilde{W}_{t}(\phi) & \Tilde{W}_{t}(\phi) \\ \Tilde{W}_{t}(\phi) & \Tilde{W}_{t}(\phi) \end{bsmallmatrix}$. Here, $\Tilde{V}_{t}(\alpha)=(\frac{1-\alpha}{\alpha})\mathbb{I}_{m}$ is the row-covariance matrix for $\Tilde{Y}_{t}$, while $\Tilde{M}_{t}=\mathcal{K}_{t}(\mathcal{U},\mathcal{S};\phi)\mathcal{K}^{-1}_{t}(\mathcal{S},\mathcal{S};\phi)$ and $\Tilde{W}_{t}(\phi)=t\big[\mathcal{K}_{t}(\mathcal{U},\mathcal{U};\phi)-\mathcal{K}_{t}(\mathcal{U},\mathcal{S};\phi)\mathcal{K}^{-1}_{t}(\mathcal{S},\mathcal{S};\phi)\mathcal{K}_{t}(\mathcal{S},\mathcal{U};\phi)\big]$ are the kriging weights and the conditional row-covariance, respectively.
%are the mean and row-covariance matrices, respectively%, specifying $p(\Tilde{\Omega}_{t}\mid \mathscr{D}_{t})$

These expressions are exact when the process at $\mathcal{U}$ follows the same random walk, $\Tilde{\Omega}_{t}=\Tilde{\Omega}_{t-1}+\Tilde{\Xi}^{\Omega}_{t}$, with innovations $\left(\Xi^{\Omega}_{t},\Tilde{\Xi}^{\Omega}_{t}\right)$ of joint row covariance $\mathcal{K}(\mathcal{S}\cup\mathcal{U},\mathcal{S}\cup\mathcal{U};\phi)$, and is initialized as the kriging extension of the initial state, $\Tilde{\Omega}_{0}=\Tilde{M}_{t}\Omega_{0}$. Then $\Tilde{\Omega}_{t}\mid\Theta_{t},\Sigma,\mathscr{D}_{t}\sim\MN(\Tilde{M}_{t}\Omega_{t},\Tilde{W}_{t}(\phi),\Sigma)$ %for any choice of $C_{0}$, and, since $\mathcal{U}$ does not enter the likelihood, this specification does not affect inference at the monitored locations
.

%This concludes the disclosure about the matrix-variate \textsc{dlm}s, and their foundational characteristics. Hereafter, we concentrate our attention on model~\eqref{eq:DLM_spatiotemporal}. 
% Nevertheless, beyond the scope of this manuscript, all the contents illustrated in the following Sections can be adapted to multivariate, or univariate, time series modeling straightforwardly.

% ################################################
% \section{Parallel adaptive spatiotemporal propagation}\label{sec:methodology}
\subsection{Parallel adaptive spatiotemporal propagation}\label{sec:propagation}

% starting section
Section~\ref{sec:dlm} augmented standard matrix-variate dynamic linear models to incorporate spatiotemporal processes. Similar approaches have been investigated for complex data analysis \citep[see e.g.][]{jimenez_assessing_2021}. Our contribution here relies on a dynamic adjustment of Bayesian predictive stacking (\textsc{bps}) that will allow the \textsc{ffbs} algorithm to exploit conjugate distribution theory and scale inference to large spatiotemporal datasets without resorting to \textsc{mcmc} algorithms.

% dynamic BPS
\textsc{bps} \citep[as formulated in][]{yao_using_2018} is a widely applicable model averaging method that has been shown to be effective for spatial data analysis \citep[see][]{zhangEtAl2025jasa,presicce_bayesian_2024}. Recently, \cite{pan_bayesian_2025} developed \textsc{bps} for hierarchical stochastic process models using spatiotemporal covariance kernels; however, their method does not scale to the massive size of our Copernicus dataset. 
% Bayesian \textsc{dlm}s treat time as discrete epochs, which allows the \textsc{ffbs} algorithm to scale to large temporal datasets. We consider \textsc{bps} an effective tool that can analyze massive volumes of spatiotemporal data. We achieve this by implementing a ``dynamic'' variation for predictive stacking that takes advantage of the predictive structure of \textsc{dlm}s.
We extend \textsc{bps} to massive spatiotemporal data through a ``dynamic'' variation that exploits the predictive structure of \textsc{dlm}s.

A key advantage of our adaptation of \textsc{bps} is that it will retain the full conjugacy of the model in Equation~\eqref{eq:DLM_spatiotemporal}. Note that if $\{\alpha,\phi\}$ is fixed, then the model in \eqref{eq:DLM_spatiotemporal} yields posterior distributions in closed form. Sampling from this posterior distribution is exact using the \textsc{ffbs} algorithm, and we do not require iterative algorithms such as \textsc{mcmc} approaches. Therefore, we fix $\{\alpha, \phi\}$ for a set of values and treat each value, $\{\alpha^{(j)}, \phi^{(j)}\}$, for $j=1,\ldots,J$ to correspond to a model $\mathscr{M}_j$. We obtain inference by stacking  \citep{breiman_stacked_1996}.   

\textsc{bps} attempts to find the optimal distribution in the convex hull $\mathcal{C}=\Big\{\sum_{j=1}^{J}w_{j}p(\cdot\mid\mathscr{M}_{j}):\sum_{j=1}^{J}w_{j}=1,w_{j}\geq 0\Big\}$ of individual posterior distributions by maximizing a proper scoring rule \citep{gneiting_strictly_2007,yao_using_2018}. Choices for a specific scoring rule include the true predictive distribution, which, however, is unknown and is computed using a leave-one-out estimate \citep[see][Section 3.1]{yao_using_2018}. This has been shown to be effective in Bayesian geostatistics \citep{zhangEtAl2025jasa}, but may not be ideal for capturing temporal dependencies. Hence, we introduce leave-future-out (\textsc{lfo}) based on gleanings from \cite{paul-christian_burkner_approximate_2020,vehtari_bayesian_2016,vehtari_practical_2017} to account for temporal dependencies in our data. \textsc{lfo} cross-validation and similar approaches that account for dependence in validation problems have also been investigated in \cite{ruiz_maraggi_using_2021}, \cite{cooper_cross-validatory_2023} and \cite{kennedy_model_2024}.

We adopt the one-step-ahead predictive distribution as our scoring function to assess the model's predictive performance in time-dependent contexts. In Section~\ref{sec:dynamicBPS}, we provide further details on deriving the optimization problem for dynamic predictive stacking. Dynamic stacking weights at time $\tau$ are then computed by solving this optimization problem
\begin{equation}\label{eq:dynamic_BPS}
    \hat{w}_{i}^{(\tau)}=\underset{w_{i}^{(\tau)}\in\mathscr{S}_{1}^{J}}{\operatorname{argmax}}\frac{1}{\tau-1}\sum_{t=1}^{\tau-1}\log\sum_{j=1}^{J} w_{i,j}^{(\tau)}\; \underbrace{p(Y_{t+1,i}\mid \mathscr{D}_{t},\mathscr{M}_{j})}_{\mT_{1,q}(Y_{t+1,i}\mid \nu_{t}^{(j)}, q_{t+1,i}^{(j)}, Q_{t+1,i}^{(j)}, \Psi_{t}^{(j)})},
\end{equation}
where $Y_{t+1,i}$ is the $i$-th row of $Y_{t+1}$, $p(Y_{t+1,i}\mid \mathscr{D}_{t},\mathscr{M}_{j})$ corresponds to the marginalization of the one-step-ahead predictive distribution in \eqref{eq:Y_FFpredictive} based on $\mathscr{M}_{j}$ fixed parameters, and $\mathscr{S}_{1}^{J}$ is the simplex of dimension $J$. The marginal predictive density $p(Y_{t+1,i}\mid \mathscr{D}_{t},\mathscr{M}_{j})$ is available in closed form as a matrix-variate t distribution $\mT_{1,q}(Y_{t+1,i}\mid \nu_{t}^{(j)}, q_{t+1,i}^{(j)}, Q_{t+1,i}^{(j)}, \Psi_{t}^{(j)} )$, %Leading Equation~\eqref{eq:dynamic_BPS} to
%\begin{equation}\label{eq:dynamic_BPS_Tdistr}
%    \hat{w}_{i}^{(\tau)}=\underset{w_{i}^{(\tau)}\in\mathscr{S}_{1}^{J}}%{\operatorname{argmax}}\frac{1}{\tau-1}\sum_{t=1}^{\tau-1}\log\sum_{j=1}^{J} %w_{i,j}^{(\tau)}\;\mT_{1,q}(Y_{t+1,i}\mid \nu_{t}^{(j)}, q_{t+1,i}^{(j)}, Q_{t+1,i}^{(j)}, \Psi_{t}^{(j)}),
%\end{equation}
where $q_{t+1,i}^{(j)} = F_{t+1,i}A_{t+1}^{(j)}$ and $Q_{t+1,i}^{(j)} = F_{t+1,i}R_{t+1}^{(j)}F_{t+1,i}^{\T} + \frac{1-\alpha^{(j)}}{\alpha^{(j)}}$, with $F_{t+1,i}$ the $i$-th row of $F_{t+1}$ and $\frac{1-\alpha^{(j)}}{\alpha^{(j)}}$ the $i$-th diagonal element of $V_{t+1}(\alpha^{(j)})$, so that $q_{t+1,i}^{(j)}$ is the $i$-th row of $q_{t+1}^{(j)}$ and $Q_{t+1,i}^{(j)}$ the $i$-th diagonal element of $Q_{t+1}^{(j)}$; $A_{t+1}^{(j)}=G_{t+1}m_{t}^{(j)}$, and $R_{t+1}^{(j)}=G_{t+1}C_{t}^{(j)}G_{t+1}^{\T}+W_{t+1}(\phi^{(j)})$. For each step $t+1$, the parameters $\{m_{t}^{(j)},C_{t}^{(j)},\Psi_{t}^{(j)},\nu_{t}^{(j)}\}$ are inherited from the previous step $t$ for each model $\mathscr{M}_{j}$. In addition, %let us recall the structure of model~\eqref{eq:DLM_matrix} (or equivalently \eqref{eq:DLM_spatiotemporal}) and its estimation. In particular, 
for any time $t$, the one-step-ahead predictive distributions are computed by default within the \textsc{ffbs} procedure (see \eqref{eq:Y_FFpredictive}). These exactly match the one-step-ahead predictive distribution required by \textsc{lfo} cross-validation. %Indeed, leave-future-out cross-validation focuses on the predictive evaluation of the next future observation given the past. Figure~\ref{fig:leavefutureout} gives a sketch of its operating principle.

\usetikzlibrary{matrix, arrows.meta, positioning, calc, decorations.pathmorphing, backgrounds, fit, petri, decorations.footprints}
\begin{figure}[t!]
    \centering
% \hspace{-1cm}
\scalebox{0.8}[0.8]{

    \begin{tikzpicture}
    % Define colors for train and test sets
    \definecolor{train}{rgb}{0.6, 0.8, 1}
    \definecolor{test}{rgb}{1, 0.8, 0.6}

    % Time series timeline, with a hole between 4 and T (represented as T)
    \draw[->] (-1, 0) -- (-1, -7) node[below] {\textbf{Time}};
    
    % Number of data points (1 to 4 and then T)
    \foreach \x in {1, 2, 3, 5, 6} {
        \filldraw (-1, -\x) circle (2pt);
    }
    \foreach \x in {1, 2, 3} {
        \node[left] at (-1, -\x) {\x};
    }
    \foreach \x in {5} {
        \pgfmathtruncatemacro{\y}{6 - \x}
        \node[left] at (-1, -\x) {T-\y};
    }
    \node[left] at (-1, -4) {\textbf{$\vdots$}}; % Represent the skipped time between 4 and T
    \node[left] at (-1, -6) {T}; % Mark time T at the end

    % Add cross-validation splits with updated labels
    \foreach \i/\trainstart/\trainend/\teststart/\testend in {
        1/0/1/2/3, 
        2/0/3/4/5, 
        3/0/5/6/7, 
        4/0/5/6/7,
        5/0/9/10/11, % This is for split T-1
        6/0/11/12/13  % This is for split T
    } {
        % Check if it is the third or last split, and replace with dots for all elements
        \ifnum\i=4
            % Instead of drawing the train and test areas, display dots for the third line
            \node at ({(\trainstart + \trainend) / 2}, -\i) {\textbf{$\vdots$}};
            \node at ({(\teststart + \testend) / 2}, -\i) {\textbf{$\vdots$}};
        \else
            
            % For Split 5, change the label to Train T-1 and Test T-1
            \ifnum\i=5
                % Draw train region for Train T-1
                \draw[rounded corners, fill=train, opacity=0.5] (\trainstart-0.5, -0.5-\i) rectangle (\trainend+0.5, 0.5-\i);
                % \node[left] at (\trainstart-1, -\i) {\textbf{Step T\(-1\)}};
                % Train T-1
                \pgfmathtruncatemacro{\prev}{\i - 1}
                \node at ({(\trainstart + \trainend) / 2}, -\i) {\textbf{\(\mathscr{D}_{T-2}\)}};

                % Draw test region for Test T-1
                \draw[rounded corners, fill=test, opacity=0.5] (\teststart-0.5, -0.5-\i) rectangle (\testend+0.5, 0.5-\i);
                % Test T-1
                \node at ({(\teststart + \testend) / 2}, -\i) {\textbf{\(\mathscr{D}_{T-1}\)}};
            \fi

            % For Split 6, change the label to Train T and Test T
            \ifnum\i=6
                % Draw train region for Train T
                \draw[rounded corners, fill=train, opacity=0.5] (\trainstart-0.5, -0.5-\i) rectangle (\trainend+0.5, 0.5-\i);
                % \node[left] at (\trainstart-1, -\i) {\textbf{Step T}};
                % Train T
                \pgfmathtruncatemacro{\prev}{\i - 1}
                \node at ({(\trainstart + \trainend) / 2}, -\i) {\textbf{$\mathscr{D}_{T-1}$}};

                % Draw test region for Test T
                \draw[rounded corners, fill=test, opacity=0.5] (\teststart-0.5, -0.5-\i) rectangle (\testend+0.5, 0.5-\i);
                % Test T
                \node at ({(\teststart + \testend) / 2}, -\i) {\textbf{$\mathscr{D}_{T}$}};
            \fi

            % Normal drawing for other splits (not split 4, 5, or 6)
            \ifnum\i<4
                % Draw train region
                \draw[rounded corners, fill=train, opacity=0.5] (\trainstart-0.5, -0.5-\i) rectangle (\trainend+0.5, 0.5-\i);
                % \node[left] at (\trainstart-1, -\i) {\textbf{Step \i}};
                % Calculate and display "Train (i-1)"
                \pgfmathtruncatemacro{\prev}{\i - 1}
                \node at ({(\trainstart + \trainend) / 2}, -\i) {\textbf{\(\mathscr{D}_{\prev}\)}};

                % Draw test region
                \draw[rounded corners, fill=test, opacity=0.5] (\teststart-0.5, -0.5-\i) rectangle (\testend+0.5, 0.5-\i);
                % Center text in the test rectangle
                \node at ({(\teststart + \testend) / 2}, -\i) {\textbf{\(\mathscr{D}_{\i}\)}};
            \fi
        \fi
    }

    % Add manual brace to indicate progression
    % \draw[thick] (0.5, 0.8) .. controls (0.5, 1.5) and (12.5, 1.5) .. (12.5, 0.8);
    % \node[above] at (6.5, 1.5) {\textbf{Leave-Future-Out Cross-Validation}};

    % Replace last bar with T instead of 12
    % \node[below, color=train, scale=1.2] at (8, -3.75) {\textbf{\textsc{train}}};
    % \node[below, color=test, scale=1.2] at (11, -3.75) {\textbf{\textsc{test}}};
    \node[below, color=train, scale=1.3] at (0.5, 0.25) {\textbf{\textsc{train}}};
    \node[below, color=test, scale=1.3] at (2.5,  0.25) {\textbf{\textsc{test}}};

    % Predictive distributions
    \draw[rounded corners, fill=train, opacity=0.2] (8, -0.75) rectangle (12, -1.25);
    \node[] at (10, -1) {\textcolor{black}{Compute $p(\;\cdot\,\mid\mathscr{D}_{t-1})$}};
    \draw[->, very thick, color = train] (8.5, -1.25) -- (8.5, -4.25);
    
    \draw[rounded corners, fill=test, opacity=0.2] (10, -2) rectangle (14, -2.5);
    \node[] at (12, -2.25) {\textcolor{black}{Evaluate $p(Y_{t}\mid\mathscr{D}_{t-1})$}};
    \draw[->, very thick, color = test] (10.5, -2.5) -- (10.5, -4.25);
    
\end{tikzpicture}}
    \caption{Leave-future-out cross-validation}
    \label{fig:leavefutureout}
\end{figure}

Note that we compute stacking weights for each location $i$ in Algorithm~\ref{alg:dynamic_bps}, which corresponds to a single time series, and for each time $t=2,\dots, T$. 
This follows from the definition of the dynamic stacking weights in \eqref{eq:dynamic_BPS}. We compare predictive performance across time %instants 
and not across observations, as standard Bayesian predictive stacking does. This originates from the difference between leave-one-out cross-validation and leave-future-out cross-validation (schematized in Figure~\ref{fig:leavefutureout}). 
For $t\le 1$, no one-step-ahead evaluation is available, and the weights are set to the starting weights $\hat{w}^{(0)}$ (uniform by default). 

A joint evaluation across all locations is infeasible because, as $n$ grows, the predictive density of the full matrix $Y_{t}$ resides in an increasingly high-dimensional space and becomes numerically unstable to evaluate.
% However, even if it is theoretically possible to design this way, a simultaneous evaluation across time and locations is infeasible. As the number of points increases, the density lies in a higher-dimensional space, where the notion of a point rapidly loses sense, and its probability goes to zero even faster. Then, as the number of locations grows, the evaluation of the predictive density for the whole matrix $Y_{t}$ results in numerical instabilities.
Nevertheless, individual evaluations allow us to reach a finer granularity of stacking weights and possibly provide individual temporal inferences for each location. Individual stacking weights are useful instruments for gleaning insights into model fitting in several areas of interest, as shown in Section~\ref{sec:sim_weights} using simulated data.

Backward sampling in Algorithm~\ref{alg:backward_sampling}, however, requires a single set of stacking weights shared across locations, and posterior inference on the parameters should not depend on any individual location. Computing and propagating location-specific weights would, moreover, be computationally demanding in big-$n$ settings, even with parallel resources. We, therefore, aggregate the individual weights across the $n$ locations into a single weight vector, for which we consider two strategies.

First, we may compute average weights across all locations, i.e. 
% \begin{equation*}
    $\{\hat{w}_{1}^{(\tau)},\dots,\hat{w}_{J}^{(\tau)}\}=\hat{w}^{(\tau)}_{G}=\frac{1}{n}\sum_{i=1}^{n}\hat{w}_{i}^{(\tau)}=\Bigg\{\frac{1}{n}\sum_{i=1}^{n}\hat{w}_{i,1}^{(\tau)},\dots,\frac{1}{n}\sum_{i=1}^{n}\hat{w}_{i,J}^{(\tau)}\Bigg\}$,
% \end{equation*}
referring to them as ``global'' weights and indexed with the subscript $G$. Alternatively, we may consider the consensus across all locations. Let the number of observations for which model $j$ has the maximum weight be denoted by $c_{j}^{(\tau)} = \Bigl|\{ i : j_{i}^{(\tau)} = j \}\Bigr|$ with $j_{i}^{(\tau)} = \underset{1 \le j \le J}{\arg\max} \;\hat{w}_{i,j}^{(\tau)}$. We then define the consensus weight vector by normalizing these counts: 
% \begin{equation*}
    $\{\hat{w}_{C,1}^{(\tau)},\dots,\hat{w}_{C,J}^{(\tau)}\}=\hat{w}^{(\tau)}_{C}=\Big\{\frac{c_{1}^{(\tau)}}{n},\ldots,\frac{c_{J}^{(\tau)}}{n}\Big\}$.
% \end{equation*}
We refer to these as ``consensus'' weights and index them with the subscript $C$. Both these solutions provide a set of weights that sum to one and, hence, is a vector of probabilities (see Section~\ref{sec:weightsproof}). We provide an empirical investigation into the impact of this choice in Section~\ref{sec:sim_weights}.

Having obtained the set of dynamic weights at time $\tau$ and $\hat{w}^{(\tau)}$ based on the preferred aggregation strategy, posterior inference follows straightforwardly using the stacked posterior distribution,
\begin{equation}\label{eq:stacked_posterior}
    \hat{p}(\cdot\mid\mathscr{D}_{\tau})=\sum_{j=1}^{J}\hat{w}_{j}^{(\tau)}\;p(\cdot\mid \mathscr{D}_{\tau},\mathscr{M}_{j}),
\end{equation}
where $\cdot$ denotes the inferential object of interest (e.g., regression parameters; the latent process; predictive values at a location or time). We sample from \eqref{eq:stacked_posterior} by first sampling one of the $J$ posteriors, $p(\cdot\mid \mathscr{D}_{\tau},\mathscr{M}_{j})$ with a probability of $\hat{w}_{j}^{(\tau)}$, and then sampling $\cdot$ from the drawn posterior. Computing the individual stacking weights within each time shard is detailed in Algorithm~\ref{alg:dynamic_bps}. We exploit parallel computing over the competitive model evaluation, as the number $J$ of competitive models in most applications allows for a light implementation even in modest computational environments.

%%%%%%%%%%%%%%%%%%%%%%%%%%%%%%%%%%%%%%%%%%%%%%%%%%%%%%%%%%%%%%%%%%%%%%%%%%%%%%%%%%%%%%%%%%%%%%%%%%%%%%%%%%

% Parallel learning adaptation
The stacked posterior in Equation~\eqref{eq:stacked_posterior} is a finite mixture, and propagating it as the prior for the next time step %would destroy 
destroys %
the %posterior-to-prior 
conjugacy on which the \textsc{ffbs} recursions rely. We resolve this by separating the two roles played by the posterior. Within each time shard $\mathscr{D}_{\tau}$, inference, prediction, smoothing, and uncertainty quantification are carried out using the stacked posterior $\hat{p}(\cdot\mid\mathscr{D}_{\tau})$, exploiting the flexibility of its mixture form. Information is propagated forward in time, instead, along $J$ parallel learning paths: each model $\mathscr{M}_{j}$ passes its own conditional posterior $p(\cdot\mid\mathscr{D}_{\tau},\mathscr{M}_{j})$, rather than the mixture, as the prior for step $\tau+1$. Since each conditional posterior is conjugate, the posterior-to-prior update is preserved exactly, and the stacking weights are recomputed at every step from the one-step-ahead predictive densities.
% A substantial problem arises when borrowing temporal posterior-to-prior information throughout \textsc{ffbs} using the content presented so far. Indeed, predictive stacking results in posterior distributions defined as finite mixtures, see Equation~\eqref{eq:stacked_posterior}. This could break the \textsc{ffbs} machinery, i.e., its conjugacy across previous instant posteriors acting as priors for the next one. A solution to the mutual information must be found. Intuitively, a solution relies on assimilating a parallel learning path for each model. Thus, the stacked distribution retains reliability for inference in each generic dataset $\mathscr{D}_{\tau}$, but information to the incoming dataset $\mathscr{D}_{\tau+1}$ only passes across $J$ parallel learning paths. Then, posterior-to-prior distribution retains the conjugacy of \textsc{ffbs} algorithm, as we do not use stacked posterior $\hat{p}(\cdot\mid\mathscr{D}_{\tau})$ but conditional posterior instead, i.e. $p(\cdot\mid\mathscr{D}_{\tau},\mathscr{M}_{j})$ for $j=1,\dots,J$. Hence, these computational adaptations allow us to preserve the conjugacy in the \textsc{ffbs} algorithm while borrowing information between time instants.

This separation is the key methodological device used in \textsc{dynbps}. It averages out the nuisance spatial hyperparameters $\{\alpha,\phi\}$ through the stacking weights while retaining the closed-form Markovian updates of the \textsc{ffbs} algorithm. The conditional posteriors act only as conduits for transferring temporal information and do not alter the within-shard inference dictated by the stacked posterior. Figure~\ref{fig:dataset_dependence} summarizes the resulting procedure within each temporal ``cell''.
% Thus, we found a solution that allows us to average out nuisance spatial parameters, and this was obtained by adapting instruments from \textsc{bps} into a dynamical environment. It is worth specifying that the conditional posterior distributions do not affect the inference within each data shard at a specific time instant. In practice, conditional posterior distributions, are only used to allow the information propagation across time, but inferences are pursued by stacked posterior. This permits us to take advantage of the finite mixture formulation of stacked posterior gained from \textsc{bps}, making it possible to fit more complex distributions, for prediction and uncertainty quantification. Figure~\ref{fig:dataset_dependence} provides a graphical representation of the entire procedure within each ``cell''.

% ################################################
\section{Computational details}\label{sec:computational}

\begin{algorithm}[!t] 
\caption{Computing individual \textsc{dynbps} weights at time $\tau$}\label{alg:dynamic_bps}
\begin{algorithmic}[1]
\Require Data $\mathscr{D}_{\tau}$, one-step-ahead predictive and filtered posterior parameters $\left\{q_{\tau}^{(j)},Q_{\tau}^{(j)},\nu_{\tau-1}^{(j)},\Psi_{\tau-1}^{(j)}\right\}_{j=1}^{J}$, one-step-ahead predictive density evaluations $\left\{pd_{i,j}^{(t)}\right\}_{t=1}^{\tau-1}$
\Ensure Individual stacking weights at time $\tau$
\For{$i=1,\dots,n$}
\State \textit{\# Evaluate one-step-ahead posterior predictive density for $Y_{\tau}$}
\State $pd_{i,j}^{(\tau)}\gets\;\dd\mT_{1,q}(Y_{\tau,i}\mid \nu_{\tau-1}^{(j)}, q_{\tau,i}^{(j)}, Q_{\tau,i}^{(j)}, \Psi_{\tau-1}^{(j)})$
\State  \textit{\# Obtain \textsc{dynbps} weights solving the convex optimization problem}
\State $\hat{w}_{i}^{(\tau)}\gets\;\underset{w_{i}^{(\tau)}\in\mathscr{S}_{1}^{J}}{\operatorname{argmax}}\frac{1}{\tau-1}\sum_{t=1}^{\tau-1}\log\sum_{j=1}^{J} w_{i,j}^{(\tau)}\;pd_{i,j}^{(t+1)}$
\EndFor
\State \Return $\left\{\hat{w}_{i}^{(\tau)}=\left\{\hat{w}_{i,1}^{(\tau)},\dots,\hat{w}_{i,J}^{(\tau)}\right\}\right\}_{i=1}^{n}$
\end{algorithmic}
\end{algorithm}

Standard \textsc{ffbs} inference requires conjugacy, which is precluded by the introduction of spatial range parameters that require simulation-based approaches such as \textsc{mcmc}. Our goal is instead to achieve a fully conjugate, simulation-free analysis for \textsc{dstm}s.
% Vanilla implementations of Bayesian predictive stacking, let alone the advantages, also introduce drawbacks. The \textsc{bps} posterior distributions are finite mixtures; this breaks down posterior-to-prior updating, which allows borrowing of information through time instants in the forward filter and backward sampling algorithm. The same happens for more sophisticated \textsc{bps} based modeling approaches \citep[e.g.,][]{presicce_bayesian_2024}. A feasible solution to retain the dynamical structure of the \textsc{ffbs} algorithm should include temporal dynamics within the Bayesian predictive stacking while preserving conjugate posterior-to-prior updates. 
We achieve this using dynamic predictive stacking (see Section~\ref{sec:propagation}). Algorithm~\ref{alg:dynamic_bps} details the computation of stacking weights.

\begin{algorithm}[!t] 
\caption{\textsc{dynbps} - parallel forward filter}\label{alg:forward_filter}
\begin{algorithmic}[1]
\Require Data $\mathscr{D}_{1:T}$, observation and state transition matrices $F_{1:T},G_{1:T}$, starting stacking weights $\hat{w}^{(0)}$, hyperparameters $\left\{m_{0}^{(j)},C_{0}^{(j)},\Psi_{0}^{(j)},\nu_{0}^{(j)},\alpha^{(j)},\phi^{(j)}\right\}_{j=1}^{J}$
\Ensure Filtering distribution parameters, stacking weights at time $t=0,\dots,T$
\MyFunction{ParallelForwFilt}{\Big($\mathscr{D}_{1:T}, F_{1:T},G_{1:T},\hat{w}^{(0)}, \{m_{0}^{(j)},C_{0}^{(j)},\Psi_{0}^{(j)},\nu_{0}^{(j)},\alpha^{(j)},\phi^{(j)}\}_{j=1}^{J}$\Big)}
\For{$t=1,\dots,T$}
\ParFor{$j=1,\dots,J$}
\State \textit{\# Compute observation and state row-covariance matrices}
\State $V_{t}^{(j)} \gets\; \left(\frac{1-\alpha^{(j)}}{\alpha^{(j)}}\right)\mathbb{I}_{n}$, $W_{t}^{(j)} \gets\;  \begin{bsmallmatrix} \mathbb{I}_{p} & 0_{p\times n} \\ 0_{n\times p} & \mathcal{K}_{t}(\mathcal{S},\mathcal{S};\phi^{(j)}) \end{bsmallmatrix}$
\State \textit{\# Compute filtered prior parameters}
\State $A_{t}^{(j)} \gets\; G_{t}m_{t-1}^{(j)}$, $R_{t}^{(j)} \gets\; G_{t}C_{t-1}^{(j)}G_{t}^{\T}+W_{t}^{(j)}$
\State \textit{\# Compute one-step ahead predictive parameters}
\State $q_{t}^{(j)} \gets F_{t}A_{t}^{(j)}$, $Q_{t}^{(j)}\gets\;F_{t}R_{t}^{(j)}F_{t}^{\T} +V_{t}^{(j)}$
\State \textit{\# Compute state matrix filtered posterior parameters}
\State ${\small m_{t}^{(j)}\gets C_{t}^{(j)}\left[(R_{t}^{(j)})^{-1}A_{t}^{(j)}+F_{t}^{\T}(V_{t}^{(j)})^{-1}Y_{t}\right]}$, ${\small C_{t}^{(j)} \gets \left[(R_{t}^{(j)})^{-1}+F_{t}^{\T}(V_{t}^{(j)})^{-1}F_{t}\right]^{-1}}$
\State \textit{\# Compute column-covariance matrix filtered posterior parameters}
\State $\nu_{t}^{(j)} \gets\; \nu_{t-1}^{(j)} + n$, $\Psi_{t}^{(j)}\gets\;\Psi_{t-1}^{(j)}+(Y_{t}-q_{t}^{(j)})^{\T}(Q_{t}^{(j)})^{-1}(Y_{t}-q_{t}^{(j)})$
\EndParFor
\State \textit{\# Obtain individual weights using Algorithm~\ref{alg:dynamic_bps} and compute global weights} 
\State $\left\{\hat{w}_{i}^{(t)}=\left\{\hat{w}_{i,1}^{(t)},\dots,\hat{w}_{i,J}^{(t)}\right\}\right\}_{i=1}^{n}$, $\hat{w}^{(t)} \gets \frac{1}{n}\sum_{i=1}^{n}\hat{w}_{i}^{(t)}=\left\{\frac{1}{n}\sum_{i=1}^{n}\hat{w}_{i,1}^{(t)},\dots,\frac{1}{n}\sum_{i=1}^{n}\hat{w}_{i,J}^{(t)}\right\}$
\EndFor
\State \Return $\left\{\hat{w}^{(t)},\{m_{t}^{(j)},C_{t}^{(j)},\Psi_{t}^{(j)},\nu_{t}^{(j)}\}_{j=1}^{J}\right\}_{t=0}^{T}$
\EndMyFunction
\end{algorithmic}
\end{algorithm}

We now present %the computational adjustment to 
the algorithms %characterizing 
adapting %
the \textsc{ffbs} procedure \citep{carter_gibbs_1994}. %This series of 
These %
algorithms %replaces 
replace %
the standard \textsc{ffbs} within the \if1\blind{ \href{https://github.com/lucapresicce/spFFBS}{\texttt{spFFBS}} }\fi\if0\blind{ \texttt{REDACTED} }\fi package. We tailored these with special consideration for spatiotemporal modeling, referring to Model~\eqref{eq:DLM_spatiotemporal}.
% 
% BPS FORWARD FILTERING
Algorithm~\ref{alg:forward_filter} adapts the forward filter (\textsc{ff}) within the \textsc{ffbs} recursion to each time step $t=1,\dots,T$. It retains the structure of the standard \textsc{ff} but runs $J$ filters in parallel, one for each model $\mathscr{M}_{j}$, that is, for each pair $\{\alpha^{(j)},\phi^{(j)}\}$, $j=1,\dots,J$ and embeds the weight computation of Algorithm~\ref{alg:dynamic_bps}.
% Once each $j$-th flow provides filtered posterior distributions, especially the one-step-ahead predictive distribution, we compute \textsc{dynbps} weights using Algorithm~\ref{alg:dynamic_bps}. 
Prior information is shared across the models $\mathscr{M}_{j}$ and propagated forward over all observed times. To fully incorporate the new piece of information, as the latest dataset occurs at $T+1$, it suffices to apply Algorithm~\ref{alg:forward_filter} once, followed by Algorithm~\ref{alg:backward_sampling} to update posterior sampling with the newest info. This gains advantage from the Markovian dependence assumption between datasets.

\begin{algorithm}[!t]
\caption{\textsc{dynbps} - weighted backward sampling}\label{alg:backward_sampling}
\begin{algorithmic}[1]
\Require Stacking weights $\left\{\hat{w}^{(t)}\right\}_{t=0}^{T}$, filtering parameters and inputs from Algorithm~\ref{alg:forward_filter}
% \Ensure R smoothed posterior samples from $\hat{p}(\Theta_{0:T}\mid \mathscr{D}_{T})$
\Ensure R smoothed posterior samples from $\hat{p}(\Theta_{0:T},\Sigma\mid \mathscr{D}_{T})$
\MyFunction{WeightedBackSamp}{\Big($\{\hat{w}^{(t)}, \{m_{t}^{(j)},C_{t}^{(j)},\Psi_{t}^{(j)},\nu_{t}^{(j)},\alpha^{(j)},\phi^{(j)}\}_{j=1}^{J}\}_{t=0}^{T}, R$\Big)}
\For{$r=1,\dots,R$}
\State  \textit{\# Draw model $\mathscr{M}_{j}$ such that} $j \sim\;\operatorname{Multinom}\left(1,J,\left\{\hat{w}_{1}^{(T)},\dots,\hat{w}_{J}^{(T)}\right\}\right)$
% \State \textit{\# Draw $\Theta_{T}$ from} $\mT_{p+n,q}\left(\nu_{T}^{(j)},h_{T}^{(j)}, H_{T}^{(j)}, \Psi_{T}^{(j)}\right)$ \textit{with} $h_{T}^{(j)}\gets\;m_{T}^{(j)}$, $H_{T}^{(j)}\gets\;C_{T}^{(j)}$
\State \textit{\# Draw $\Sigma$ from} $\IW\left(\nu_{T}^{(j)},\Psi_{T}^{(j)}\right)$
\State \textit{\# Draw $\Theta_{T}$ from} $\MN\left(h_{T}^{(j)}, H_{T}^{(j)}, \Sigma\right)$ \textit{with} $H_{T}^{(j)}\gets\;C_{T}^{(j)}$, $h_{T}^{(j)}\gets\;m_{T}^{(j)}$
\For{$t=T-1,\dots,0$}
\State  \textit{\# Draw model $\mathscr{M}_{j}$ such that} $j \sim\;\operatorname{Multinom}\left(1,J,\left\{\hat{w}_{1}^{(t)},\dots,\hat{w}_{J}^{(t)}\right\}\right)$
% \State \textit{\# Draw $\Theta_{t+1}$ from} $\mT_{p+n,q}\left(\nu_{t+1}^{(j)}, m_{t+1}^{(j)}, C_{t+1}^{(j)}, \Psi_{t+1}^{(j)}\right)$
\State \textit{\# Compute smoothed posterior parameters}
\State $H_{t}^{(j)}\gets\;\left[(C_{t}^{(j)})^{-1}+G_{t+1}^{\T}(W_{t+1}^{(j)})^{-1}G_{t+1}\right]^{-1}$
\State $h_{t}^{(j)}\gets\;H_{t}^{(j)}\left[(C_{t}^{(j)})^{-1}m_{t}^{(j)}+G_{t+1}^{\T}(W_{t+1}^{(j)})^{-1}\Theta_{t+1}\right]$
% \State \textit{\# Draw $\Theta_{t}$ from} $\mT_{p+n,q}\left(\nu_{T}^{(j)}, h_{t}^{(j)}, H_{t}^{(j)}, \Psi_{T}^{(j)} \right)$
\State \textit{\# Draw $\Theta_{t}$ from} $\MN\left(h_{t}^{(j)}, H_{t}^{(j)}, \Sigma\right)$
\EndFor
\EndFor
% \State \Return R smoothed posterior samples of $\Theta_{0:T}$
\State \Return R smoothed posterior samples of $(\Theta_{0:T},\Sigma)$
\EndMyFunction
\end{algorithmic}
\end{algorithm}

% BPS BACKWARD SAMPLING
In \textsc{ffbs} algorithm presented in \cite{carter_gibbs_1994}, backward sampling (\textsc{bs}) passes the information assimilated from all observed $T$ states backward to each time point $t=T-1,\dots,0$. This strategy serves to obtain smoothed posterior samples, improving inference by exploiting available future pieces of information. With Algorithm~\ref{alg:backward_sampling} we propose a weighted alternative to the standard backward sampling procedure.
For each desired posterior draw $r=1,\dots,R$, we first sample a model $\mathscr{M}_{j}$, with $j=1,\dots,J$, by giving models weights $\hat{w}^{(T)}$ obtained at last observed time.
% Then, we draw the posterior sample from the corresponding conditional posterior for $\Theta_{T}$, i.e. $p(\Theta_{T}\mid \mathscr{D}_{T},\mathscr{M}_{j})$. 
Then, we draw $\Sigma$ and $\Theta_{T}$ from the corresponding conditional posterior, i.e. $p(\Sigma,\Theta_{T}\mid \mathscr{D}_{T},\mathscr{M}_{j})$. The same $\Sigma$ is then shared by all the backward steps, as in the standard \textsc{ffbs} algorithm, so that the temporal dependence of the sampled path is preserved.
In doing so, we obtain a sample from the approximated stacked posterior $\hat{p}(\Theta_{T},\Sigma\mid \mathscr{D}_{T})=\sum_{j=1}^{J}\hat{w}_{j}^{(T)}\;p(\Theta_{T},\Sigma\mid \mathscr{D}_{T},\mathscr{M}_{j})$.
For $r=1,\dots,R$, we use standard \textsc{bs} algorithm backwards, repeating ``model sampling'' given correspondent weights $\hat{w}^{(t)}$. 
At the end of the procedure depicted in Algorithm~\ref{alg:backward_sampling}, we obtain a smoothed posterior distribution sample of size $R$. In particular, with Algorithm~\ref{alg:backward_sampling}, we obtain samples from the stacked (approximated) joint posterior, which can be derived as
\begin{equation}\label{eq:stacked_posterior_jointsmooth}
    \hat{p}(\Theta_{0:T},\Sigma\mid \mathscr{D}_{T})=\hat{p}(\Theta_{T},\Sigma\mid \mathscr{D}_{T})\prod_{t=0}^{T-1}\hat{p}(\Theta_{t}\mid\Theta_{t+1},\Sigma,\mathscr{D}_{T}),
\end{equation}
where $\hat{p}(\Theta_{T},\Sigma\mid \mathscr{D}_{T})=\sum_{j=1}^{J}\hat{w}_{j}^{(T)}\;p(\Theta_{T},\Sigma\mid \mathscr{D}_{T},\mathscr{M}_{j})$, and each $\hat{p}(\Theta_{t}\mid\Theta_{t+1},\Sigma,\mathscr{D}_{T})=\sum_{j=1}^{J}\hat{w}_{j}^{(t)}\;p(\Theta_{t}\mid\Theta_{t+1},\Sigma,\mathscr{D}_{T},\mathscr{M}_{j})$.
Marginalizing \eqref{eq:stacked_posterior_jointsmooth} over $\Theta_{0:T}$, the column covariance has stacked posterior $\hat{p}(\Sigma\mid\mathscr{D}_{T})=\sum_{j=1}^{J}\hat{w}_{j}^{(T)}\,\IW\big(\Sigma\mid\nu_{T}^{(j)},\Psi_{T}^{(j)}\big)$, and Algorithm~\ref{alg:backward_sampling} returns its draws together with the smoothed states.

% \begin{equation}\label{eq:stacked_posterior_jointsmooth}
%     \hat{p}(\Theta_{0},\dots,\Theta_{T}\mid \mathscr{D}_{T})=\hat{p}(\Theta_{0}\mid\Theta_{1}, \mathscr{D}_{T})\hat{p}(\Theta_{1}\mid\Theta_{2}, \mathscr{D}_{T})\cdots\hat{p}(\Theta_{T-1}\mid\Theta_{T}, \mathscr{D}_{T})\hat{p}(\Theta_{T}\mid \mathscr{D}_{T}),
% \end{equation}
% where each $\hat{p}(\Theta_{t}\mid\Theta_{t+1}, \mathscr{D}_{T})=\sum_{j=1}^{J}\hat{w}_{j}^{(t)}\;\hat{p}(\Theta_{t}\mid\Theta_{t+1}, \mathscr{D}_{T},\mathscr{M}_{j})$.
We clarify the distributional nature of these draws. The stacking weights $\hat{w}_{j}^{(t)}$, by themselves, are not posterior model probabilities. They are obtained by maximizing a leave-future-out log-score (Section~\ref{sec:dynamicBPS}), and no prior is placed on the models or on the weights. 
% Consequently, the samples produced by Algorithm~\ref{alg:backward_sampling} are not draws from the joint posterior $p(\Theta_{0:T},\mathscr{M}_{1},\dots,\mathscr{M}_{J}\mid\mathscr{D}_{T})$, but are draws from the stacked posterior in \eqref{eq:stacked_posterior_jointsmooth}, where $\hat{p}(\Theta_{0},\dots,\Theta_{T}\mid\mathscr{D}_{T})$ is defined as the distribution induced by the sequential backward sampler. 
% Since each conditional $\hat{p}(\Theta_{t}\mid\Theta_{t+1},\Sigma,\mathscr{D}_{T})$ is a finite mixture of proper matrix normal densities with weights in the simplex, the induced joint law is well defined, and marginalizing over $\Sigma$ gives $\hat{p}(\Theta_{0:T}\mid\mathscr{D}_{T})$.
% For $J=1$, \eqref{eq:stacked_posterior_jointsmooth} reduces to the exact joint smoothing posterior of the \textsc{ffbs} algorithm.
Consequently, the samples produced by Algorithm~\ref{alg:backward_sampling} are not draws from the joint posterior $p(\Theta_{0:T},\Sigma,\mathscr{M}_{1},\dots,\mathscr{M}_{J}\mid\mathscr{D}_{T})$, but from the stacked distribution $\hat{p}(\Theta_{0:T},\Sigma\mid\mathscr{D}_{T})$ in \eqref{eq:stacked_posterior_jointsmooth}, defined as the law induced by the sequential backward sampler. Since $\hat{p}(\Theta_{T},\Sigma\mid\mathscr{D}_{T})$ is a finite mixture of proper matrix-normal inverse-Wishart densities and each conditional $\hat{p}(\Theta_{t}\mid\Theta_{t+1},\Sigma,\mathscr{D}_{T})$ is a finite mixture of proper matrix-normal densities, with weights in the simplex, this law is well defined, and marginalizing over $\Sigma$ gives $\hat{p}(\Theta_{0:T}\mid\mathscr{D}_{T})$. 
For $J=1$, \eqref{eq:stacked_posterior_jointsmooth} reduces to the exact joint smoothing posterior $p(\Theta_{0:T},\Sigma\mid\mathscr{D}_{T})$ of the \textsc{ffbs} algorithm.
Together with Algorithm~\ref{alg:forward_filter}, we achieve posterior inference and uncertainty quantification for Model~\eqref{eq:DLM_spatiotemporal}.
The complete \textsc{dynbps} variant of \textsc{ffbs} runs the parallel forward filter (Algorithm~\ref{alg:forward_filter}) followed by the weighted backward sampler (Algorithm~\ref{alg:backward_sampling}).

We conclude this section with the two predictive tasks central to spatiotemporal modeling: temporal forecasting and spatial interpolation. Both reuse the filtered output of Algorithm~\ref{alg:forward_filter} and, as in Algorithm~\ref{alg:backward_sampling}, draw a model $\mathscr{M}_{j}$ at each iteration based on the stacking weights before sampling from the corresponding conditional predictive distribution.
For temporal forecasting (Algorithm~\ref{alg:temporal_forecasting}), we choose $\mathscr{M}_{j}$ according to the set of weights in $\hat{w}^{(T)}$ and then apply the $k$-step-ahead recursion of Equation~\eqref{eq:Joint_forecast}, in which the $k$-step predictive equation depends on the $(k-1)$-th step. This yields samples from the stacked predictive $\hat{p}(\Theta_{T+k},Y_{T+k}\mid \mathscr{D}_{T})=\sum_{j=1}^{J}\hat{w}_{j}^{(T)}\,p(\Theta_{T+k},Y_{T+k}\mid \mathscr{D}_{T},\mathscr{M}_{j})$. 
As the horizon $k$ grows, the predictive distribution diffuses, mirroring the predictive behavior of autoregressive integrated moving average (\textsc{arima}) models: with $G_{t}=\mathbb{I}_{p+n}$, the predictive mean remains at the last filtered value and the variance grows linearly in $k$, whereas $G_{t}=\rho_{t}\mathbb{I}_{p+n}$ with $|\rho_{t}|<1$ would yield a mean-reverting predictive mean and a bounded variance.

For spatial interpolation at a fixed time $t$ (Algorithm~\ref{alg:spatial_prediction}), the same model-sampling step is combined with the closed-form joint predictive of Equation~\eqref{eq:Predictive_posterior} over the unobserved locations $\mathcal{U}=\{u_{1},\dots,u_{m}\}$, producing draws from $\hat{p}(\Tilde{Y}_{t},\Tilde{\Omega}_{t}\mid \mathscr{D}_{t})=\sum_{j=1}^{J}\hat{w}_{j}^{(t)}\,p(\Tilde{Y}_{t},\Tilde{\Omega}_{t}\mid \mathscr{D}_{t},\mathscr{M}_{j})$. Repeating over $R$ draws delivers full uncertainty quantification for both responses and latent spatial processes across the region of interest, enabling the generation of interpolated spatial maps.

\subsection{Computational complexity and scalability}\label{sec:complexity}

We evaluate the computational cost of \textsc{dynbps} in terms of $n$ (spatial locations), $T$ (time points), $p$ (predictors), $q$ (outcomes), $J$ (candidate models), and $R$ (posterior draws); in the spatiotemporal settings of interest, $p \ll n$. Table~\ref{tab:complexity} summarizes the main results.

For a fixed model $\mathscr{M}_{j}$ and time step $t$, the dominant cost in the forward filter arises from the filtered posterior covariance update. Because $F_t = [X_t : \mathbb{I}_{n}]$ is $n \times (p+n)$, the one-step-ahead predictive covariance $Q_t = F_t R_t F_t^\T + V_t(\alpha^{(j)})$ is $n \times n$, regardless of the state dimension $p + n$. The Woodbury identity yields $C_t = R_t - R_t F_t^\T Q_t^{-1} F_t R_t$, so the update requires a single $n \times n$ Cholesky factorization of $Q_t$ and associated triangular solves, at a cost of $O(n^3)$. 
% The block-diagonal structure $W_{t}(\phi)=\begin{bsmallmatrix} W^{B}_{t} & 0_{p\times n} \\ 0_{n\times p} & \mathcal{K}_{t}(\mathcal{S},\mathcal{S};\phi) \end{bsmallmatrix}$ and the diagonal form $V_t(\alpha^{(j)}) = \tfrac{1-\alpha^{(j)}}{\alpha^{(j)}}\mathbb{I}_{n}$ ensure that the prediction step costs $O(p^3 + n^2)$ rather than $O((p+n)^3)$.
The block-diagonal structure $W_{t}(\phi)=\begin{bsmallmatrix} W^{B}_{t} & 0_{p\times n} \\ 0_{n\times p} & \mathcal{K}_{t}(\mathcal{S},\mathcal{S};\phi) \end{bsmallmatrix}$ and the diagonal form $V_t(\alpha^{(j)}) = \tfrac{1-\alpha^{(j)}}{\alpha^{(j)}}\mathbb{I}_{n}$, together with $G_{t}=\mathbb{I}_{p+n}$, which reduces the prediction step to $R_{t}=C_{t-1}+W_{t}(\phi)$, ensure that the prediction step costs $O(n^{2})$ and forming $Q_{t}$ costs $O(n^{2}p)$, rather than $O((p+n)^3)$ for dense $(p+n)\times(p+n)$ operations.
Moreover, $\mathcal{K}_{t}(\mathcal{S},\mathcal{S};\phi^{(j)})$ and its inverse are precomputed once prior to the time loop at a cost of $O(J \cdot n^3)$, yielding a $T$-fold reduction over a naive per-step recomputation. 
The covariance blocks $(C_{t,BB},\,C_{t,B\Omega},\,C_{t,\Omega\Omega})$ are updated at costs $O(pn^2)$, $O(pn^2)$, and $O(n^3)$, respectively, with the last term dominant; the full $(p+n)\times(p+n)$ matrix $C_t$ is never assembled, and only its three blocks are stored at a memory cost of $O(p^2 + pn + n^2)$. The forward filter, therefore, costs $O(n^3)$ per model per time step.

The smoothing covariance $H_t = \bigl[C_t^{-1} + G_{t+1}^\T W_{t+1}^{-1} G_{t+1}\bigr]^{-1}$ is $(p+n)\times(p+n)$, which naively costs $O((p+n)^3)$. Exploiting the block-diagonal structure of $W_{t+1}$ and the block storage of $C_t$, the inverse $C_t^{-1}$ is recovered block-wise through the Schur complement of $C_{t,BB}$ within $C_t$,
% \begin{equation*}
  $S_t = C_{t,\Omega\Omega} -
        C_{t,B\Omega}^\T C_{t,BB}^{-1} C_{t,B\Omega}$,
% \end{equation*}
which is $n \times n$ at a cost of $O(n^3)$, with the $p\times p$ contribution $O(p^3)$ being negligible. A second Schur's complement step then recovers $H_t$ through two $n\times n$ and two $p\times p$ sub-problems, fully avoiding the $O((p+n)^3)$ full matrix inversion. The Cholesky factorization of the assembled $(p+n)\times(p+n)$ matrix $H_t$, required for posterior draws, costs $O((p+n)^3)$; in the regime $p \ll n$ this reduces to $O(n^3)$ with a multiplicative overhead $(1+p/n)^3$ bounded by a small constant. Each smoothed posterior draw then costs $O(n^2 q)$, with the factorization of $H_t$ reused across all $R$ draws.

\begin{table}[t!]
  \centering
  \begin{tabular}{@{}p{4cm}p{6.5cm}c@{}}
    \toprule
    Phase & Operation & Cost \\
    \midrule
    \textit{Precomputation}
      & Spatial kernels $\mathcal{K}_j$, $\mathcal{K}_j^{-1}$%, $j = 1, \ldots, J$ (once)
      & $O(Jn^3)$ \\
    \addlinespace
    \midrule
    \textit{Forward filter}
      & Per model, per time step
      & $O(n^3)$ \\
      & Total work, $J$ models over $T$ steps
      & $O(JTn^3)$ \\
      & Wall-clock, $J$ parallel workers
      & $O(Tn^3)$ \\
    \addlinespace
    \midrule
    \textit{Stacking weights}
      & $n$ convex programs over $T$ steps
      & $O(nT^2J)$ \\
    \addlinespace
    \midrule
    \textit{Backward sampler}
      & Per model, per time step
      & $O(n^3)$ \\
      & $R$ posterior draws
      & $O(Rn^2q)$ \\
    \addlinespace
    \midrule
    \textit{Spatial prediction}
      & $m$ unobserved locations
      & $O(n^3 + m^3)$ \\
    \bottomrule
  \end{tabular}
  \caption{Computational cost of \textsc{dynbps} by phase. Wall-clock costs assume $J$ parallel workers for the forward filter and $R$ parallel workers for backward draws.}
  \label{tab:complexity}
\end{table}

The \textsc{dynbps} stacking weights at time $\tau$ require solving $n$ separate convex optimization programs of dimension $J$ over $\mathscr{S}_{1}^{J}$, each using the $\tau \cdot J$ predictive density evaluations already produced by the forward pass. Standard convex solvers scale as $O(\tau \cdot J)$ per objective evaluation, giving $O(n \cdot T^2 \cdot J)$ total over all $T$ steps; alternative methods can reduce the constant factor without altering this asymptotic order. The $J$ forward filters are executed in parallel via \texttt{OpenMP} within the \texttt{spFFBS} package, reducing the wall-clock cost of the forward pass from $O(J \cdot T \cdot n^3)$ to $O(T \cdot n^3)$. The dominant total work is therefore $O(J \cdot T \cdot n^3)$.

We note that the $O(n^3)$ per-step scaling is not specific to \textsc{dynbps}. Any method requiring the inversion of $Q_t$ or $\mathcal{K}_{t}(\mathcal{S},\mathcal{S};\phi)$ shares this cost. The computational advantage of \textsc{dynbps} does not arise from reducing this fundamental scaling but from replacing iterative sampling with closed-form evaluations. 
Standard \textsc{mcmc} requires $O(K_{\mathrm{iter}} \times T \times n^3)$ operations, with $K_{\mathrm{iter}}$ typically of order $10^3$--$10^4$ iterations to achieve convergence, whereas \textsc{dynbps} substitutes $J$ closed-form evaluations per time step; Section~\ref{sec:sim_mcmc} quantifies this as a speedup of approximately $750\times$ per dataset fit, at comparable predictive accuracy.
A further advantage follows from the Markovian structure: incorporating a new observation at time $T+1$ requires only a single forward pass per model at a cost of $O(J \cdot n^3)$, independent of $T$, with an online memory footprint of $O(J \cdot n^2)$ for the current filtered state. 

Should scaling inference to a massive number of spatial locations be desired, the dense spatial covariance can be modeled using scalable, low-rank, or sparse approximations, such as predictive process models \citep[e.g.,][]{banerjee_gaussian_2008,finley_improving_2011}, nearest-neighbor, meshed, or other Vecchia-approximation-based Gaussian processes \citep{datta_hierarchical_2016, peruzzi_highly_2022, katzfuss_general_2021}; see \cite{banerjee_high-dimensional_2017} for a review of Bayesian hierarchical spatial models for high-dimensional spatial data analysis. Spatial scalability can also be achieved through spatial streaming, as in the ``season-episode'' framework of \citep[][Section~4]{banerjee_dynamic_2025}. The double Bayesian predictive stacking method devised in \cite{presicce_bayesian_2024} offers a further route to extend \textsc{dynbps} to settings where both $n$ and $T$ are large.

% ################################################
\section{Simulation experiments}\label{sec:simulations}

We illustrate dynamic Bayesian predictive stacking under different simulation experiments, while posing challenges and exploring features of multivariate spatiotemporal analysis.
Section~\ref{sec:sim_Msettings} assesses model class influence under $\mathscr{M}$-closed and $\mathscr{M}$-open settings, including comparisons with equal weighting, static weights, and top model selection ablations. Two complementary studies are reported in the Supplementary Material: Section~\ref{sec:sim_mcmc} directly compares \textsc{dynbps} with standard \textsc{mcmc} in terms of computational cost and predictive accuracy, and Section~\ref{sec:sim_weights} investigates the behavior and dynamics of \textsc{dynbps} weighting strategies across space and time.
    
\textsc{dynbps} is implemented in the \if1\blind{\href{https://github.com/lucapresicce/spFFBS}{\texttt{spFFBS} }}\fi\if0\blind{\texttt{REDACTED }}\fi package.
% We have developed an efficient and effective approach to managing online spatiotemporal learning for massive georeferenced historical data within the \if1\blind{\texttt{Rcpp}-based \href{https://github.com/lucapresicce/spFFBS}{\texttt{spFFBS} }}\fi\if0\blind{\texttt{REDACTED}}\fi package.
The simulations and analysis were implemented using native \texttt{R} and \texttt{C++} programming languages.
Reproducibility is ensured by access to the public repository \if1\blind{\href{https://github.com/lucapresicce/Markovian-Spatiotemporal-Propagation}{https://github.com/lucapresicce/Markovian-Spatiotemporal-Propagation}}\fi\if0\blind{\texttt{REDACTED}}\fi, which provides all the programs required to replicate results.
We obtain the results by using just a standard laptop running an Intel Core i7-8750H CPU with 5 cores for parallel computation, with 16 GB of \textsc{ram}, representing the efficiency of \textsc{dynbps} within scarce computational resources frameworks.
% 
% Details on programs and Algorithms are explored extensively in Section~\ref{sec:computational}. 

% We implemented the dynamic \textsc{dynbps} in native \texttt{R} and \texttt{c++} deploying the \texttt{spFFBS} package. All programs required to reproduce the analysis are publicly accessible from the GitHub repository \href{https://github.com/lucapresicce/Markovian-Spatiotemporal-Propagation}{lucapresicce/Markovian-Spatiotemporal-Propagation} that links the \texttt{Rcpp}-based \href{https://github.com/lucapresicce/spFFBS}{\texttt{spFFBS}} package. The reported results are from a standard laptop running an Intel Core i7-$8750$H CPU with $5$ cores for parallel computation and $16$ GB of \textsc{ram}.

% %%%%%%%%%%%%%%%%%%%%%%%%%%%%%%%%%%%%%%%%%%%%%%%%
\subsection{Model class influence}\label{sec:sim_Msettings}
We explore how \textsc{dynbps} behaves when the data-generating model is included in, or excluded from, the set of candidate models. In the $\mathscr{M}$-closed setting, the data-generating model is one of the candidates.
In the $\mathscr{M}$-open setting, following the usage of \citet[][Abstract and Sections~4.1--4.2]{yao_using_2018}, the data-generating model is not among the candidates, although it belongs to the same parametric family. Under the stricter taxonomy of \citet{bernardo_bayesian_1994}, formally adopted in \citet[][Section~2]{yao_using_2018}, $\mathscr{M}$-open additionally requires that the true model cannot be explicitly specified, and this setting may also be classified as $\mathscr{M}$-complete. As noted by \citet{clarke_comment_2018}, the original definitions allow the two classes to overlap. We retain the $\mathscr{M}$-open label because stacking, as a predictive approach, does not construct or use an explicit reference model, which characterizes the $\mathscr{M}$-open view in the sense of \citet[][Section~4.1]{vehtari_survey_2012} and \citet[][Section~2]{piironen_comparison_2017} \citep[see also][Section~3.2]{yao_using_2018}.
% We explore how dynamic \textsc{bps} behaves in the $\mathscr{M}$-closed and $\mathscr{M}$-open settings. Generally, these two classes of model misspecifications relate to opposite scenarios. The $\mathscr{M}$-closed refers to the case when the true model can be identified within a finite set of candidate models. $\mathscr{M}$-open class admits the existence of the true models, but this cannot be directly specified within candidate models.

Hereafter, model specification is characterized by values of $\alpha$ and $\phi$, where the data generating values $\{\alpha=0.8, \phi=4\}$ represent the true model (\textsc{true}).
The dynamic \textsc{bps} was tested over $\mathscr{M}$-closed and $\mathscr{M}$-open settings. For \textsc{dynbps} under closed settings (\textsc{dynbps-c}), we specify $J=9$ competitive models with $\alpha \in \{0.7, 0.8, 0.9\}$ and $\phi \in \{2, 4, 6\}$ that yield effective spatial ranges as a percentage of the maximum point inter-distance of $105\%, 53\%, 35\%$, respectively, including the true model as one of the possible candidates. Conversely, in the $\mathscr{M}$-open setting, the true model exists and belongs to the same family, but it is not included among the candidates. 
Thus, for dynamic \textsc{bps} under the open setting (\textsc{dynbps-o}), we randomly define $J=9$ candidate models, so that the true model is excluded with probability one. In particular, we uniformly sample 3 values of $\alpha \in (0.5, 1)$ and 3 values of $\phi \in [1, 50]$. 

In addition to \textsc{dynbps-c} and \textsc{dynbps-o}, we compare three ablation baselines: Equal Weighting (\textsc{ew}), assigning uniform weights $w_{j}^{(t)}=1/J$ to all $J$ candidate models at each time $t$; Static Weights (\textsc{sw}), which computes the stacking weights once, in batch, over the whole training window, and holds them fixed at every $t$; 
and Top Model Selection (\textsc{tms}), which applies hard weighting at each time $t$: it assigns unit weight to the candidate model with the largest weight and zero weight to all the others.
% Static Weights (\textsc{sw}), computing Bayesian predictive stacking weights in batch using all time-point data at $t=1$ and keeping them fixed thereafter; and Top Model Selection (\textsc{tms}), applying a hard threshold to the stacking weights at each $t$ by assigning full weight to the highest-weighted candidate model. 
We perform the experiment using 50 replications.

\begin{figure}[t!]
    \centering
    \includegraphics[width=\linewidth]{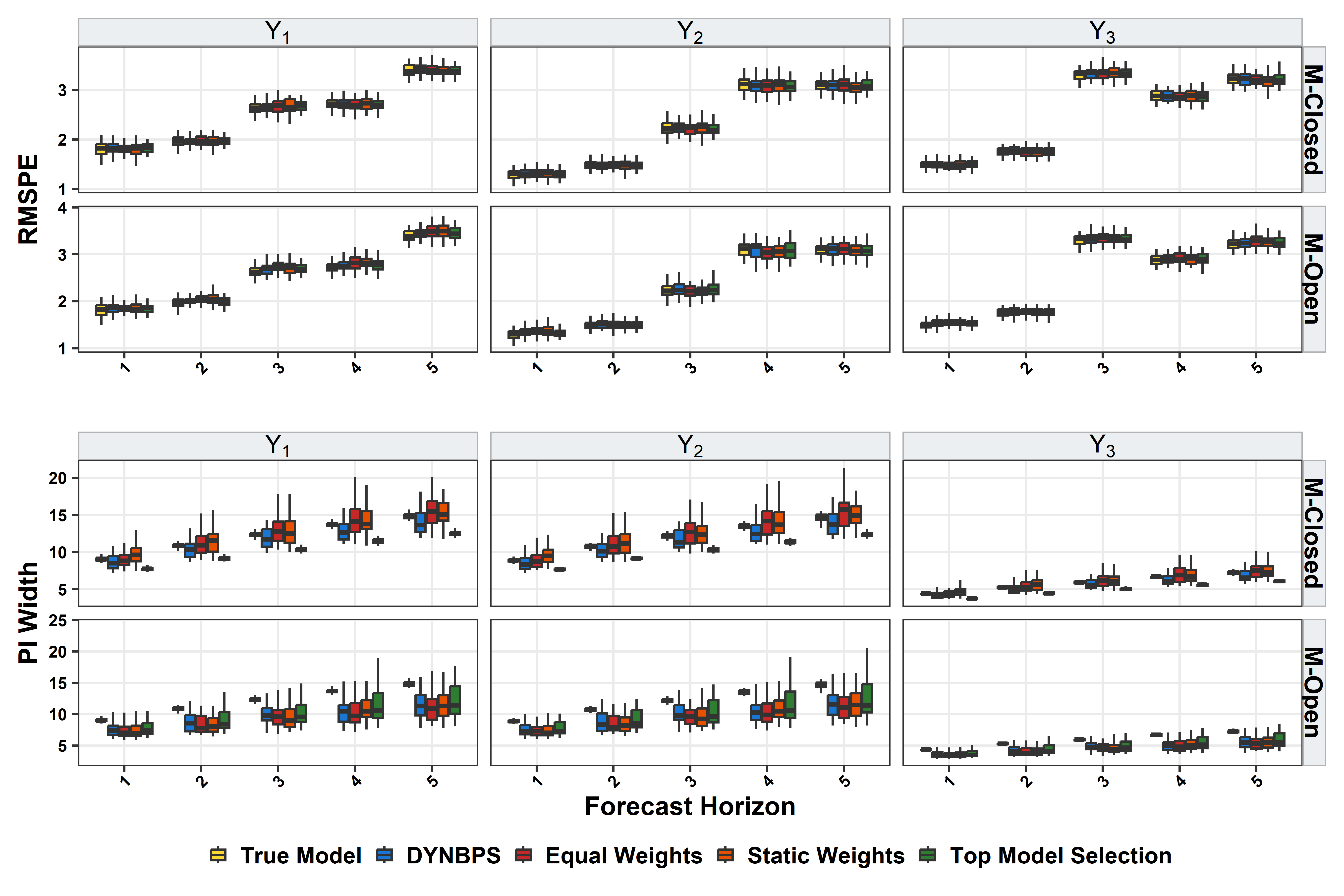}
    \caption{ Root mean squared prediction error (\textsc{rmspe}) and predictive interval width at each future time point for each component of the response matrix $Y$, under the considered model configurations and 50 replications.}
    \label{fig:Msettings_pred}
\end{figure}

Each replicate consists of values of $n\times q\times T$ outcome $Y$ generated from \eqref{eq:DLM_spatiotemporal} with $n=100$, $q=3$, $T=20$ and $p=2$, $X$ includes two predictors generated from a standard uniform distribution over $[0,1]$, $\Sigma = \begin{bsmallmatrix} 1 & -0.3 & 0.6 \\ -0.3 & 1.2 & 0.4 \\ 0.6 & 0.4 & 1 \end{bsmallmatrix}$. $\Theta=[B^{\T}:\Omega^{\T}]^{\T}$ was defined by allowing the system to evolve in \eqref{eq:DLM_spatiotemporal}, starting with a realization from the marginal prior distribution. We defined the joint prior distribution for all models $\mathscr{M}_{j}$ regarding $\Theta, \Sigma$ from the matrix-normal inverse Wishart family, with parameters $m_{0}=0_{(p+n)\times q}$,  $C_{0}=\begin{bsmallmatrix} 0.05\mathbb{I}_{p} & 0 \\ 0 & \mathcal{K}(\mathcal{S}, \mathcal{S}; \phi=1) \end{bsmallmatrix}$, $\Psi_{0}=10\mathbb{I}_{q}$, and $\nu_{0}=q+1$.
The spatial correlation matrix is specified using an exponential correlation function with $\phi=4$, and we set $\alpha=0.8$.

% \begin{figure}[p]
%     \centering
%     \begin{minipage}{\textwidth}
%         \centering
%         \includegraphics[width=\textwidth]{images/review/plot_theta_C.png}
%         \small (a) $\mathscr{M}$-Closed
%     \end{minipage}
%     \vspace{1em}
%     \begin{minipage}{\textwidth}
%         \centering
%         \includegraphics[width=\textwidth]{images/review/plot_theta_O.png}
%         \small (b) $\mathscr{M}$-Open
%     \end{minipage}
%     \caption{Coverage ($95\%$) and Frobenius norm at each time point for regression coefficients $B$, i.e. $\Theta_{1:p,1:q}$, under the considered model configurations and 50 replications in (a) $\mathscr{M}$-closed and (b) $\mathscr{M}$-open settings.}
%     \label{fig:Msettings_theta}
% \end{figure}

Figure~\ref{fig:Msettings_pred} reports root mean squared prediction error (\textsc{rmspe}) and predictive interval (\textsc{pi}) width across forecast horizons for each of the $q=3$ outcomes, under both $\mathscr{M}$-closed and $\mathscr{M}$-open settings. Boxplots display the distribution of each metric across the 50 replicates at any instant in the forecast horizon, i.e., future time points. In terms of \textsc{rmspe}, all methods achieve equivalent performance, with \textsc{dynbps} closely matching \textsc{true} across outcomes and horizons in both settings. Differences emerge for \textsc{pi} width: \textsc{dynbps} produces narrower, more calibrated intervals compared to \textsc{ew} and \textsc{sw}, which tend to be more conservative as the forecast horizon grows. \textsc{tms} lies between the two extremes. The $\mathscr{M}$-open setting induces greater predictive variability across all methods, though results remain broadly comparable with the $\mathscr{M}$-closed case, confirming the predictive aptitude of dynamic \textsc{bps} regardless of the model class.

\begin{figure}[t!]
    \centering
    \includegraphics[width=\linewidth]{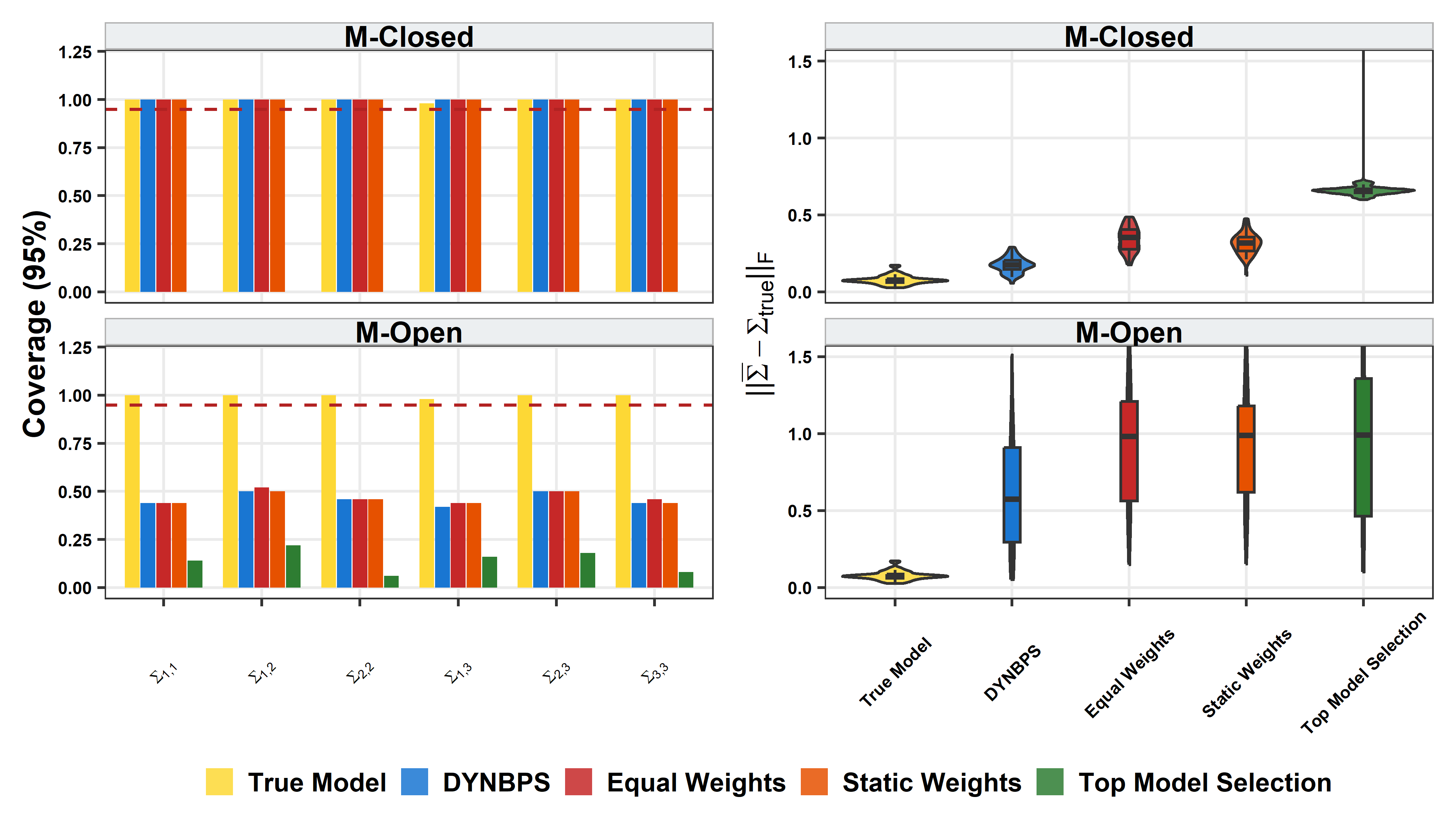}
    \caption{ Coverage ($95\%$) and Frobenius norm for $\Sigma$, under the considered model configurations and 50 replications.}
    \label{fig:Msettings_sigma}
\end{figure}

% Figures~\ref{fig:Msettings_theta}, \ref{fig:Msettings_omega}, and~\ref{fig:Msettings_sigma} provide insight into the posterior inference achieved by all methods under $\mathscr{M}$-closed and $\mathscr{M}$-open settings. 
Figures~\ref{fig:Msettings_theta}, \ref{fig:Msettings_omega}, reported in Section~\ref{sec:appendixD_graphics} of the Supplementary Material due to space constraints, and Figure~\ref{fig:Msettings_sigma} provide insight into the posterior inference achieved by all methods under $\mathscr{M}$-closed and $\mathscr{M}$-open settings.
Each figure reports empirical coverage at the $95\%$ nominal level and the distribution of the Frobenius norm across the 50 replicates.

% 
% \begin{figure}[p]
%     \centering
%     \begin{minipage}{\textwidth}
%         \centering
%         \includegraphics[width=\textwidth]{images/review/plot_omega_C.png}
%         \small (a) $\mathscr{M}$-Closed
%     \end{minipage}
%     \vspace{1em}
%     \begin{minipage}{\textwidth}
%         \centering
%         \includegraphics[width=\textwidth]{images/review/plot_omega_O.png}
%         \small (b) $\mathscr{M}$-Open
%     \end{minipage}
%     \caption{Coverage ($95\%$) and Frobenius norm at each time point for each component of the multivariate spatial process $\Omega$, under the considered model configurations and 50 replications.}
%     \label{fig:Msettings_omega}
% \end{figure}

Starting from time-dependent parameters, Figure~\ref{fig:Msettings_theta} shows the identification of the regression coefficient submatrix $B_{t}$ across methods. The bottom panel reports the Frobenius norm $\parallel\hat{B_{t}} - B_{t}\parallel_{F}=\sqrt{\sum_{i=1}^{p}\sum_{j=1}^{q}\mid \hat{\beta}_{i,j,t} - \beta_{i,j,t} \mid^2}$, where $\hat{B}_t$ and $\hat{\beta}_t$ represent the maximum a posteriori (\textsc{map}) estimate of $B$ and its element $\beta_{i,j}$; a colored line represents the median of the Frobenius norm distribution across time points for each method. 
In the $\mathscr{M}$-closed setting, coverage and the Frobenius norm are nearly identical across all methods. If the true model belongs to the candidate set, the specific weighting scheme does not seem to strongly affect inference. The picture changes in the $\mathscr{M}$-open setting, where \textsc{dynbps} consistently achieves the lowest Frobenius norm among weighting strategies, \textsc{ew} and \textsc{sw} exhibit pronounced increases in estimation error around $t=8$-$12$, and \textsc{tms} shows the highest variability. Coverage remains broadly comparable across methods in both settings, with greater fluctuations observed for the off-diagonal elements $\Theta_{1,3}$ and $\Theta_{2,3}$ in the open case. This is expected as sampling widely different $\{\alpha,\phi\}$ values across replications inflates variability in diffusion metrics, and only \textsc{dynbps} dynamically adapts its weights smoothly to track these changes.

Figure~\ref{fig:Msettings_omega} collects coverage and Frobenius norm for each of the $q=3$ components of the multivariate spatial process $\Omega$. Under the $\mathscr{M}$-closed setting, all methods achieve nearly identical results. Coverage is near the nominal $95\%$ level for $\Omega_{1}$ and $\Omega_{2}$, with mild under-coverage for $\Omega_{3}$, and Frobenius norm medians are indistinguishable across methods. The $\mathscr{M}$-open setting reveals a net separation in that \textsc{dynbps} achieves the lowest Frobenius norm among weighting strategies throughout the observation window, \textsc{ew} exhibits a pronounced peak in estimation error around $t=8$-$10$, \textsc{sw} follows a similar trajectory, and \textsc{tms} shows the greatest variability overall. Coverage for $\Omega_{1}$ and $\Omega_{2}$ remains broadly comparable across methods, while $\Omega_{3}$ shows greater sensitivity to model misspecification for all methods. 
% Overall, Figure~\ref{fig:Msettings_omega} confirms that \textsc{dynbps} provides the most effective recovery of multivariate spatial dependence, with superior uncertainty quantification compared to the ablation baselines under $\mathscr{M}$-open conditions.

For the dense covariance matrix $\Sigma$ (the non-spatial dependence among the $q$ variables), coverage is reported in Figure~\ref{fig:Msettings_sigma} per element $\Sigma_{i,j}$, $i,j=1,\dots,q$, and the Frobenius norm at the last observed time $T$. Under $\mathscr{M}$-closed, all methods achieve near-nominal or slightly above nominal results, with a clear hierarchy in Frobenius norm: \textsc{true} $\ll$ \textsc{dynbps} $<$ \textsc{ew} $\approx$ \textsc{sw} $<$ \textsc{tms}. The $\mathscr{M}$-open setting reveals a sharp deterioration for all stacking methods. Coverage drops below the nominal levels for \textsc{dynbps}, \textsc{ew}, and \textsc{sw} across all elements of $\Sigma$, while \textsc{tms} collapses to near-zero coverage. 
% This sharpens the pattern already observed for $B$ and $\Omega$. 
In the open setting, weight misspecification translates directly into inferential failure for $\Sigma$, with \textsc{tms} paying the steepest price for its hard model-selection strategy. Nevertheless, as confirmed in Figure~\ref{fig:Msettings_pred}, the lower coverage for $\Sigma$ does not substantially affect predictive inference. 
It is worth noting that the $\mathscr{M}$-open simulation is deliberately challenging: candidate parameters are redrawn uniformly over $\alpha\in(0.5,1)$ and $\phi\in[1,50]$, so that across replications some models may be placed far from the true values $\{\alpha=0.8,\,\phi=4\}$. 
% Under such conditions, the degree of inferential degradation observed for $\Sigma$ remains moderate.

We are not surprised that the worst inferential performances are related to the common-covariance matrix $\Sigma$. Indeed, when working with georeferenced data, it is well known that covariance parameters are not consistently estimable under fixed-domain asymptotics \citep{zhang_inconsistent_2004}, especially when the data exhibit multidimensional dependencies. These findings are consistent with \cite{stein_asymptotically_1988} and \cite{stein_asymptotic_1989}, which theoretically established that Gaussian processes can deliver good predictive performance even with misspecified covariance functions in fixed domains.

% ################################################
\section{Case study results}\label{sec:dataappl}

% The \href{https://www.copernicus.eu}{Copernicus Data Space Ecosystem (CDSE)} is an open, cloud-based ecosystem for Earth observation data, providing free and instant access to data from the Copernicus Program, especially the Sentinel satellites, along with tools for processing and analysis. It serves as a central platform that enables users, from researchers to developers, to explore, access, and process Earth data for various applications, supporting the development of new value-added services. 
% The \textsc{cdse} provides standardized access to Copernicus climate archives, offering a reproducible large-scale environment well-suited for evaluating spatiotemporal statistical methodology at scale.

% Easy accessibility to such data generates scientific questions relevant to how underlying processes manifest themselves in observed data and, more specifically, the role of different types of associations in such manifestations. Access to massive volumes of data, along with accompanying machine learning tools, requires statistical models that are congruent with frameworks driven by Artificial Intelligence (\textsc{ai}). We analyze four key Copernicus variables that are associated with each other and vary over space and time: 2-meter air temperature (degrees Celsius), total monthly precipitation ($mm$), 10-meter wind speed ($ms^{-1}$), and monthly evaporation ($mm$). These variables collectively represent essential surface climate processes that are relevant to hydrology, energy balance, and atmospheric dynamics.

The \href{https://www.copernicus.eu}{Copernicus Data Space Ecosystem (CDSE)} provides free, standardized access to Earth-observation data from the Copernicus Programme. It offers a reproducible, large-scale environment well-suited for evaluating spatiotemporal statistical methodology at scale. The sheer volume of such data calls for models that remain accurate and scalable under massive spatiotemporal dimensions. We analyze four key Copernicus variables that are associated with each other and vary over space and time: 2-meter air temperature (degrees Celsius), total monthly precipitation ($mm$), 10-meter wind speed ($ms^{-1}$), and monthly evaporation ($mm$). These variables collectively represent essential surface climate processes that are relevant to hydrology, energy balance, and atmospheric dynamics.

Our statistical modeling approach incorporated a spatiotemporal regression structure in which we included monthly cloud-coverage percentage (in~\%), monthly surface solar radiation (in~W\,m$^{-2}$, representing the monthly mean incident shortwave radiation on a horizontal surface), monthly sea-level pressure (in~hPa), and monthly near-surface specific humidity (in~g\,kg$^{-1}$, representing the monthly amount of moisture in the air near the surface divided by the total mass of air plus moisture) as predictors to control for broad-scale environmental variability. 
We focus our analysis of four key climate variables of interest on the European region bounded by latitudes 30\textdegree, 60\textdegree N and longitudes -10\textdegree,40\textdegree E, using complete monthly records from January 2003 to December 2014, which is the latest available data point with full variable coverage. This yields $T=144$ time points, of which the last $h=24$ months are held for predictive evaluation, leaving $120$ months to comprise a learning set. From this spatiotemporal domain, we use $n=500$ regularly distributed locations for training and keep an additional $100$ locations to assess spatial interpolation, aggregated for $q=4$ response variables. The resulting data consist of $500\times 120\times 4=240,000$ spatiotemporal observations. The maximum inter-site distance among the monitored locations is approximately $6,320$ kilometers.
An extensive exploratory data analysis accompanies this dataset, offering a detailed characterization of its spatiotemporal dependence structure (see Section~\ref{sec:appendix_eda}).

\begin{figure}[t!]
\begin{subfigure}{1\textwidth}
\centering
\includegraphics[width=0.6\linewidth]{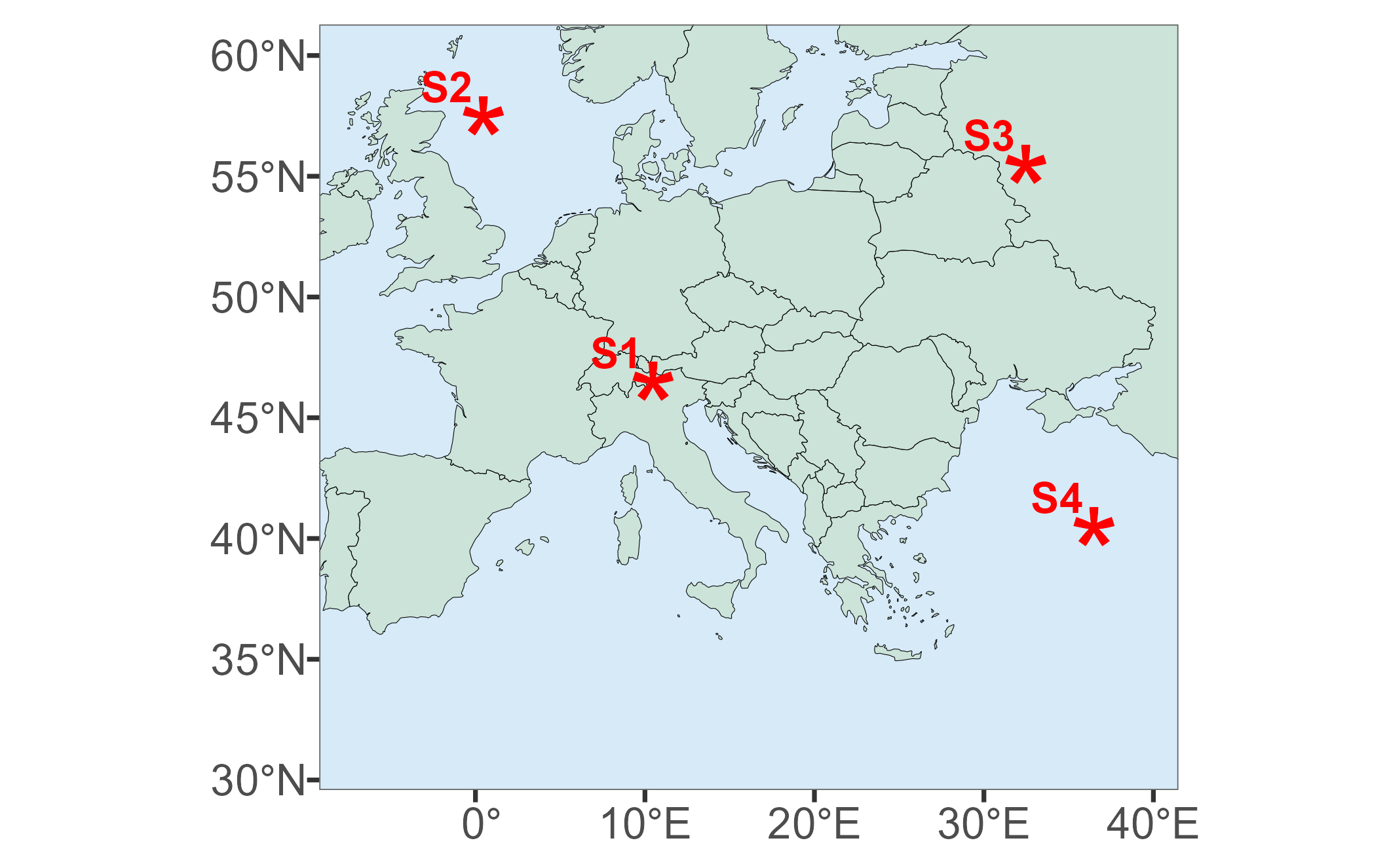} 
\caption{Randomly selected locations over the central Europe region}
\end{subfigure}
\begin{subfigure}{1\textwidth}
\centering
\includegraphics[width=\linewidth]{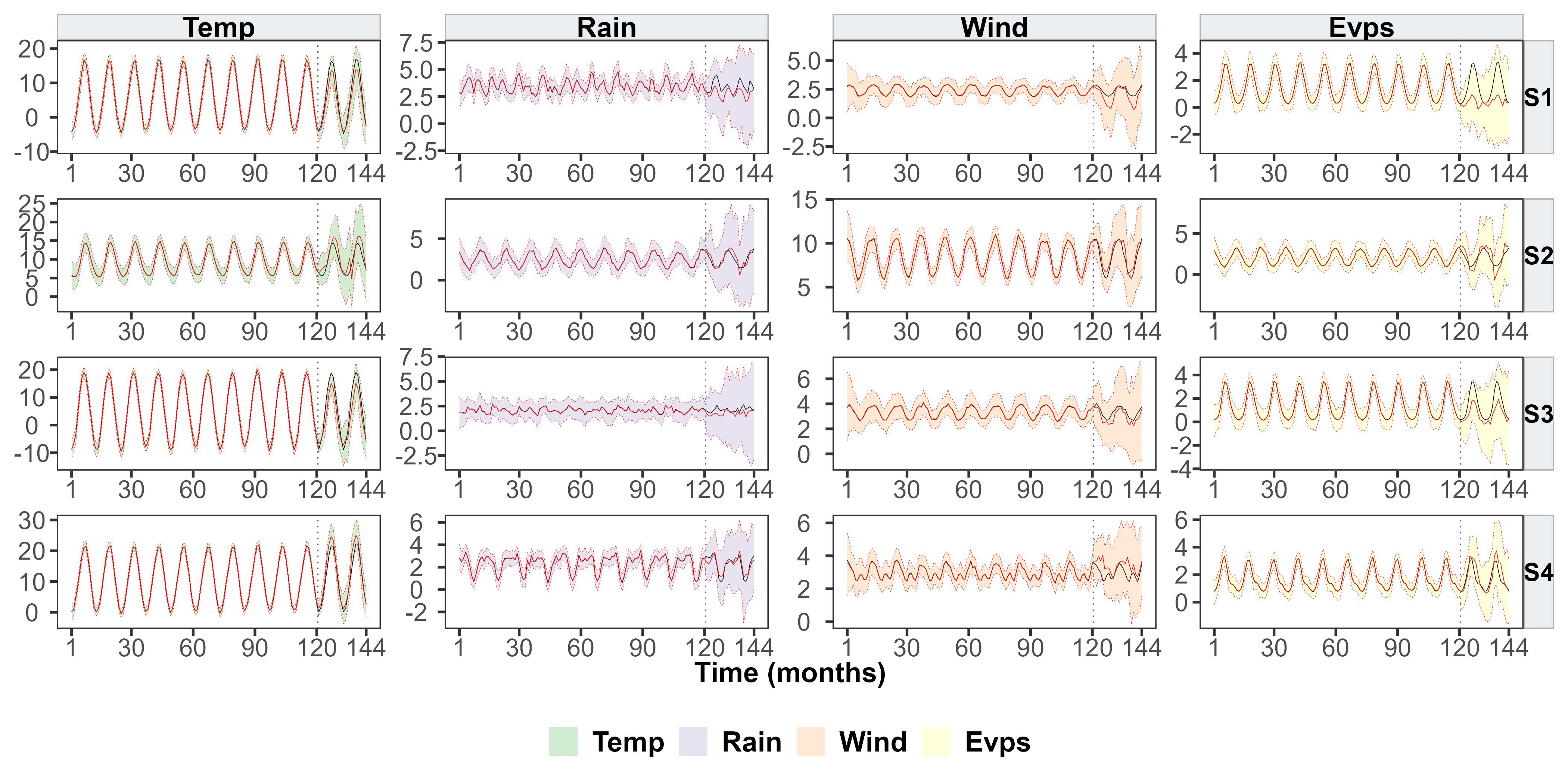}
\caption{Forecasts reporting: \textsc{map} (red solid line), $95\%$ credible interval (red dashed line), and ground truth (black solid line), for selected points}
\end{subfigure}
\caption{Average monthly forecasts at future time points for selected spatial points}
\label{fig:forecast_points}
\end{figure}

The use of these covariates to control for the large-scale atmospheric driver factors is widely supported in the literature for modeling climate response variables \citep[see e.g.,][]{jacobs_goes_2009,tramblay_brief_2011,van_osnabrugge_contribution_2019,felsche_applying_2021,chen_machine_2024}. We omit an intercept term deliberately, as we introduce seasonal dummy variables.
To account for seasonal effects, we apply structural modifications to the matrices $F_{t}$ and $G_{t}$. As we observed monthly seasonality in the outcomes, at any time instant $t$ we define $F_{t}=[X_{t}:\Delta_{t}:\mathbb{I}_{n}]$, where $\Delta_{t}=[\delta_{1,t},\dots,\delta_{11,t}]$, and each of $\delta_{i,t}=\mathds{1}_{(i=t)}(i)$, i.e., $\delta_{i,t}$ is a vector of ones when the month at time point $t$ is the same as $i$. Then, we removed the intercept from $X_{t}$, and we incorporate monthly dummy variables, such that $\tilde{X}_{t}=[X_{t}:\Delta_{t}]=[x_{1,t},x_{2,t},x_{3,t},x_{4,t},\delta_{1,t},\dots,\delta_{11,t}]$, specifying $11$ variables to ensure full rank for $\tilde{X}_{t}$ and $F_{t}=[\tilde{X}_{t}:\mathbb{I}_{n}]$, corresponding to total number of predictors $p=15$. This led to the formulation $\Theta_{t}=[B_{t}^{\T}:S_{t}^{\T}:\Omega_{t}^{\T}]^{\T}$, where $S_{t}$ is the $11\times q$ matrix of seasonal effects. The evolution matrix $G_{t}$ only requires an augmentation of its dimension, induced by passing from $X_{t}$ to $\tilde{X}_{t}$ in Model~\eqref{eq:DLM_spatiotemporal}.
For the spatial covariance function at any month, we choose to use an exponential kernel, which reflects the spatial process smoothness typically observed over large-scale domains.
The temporal structure via monthly indexing enables the model to learn seasonal and interannual dynamics; this setup facilitates both interpolation and prediction tasks within a coherent statistical framework. 
We implement a fully automated procedure setting the grid of values for $\{\alpha,\phi\}$, which makes use of spatiotemporal variograms, see Section~\ref{sec:appendix_eda}. The process ends by providing $J=4$ candidate models: the variogram estimate missing nugget for the outcomes, then it automatically fixes $\alpha=0.999$, while letting the range parameter vary as $\phi\in\{0.215,0.284,0.304,0.324\}$. We set the prior distribution to non-informative choices typically used for matrix-variate models in spatial analysis \citep[see e.g.,][]{zhang_spatial_2022,presicce_bayesian_2024}. We give on $\Theta,\Sigma$ a matrix-normal inverse Wishart prior, where parameters were fixed as: $m_{0}=0_{(p+n)\times q}, C_{0}=\begin{bsmallmatrix}\mathbb{I}_{p} & 0_{p\times n} \\ 0_{n\times p} & \mathcal{K}_{0}(\mathcal{S},\mathcal{S};\phi=1) \end{bsmallmatrix}$, opting for an exponential spatial correlation function, and $\nu_{0}=q,\Psi_{0}=\mathbb{I}_{q}$.

For a comprehensive comparison, we benchmark \textsc{dynbps} against four alternative methods, all applied to the same \textsc{dlm} formulation in Equation~\eqref{eq:DLM_spatiotemporal}, design matrices, spatial structure, and prior specification described above. 
(\textit{i})~ Empirical Bayes (\textsc{eb}): optimizes marginal likelihood to estimate $\{\alpha,\phi\}$ via a batch implementation using all available training data. 
(\textit{ii})~\textsc{mcmc}: a standard batch Markov chain Monte Carlo sampler targeting the full \textsc{dlm} posterior over the entire training set, run for $2000$ iterations, discarding the first $1000$ as burn-in, as in the simulation study of Section~\ref{sec:sim_mcmc}. The chain explores $\{\alpha,\phi\}$, with $(\Theta,\Sigma)$ available in closed form conditionally. The effective sample sizes ($247$ for $\phi$ and $118$ for $\alpha$, out of $1000$ retained draws) show no evidence of convergence issues. Longer chains were not considered given the running time reported in Table~\ref{tab:comparison}.
% (\textit{ii})~\textsc{mcmc}: a standard batch Markov chain Monte Carlo sampler targeting the full \textsc{dlm} posterior over the entire training set, run for $2000$ iterations. 
(\textit{iii})~Empirical Orthogonal Function (\textsc{eof}): applies an empirical orthogonal function decomposition to the multivariate response matrix at the monitored locations to reduce its dimension, and then fits a \textsc{dlm}. We opt for $K=10$ components, explaining more than $90\%$ of the variance. The loadings are kriged to the withheld locations for spatial interpolation, and the predictive intervals account for the observation error of the scores.
% (\textit{iii})~Empirical Orthogonal Function (\textsc{eof}): applies an empirical orthogonal function decomposition to the multivariate response matrix to reduce its dimension, and then fits a \textsc{dlm} on leading components.
(\textit{iv})~\textsc{inla}: uses integrated nested Laplace approximations for fast approximate Bayesian inference on the multivariate \textsc{dlm}, approximating the spatial random field using a triangulation mesh with $79$ nodes. Finer meshes did not improve predictive performance: with $201$ nodes, it slightly deteriorated, possibly because the additional flexibility of the latent field is weakly identified by the data, while the running time increased from $691$ to $4,526$ seconds. Using $327$ nodes, the fit could not be completed due to memory limitations instead.
% (\textit{iv})~\textsc{inla}: uses integrated nested Laplace approximations for fast approximate Bayesian inference on the multivariate \textsc{dlm}, approximating the spatial random field using a triangulation mesh with $79$ nodes.

To isolate the contribution of time-varying weights, we also consider three ablations of \textsc{dynbps}: (\textit{v})~equal weights (\textsc{dynbps-ew}), which assigns $w_{j}=1/J$ to every candidate model at every time; (\textit{vi})~static weights (\textsc{dynbps-sw}), a single time-invariant weight vector obtained by optimizing the \textsc{bps} objective once over the whole learning set; and (\textit{vii})~top model selection (\textsc{dynbps-tms}), which at each time point assigns unit weight to the candidate model with the largest weight and zero weight to the others. All three reuse the same forward-filtering output, grid of $\{\alpha,\phi\}$, prior, and predictive samplers as \textsc{dynbps}, so they differ only in how the candidate models are weighted afterwards.
Predictive performance is measured using root mean square prediction error (\textsc{rmspe}), empirical coverage at the $95\%$ nominal level, interval score (\textsc{is}), energy score (\textsc{es}), and total running time.

\begin{table}[t!]
\centering
\smallskip
\begin{subtable}[t]{\linewidth}
\centering\small
\begin{tabular}{>{\raggedright\arraybackslash}p{3.0cm}
                >{\raggedleft\arraybackslash}p{1.5cm}
                >{\raggedleft\arraybackslash}p{1.8cm}
                >{\raggedleft\arraybackslash}p{2.0cm}
                >{\raggedleft\arraybackslash}p{1.8cm}
                >{\raggedleft\arraybackslash}p{2.2cm}}
\toprule
& \multicolumn{4}{c}{Predictive metrics} & \\
\cmidrule(lr){2-5}
Method
  & \textsc{rmspe} & Coverage & \textsc{is} & \textsc{es}
  & Time (s) \\
\midrule
\textsc{dynbps}
  & \textbf{1.918} & \textbf{0.947} & \textbf{10.053} & \textbf{2.247}
  & 36 \\
\textsc{dynbps-ew}
  & 1.930 & 0.940 & 10.266 & 2.493
  & 35 \\
\textsc{dynbps-sw}
  & 1.930 & 0.942 & 10.160 & 2.302
  & 36 \\
\textsc{dynbps-tms}
  & 1.919 & 0.940 & 10.060 & 2.277
  & 35 \\
\textsc{eb}
  & 2.341 & 0.996 & 18.277 & 3.563
  & $2,889$ \\
\textsc{mcmc}
  & 2.215 & 0.979 & 13.694 & 2.949
  & $231,041$ \\
\textsc{eof}
  & 2.782 & 0.966 & 16.595 & 3.425
  & \textbf{5} \\
\textsc{inla}
  & 3.192 & 0.854 & 21.406 & 4.154
  & $691$ \\
\bottomrule
\end{tabular}
\caption{Temporal forecast (averaged over $n=500$ locations, $h=24$ future time points, and $q=4$ variables)}
\label{tab:temporal}
\end{subtable}

\vspace{0.5cm}

\begin{subtable}[t]{\linewidth}
\centering\small
\begin{tabular}{>{\raggedright\arraybackslash}p{3.0cm}
                >{\raggedleft\arraybackslash}p{1.5cm}
                >{\raggedleft\arraybackslash}p{1.8cm}
                >{\raggedleft\arraybackslash}p{2.0cm}
                >{\raggedleft\arraybackslash}p{1.8cm}
                >{\raggedleft\arraybackslash}p{2.2cm}}
\toprule
& \multicolumn{4}{c}{Predictive metrics} & \\
\cmidrule(lr){2-5}
Method
  & \textsc{rmspe} & Coverage & \textsc{is} & \textsc{es}
  & Time (s) \\
\midrule
\textsc{dynbps}
  & \textbf{1.336} & \textbf{0.963} & \textbf{8.417} & \textbf{1.491}
  & 36 \\
\textsc{dynbps-ew}
  & 1.373 & 0.973 & 8.687 & 1.647
  & 35 \\
\textsc{dynbps-sw}
  & 1.372 & 0.971 & 8.692 & 1.646
  & 36 \\
\textsc{dynbps-tms}
  & 1.367 & 0.975 & 8.739 & 1.649
  & 35 \\
\textsc{eb}
  & 1.655 & 0.930 & 13.077 & 2.452
  & $2,889$ \\
\textsc{mcmc}
  & 1.500 & 0.918 &  9.848 & 1.937
  & $231,041$ \\
\textsc{eof}
  & 4.265 & 0.894 & 24.989 & 5.450
  & \textbf{5} \\
\textsc{inla}
  & 2.973 & 0.835 &  20.347 &  3.475
  & $691$ \\
\bottomrule
\end{tabular}
\caption{Spatial interpolation (averaged over $u=100$ withheld locations, multiple in-sample and out-of-sample time points, and $q=4$ variables)}
\label{tab:spatial}
\end{subtable}
\caption{Predictive evaluation metrics for the Copernicus case study, comparing \textsc{dynbps} with three ablations (\textsc{dynbps-ew} = equal weights, \textsc{dynbps-sw} = static weights, \textsc{dynbps-tms} = top model selection) and four competitors. \textsc{rmspe}: root mean square prediction error; Coverage: empirical, at $95\%$ nominal level (closest to $0.95$ is best); \textsc{is}: interval score; \textsc{es}: energy score. Running time in seconds. Bold: best per column.}
\label{tab:comparison}
\end{table}

% We train the model over the entire observed multivariate time series and subsequently conduct one-step-ahead forecasting for the full temporal sequence, including both observed and unobserved time points. 
All methods are fitted on the learning set of the first $120$ months at the $n=500$ monitored locations. Temporal forecasts are evaluated over the last $h=24$ months (from $1$ to $24$ steps ahead of the end of the learning set) at the monitored locations, whereas spatial interpolation is evaluated at the $u=100$ withheld locations, at the last month of the learning set and at $1$, $6$, $12$ and $24$ months ahead.
Tables~\ref{tab:temporal} and~\ref{tab:spatial} report evaluation metrics for all methods. 
We see that \textsc{dynbps} achieves the best performance on every predictive criterion in both tasks.
It attains the lowest \textsc{rmspe}, interval score, and energy score, with empirical coverage close to the nominal $95\%$ level ($0.947$ for temporal prediction, $0.963$ for spatial interpolation).
Relative to its ablations, \textsc{dynbps} attains comparable \textsc{rmspe}, \textsc{is} and \textsc{es}, and the empirical coverage closest to the nominal level, in both tasks (Tables~\ref{tab:temporal} and~\ref{tab:spatial}). The margins are moderate, which is expected because the variogram-informed grid yields $J=4$ candidates that share $\alpha=0.999$ and differ only through $\phi$ within a narrow range, so the candidate models are similar. 
Overall, \textsc{dynbps} weights deliver consistently, though modestly, better-calibrated predictions on this dataset, while their advantage is larger in the $\mathscr{M}$-open simulations of Section~\ref{sec:sim_Msettings}, where the candidate models differ substantially.
In contrast, \textsc{eof} has a larger \textsc{rmspe} in both tasks ($2.78$ against $1.92$ in temporal forecasting and $4.27$ against $1.34$ in spatial interpolation) and presents undercoverage in spatial interpolation ($0.894$), while \textsc{eb} yields overly conservative intervals in temporal forecasting ($0.996$).
% In contrast, \textsc{eof} yields erroneously calibrated intervals ($0.241$ approximate temporal coverage), while \textsc{eb} overestimates coverage ($0.996$), which is indicative of overfitting. 
The computational advantage of \textsc{dynbps} is clear: full fitting and predictive sampling in $\sim 36$ seconds, compared to $48$ minutes for \textsc{eb} and over $64$ hours for \textsc{mcmc}, with no discernible loss in predictive performance. 
% Forecasting was evaluated over the held-out future points of the time series for each of the four variables. 
Figure~\ref{fig:forecast_points} demonstrates accurate predictive behavior, with forecast trajectories closely aligned with ground-truth values for a random selection of equally spaced points over Europe.

\begin{figure}[t!]
    \centering
    \includegraphics[width=\linewidth]{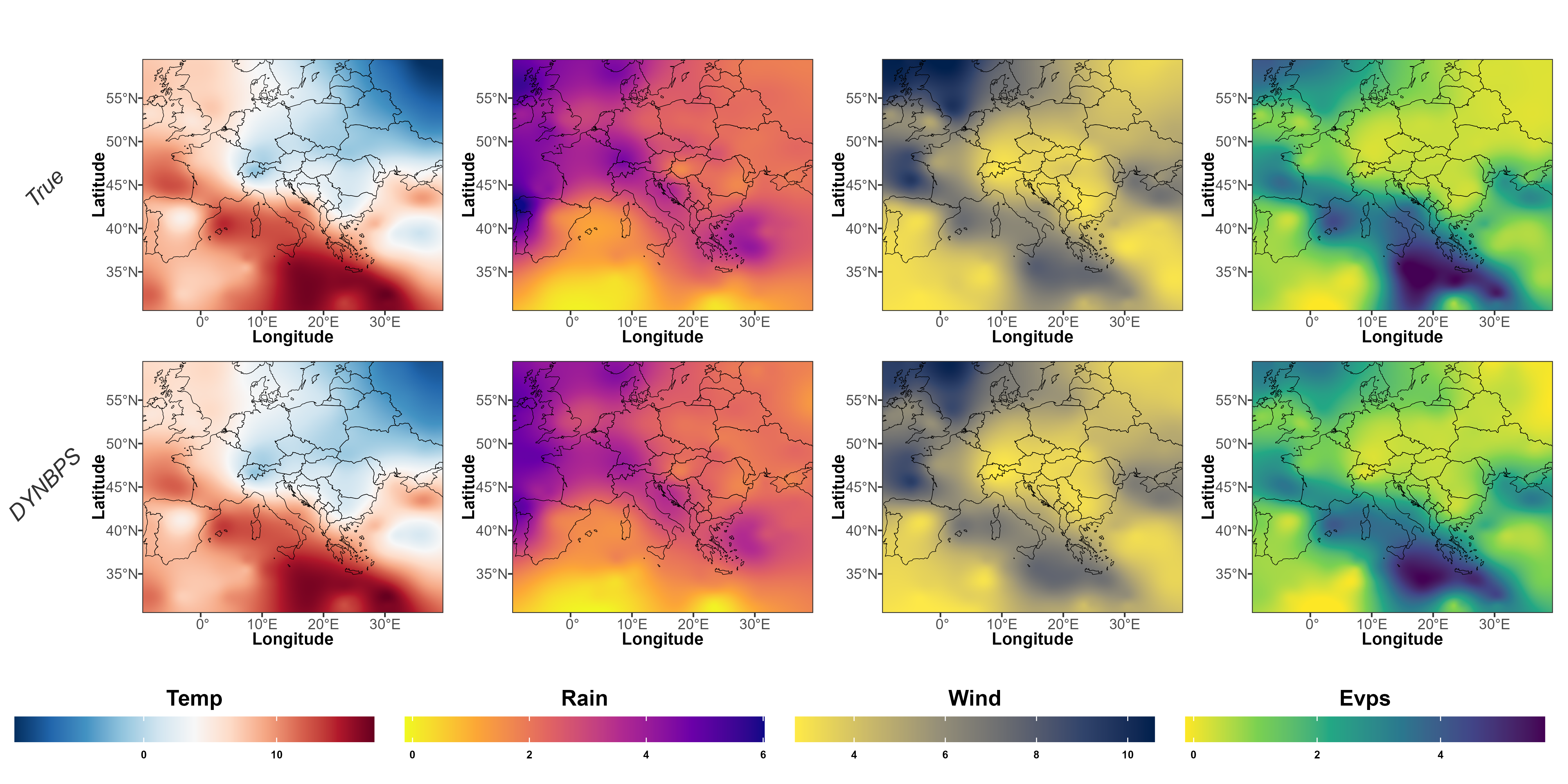}
    \caption{Spatial surface interpolation at unobserved time (out of sample, December~2014, $h=24$): true spatial surfaces (top row) and \textsc{dynbps} predictions (bottom row) for all four climate variables.}
    \label{fig:interpolation_out}
\end{figure}

Forecast surfaces for each variable are provided in Figures~\ref{fig:forecast_temp}-\ref{fig:forecast_evps} of the Supplemental Material.
% In addition to temporal forecasting, spatial interpolation was assessed at $u=100$ withheld test locations at both observed and future unobserved times. We conducted spatial prediction at observed time points (in-sample) and at future unobserved times (out-of-sample), providing a robust assessment of the model’s ability to generalize spatial interpolation beyond its learning set. 
Multivariate interpolated maps exhibited smooth gradients and high alignment with known climatological patterns, suggesting that the model accurately captures both temporal variability and spatial structure. 
Figure~\ref{fig:interpolation_in} (Supplemental Material) presents interpolated surfaces for a selected observed time slice (December~2012, $t=120$) for all four variables, comparing the true spatial fields (top row) with \textsc{dynbps} predictions (bottom row). Similarly, Figure~\ref{fig:interpolation_out} reports spatial predicted surfaces for a selected out-of-sample month (December~2014, $h=24$) in the same layout.
 
The top panel in Figure~\ref{fig:forecast_points} shows the exact georeferencing for each of the points included in the temporal series forecast in the bottom panel. As expected, out-of-sample predictions show greater variability for predictive intervals, although the empirical coverage is close to the nominal $95\%$ level, as confirmed by Table~\ref{tab:temporal}. 
For any of the considered points, the model retraces the truth very closely, capturing the seasonality and providing good extrapolations.

% Table~\ref{tab:spatial} evaluates spatial prediction for \textsc{dynbps}. It achieves the lowest \textsc{rmspe} ($1.34$), interval score ($8.50$), and energy score ($1.49$) across all competing methods and evaluation times. In addition, w
We compare Figure~\ref{fig:interpolation_in} with Figure~\ref{fig:interpolation_out} for in-sample and out-of-sample interpolation corresponding to the month of December. These figures depict strong similarities, implying the model's ability to learn spatial patterns without losing seasonal coherence.

Overall, these results demonstrate the scalability of \textsc{dynbps} for joint multivariate forecasting and interpolation across the European continent.
% Dynamic \textsc{bps} brings the feasibility of combining physically interpretable multidimensional factors with data-driven spatiotemporal modeling to a new level, supporting climate monitoring and decision-making within \textsc{geoai} systems. 
% As reported above and detailed in Section~\ref{sec:computational}, the entire estimation and prediction pipeline runs on a standard laptop with very limited resources, enabling almost-automated multivariate spatiotemporal Bayesian modeling for large-scale settings in constrained \textsc{geoai} systems.

% ################################################
\section{Discussion}\label{sec:discuss}

We have devised a spatiotemporal statistical model tailored to readily analyze large-scale problems, minimizing user effort. The contribution harnesses analytically accessible matrix-variate statistical distributions in conjunction with a novel introduction: the dynamic Bayesian predictive stacking. The \textsc{dynbps} delivers prompt inference by avoiding simulation-based algorithms that often require extensive human tuning. Our proposed spatiotemporal dynamic modeling approach relies on Markovian propagation across spatial datasets through a modification of the standard \textsc{ffbs} procedure to process massive amounts of online spatial data on high-performance CPU architectures. Nevertheless, \textsc{dynbps} demonstrates extremely useful computational features even for limited architectures.
Some additional remarks are warranted. The development here has been elucidated with a hierarchical matrix-variate spatiotemporal process within dynamic linear model frameworks. Although modeling simplifications have been introduced to minimize human intervention, we emphasize that \textsc{dynbps} seamlessly applies to more versatile but analytically intractable models. 
 
Future research can build upon \cite{tallman_bayesian_2023} and \cite{cabel_bayesian_2025} to enrich spatiotemporal dependence structures within \textsc{dynbps}. We are also planning to study and introduce concepts from Bayesian predictive synthesis \citep{mcalinn_dynamic_2019,mcalinn_multivariate_2020} into dynamic Bayesian predictive stacking, thereby enlarging the class of solvable problems without compromising uncertainty quantification on any spatial hyperparameters.
Disseminating our proposed product, along with our accompanying software, available as an \texttt{R} package on CRAN, is expected to significantly boost dynamic spatiotemporal modeling.
% Future directions will also explore the perceived potential of \textsc{dynbps} as a feeder for emerging amortized inference methods \citep{ganguly_amortized_2023,zammit-mangion_neural_2024, sainsbury-dale_neural_2024} to achieve Bayesian inference. Rapid delivery of posterior estimates of the entire spatiotemporal process from \textsc{dynbps} will provide more training data for amortized neural learners, which can result in accelerated tuning for subsequent Bayesian inference. We do not see our proposed approach as a competitor to, but rather as supplementary to, amortized neural inference. 

% Such developments will be pursued as future research. We also seek to expand and fully investigate automated \textsc{dynbps} using Markovian graphical structures across data subsets to further expedite and improve Geospatial \textsc{ai} systems. 

% ################################################
\section*{Supplementary Materials}
The online supplement includes derivations and theoretical details on dynamic Bayesian predictive stacking (Section~\ref{sec:appendixA_theory}), computational methods and algorithms (Section~\ref{sec:appendixB_algs}), supplementary simulation experiment results (Section~\ref{sec:appendixC_sim}), exploratory data analysis, and automated hyperparameter settings (Section~\ref{sec:appendix_eda}), and additional graphics (Section~\ref{sec:appendixD_graphics}).
All computer programs required to reproduce the analysis are publicly accessible at the GitHub repository
%\href{https://github.com/lucapresicce/Markovian-Spatiotemporal-Propagation}{https://github.com/lucapresicce/Markovian-Spatiotemporal-Propagation}
\if1\blind{\url{https://github.com/lucapresicce/Markovian-Spatiotemporal-Propagation}}\fi\if0\blind{\texttt{REDACTED}}\fi
.

\section*{Generative Artificial Intelligence (AI)} 
No generative AI tools were used in the production of this manuscript.

% ################################################
%% ** The bibliograhy **
\bibliographystyle{chicago}
\bibliography{reference}% place <bib-data-file> 

% --------------------------------------------------
\end{document}

\def\spacingset#1{\renewcommand{\baselinestretch}%
		{#1}\small\normalsize} \spacingset{1}
\maketitle
\spacingset{1.9}
\section*{Organization of the supplemental material}

Section~\ref{sec:appendixA_theory} provides additional results and theoretical derivations for dynamic Bayesian predictive stacking. In particular, Section~\ref{sec:dynamicBPS} presents the analytical derivation of the \textsc{dynbps} optimization problem in Equation~\eqref{eq:dynamic_BPS} (Section~\ref{sec:propagation}), while Section~\ref{sec:weightsproof} investigates the aggregated weighting strategies for \textsc{dynbps} introduced in Section~\ref{sec:propagation}.
Section~\ref{sec:appendixB_algs} describes the complementary methods and algorithms presented in Section~\ref{sec:computational}, and Section~\ref{sec:appendixC_sim} reports supplementary simulation results, including a direct comparison of computational cost and predictive accuracy between \textsc{dynbps} and \textsc{mcmc} (Section~\ref{sec:sim_mcmc}) and the impact of the proposed weighting strategies.
Section~\ref{sec:appendix_eda} provides details on non-spatial and spatial exploratory data analysis related to the case study in Section~\ref{sec:dataappl}. Finally, Section~\ref{sec:appendixD_graphics} presents complementary graphics supporting the main manuscript.

% ------------------------------------------------
\section{Theoretical derivations}\label{sec:appendixA_theory}

% %%%%%%%%%%%%%%%%%%%%%%%%%%%%%%%%%%%%%%%%%%%%%%%%
\subsection{Dynamic Bayesian predictive stacking}\label{sec:dynamicBPS}

As presented in \citep[][Section 3.1]{yao_using_2018}, and summarized in Section~\ref{sec:methodology}, the Bayesian predictive stacking problem arose as
{\small
\begin{equation}\label{eq:stacking_problem}
    \min_{w \in \mathscr{S}_1^{J}} d \left( \sum_{j=1}^{J} w_j p(\cdot \mid \mathscr{D}_{t}, \mathscr{M}_{j}), p^{\star}(\cdot \mid \mathscr{D}_{t}) \right) \;\text{ or }\;\max_{w \in \mathscr{S}_1^{J}} S \left( \sum_{j=1}^{J} w_j p(\cdot \mid \mathscr{D}_{t}, \mathscr{M}_{j}), p^{\star}(\cdot \mid \mathscr{D}_{t}) \right)\!\!.
\end{equation}}
Here, $S(\cdot\,,\cdot)$ is a proper scoring rule and $d(\cdot\,,\cdot)$ its induced divergence.
The empirical approximation to Equation~\eqref{eq:stacking_problem} proposed in the original manuscript of \cite{yao_using_2018} relies on replacing the full predictive evaluated at a new data point with its corresponding \textsc{leave-one-out} predictive distribution. In doing so, they aim to approximate the expectation of the score through a leave-one-out cross-validation estimator, avoiding overly optimistic estimates. 

Here, we define a renewed stacking problem ad hoc for dynamic contexts. This is specially made for these scenarios where time-dependence is relevant, and one-step-ahead predictive distribution is more informative than the usual predictive distribution, while \textsc{loo} fails \citep{snijders_cross-validation_1988,paul-christian_burkner_approximate_2020}. In this matter, the proper scoring rule must be assessed with respect to the true one-step-ahead predictive distribution at a generic time instant $\tau$, i.e. $p^{\star}(Y_{\tau+1} \mid \mathscr{D}_{\tau})$. We define the ``dynamic'' stacking problem as
{\small
\begin{equation}\label{eq:dynamic_stacking_problem}
     \min_{w^{(\tau)} \in \mathscr{S}_1^{J}} d \left( \sum_{j=1}^{J} w_{j}^{(\tau)} p(\cdot \mid \mathscr{D}_{\tau}, \mathscr{M}_{j}), p^{\star}(\cdot \mid \mathscr{D}_{\tau}) \right)\;\text{ or }\;\max_{w^{(\tau)} \in \mathscr{S}_1^{J}} S \left( \sum_{j=1}^{J} w_{j}^{(\tau)} p(\cdot \mid \mathscr{D}_{\tau}, \mathscr{M}_{j}), p^{\star}(\cdot \mid \mathscr{D}_{\tau}) \right)\!\!.
\end{equation}}
Since $p^{\star}(\,\cdot\,)$ is again unknown, an approximation is required. In this situation, \textsc{loocv} does not provide an accurate approximation, as it does not account for the time ordering naturally present in time series data \citep{snijders_cross-validation_1988}. Conversely, \textsc{lfo} arises in the literature to solve precisely this issue. We then pursue an approximation to the expectation of the proper scoring rule. Indeed, proper scoring rules for two probabilistic forecasts can also be written as expectation \cite[][for further details]{gneiting_strictly_2007,yao_using_2018}.
\begin{equation}\label{eq:expectation_dynamicBPS}
    S \left( \sum_{j=1}^{J} w_{j}^{(\tau)} p(\cdot \mid \mathscr{D}_{\tau}, \mathscr{M}_{j}), p^{\star}(\cdot \mid \mathscr{D}_{\tau}) \right) = \mathbb{E}_{p^{\star}}\left[  S \left( \sum_{j=1}^{J} w_{j}^{(\tau)} p(Y_{\tau+1} \mid \mathscr{D}_{\tau}, \mathscr{M}_{j}), Y_{\tau+1} \right) \right]\!\!.
\end{equation}
Implementing the leave-future-out cross-validation, we derive the subsequent approximation to the expectation in \eqref{eq:expectation_dynamicBPS}, which takes into account temporal dynamics
\begin{equation}\label{eq:approximation_dynamicBPS}
     (\hat{w}_{1}^{(\tau)},\dots,\hat{w}_{J}^{(\tau)})^{\T}=\underset{w^{(\tau)}\in\mathscr{S}_1^{J}}{\arg\max}\;\frac{1}{\tau-1}\sum_{t=1}^{\tau-1} S \left( \sum_{j=1}^{J} w_{j}^{(\tau)} p(Y_{t+1} \mid \mathscr{D}_{t}, \mathscr{M}_{j}), Y_{t+1} \right)\!\!.
\end{equation}
When using the logarithmic score, \cite{yao_using_2018} refer to it as the stacking of predictive distributions. We opt for the logarithmic score since solving~\eqref{eq:dynamic_BPS} again minimizes the Kullback-Leibler divergence from the true predictive distribution. In addition, it can be easily executed using convex optimization \citep{grant_disciplined_2005,grant_graph_2008}. This leads to the optimization problem reported in Equation~\eqref{eq:dynamic_BPS}. Similarly, we refer to it as dynamic stacking of predictive distributions.

% %%%%%%%%%%%%%%%%%%%%%%%%%%%%%%%%%%%%%%%%%%%%%%%%
\subsection{Properties of aggregated dynamic Bayesian predictive weights}\label{sec:weightsproof}

Let $W^{(\tau)}=\big\{\hat{w}_{i,j}^{(\tau)}\big\}\in[0,1]^{n\times J}$ be a row-stochastic matrix, where each row sums to one, i.e., 
\begin{equation*}
    \sum_{j=1}^{J}\hat{w}_{i,j}^{(\tau)}=1, \quad \forall i=1,\ldots,n
\end{equation*}
This matrix is defined by collecting the individual-wise dynamic Bayesian predictive stacking weights computed using Equation~\eqref{eq:dynamic_BPS}. Without loss of generality, we consider a generic time point $\tau\in\mathcal{T}$, but the same arguments are valid for each element of $\mathcal{T}$.
Each row corresponds to a spatial observation, and the entries $\hat{w}_{i,j}^{(\tau)}$ are the \textsc{dynbps} weights assigned to the $J$ model configurations $\mathscr{M}_{j}$, with $j=1,\ldots,J$. These entries represent a probability distribution over $J$ candidate models for that observation. 

To obtain a single distribution over the models, i.e., a weights vector, valid for all the observations, we consider two aggregation strategies: (i) global weights and (ii) consensus weights, as presented in Section~\ref{sec:propagation}. Both are shown here to yield valid probability vectors, thereby preserving the probabilistic interpretation of the original individual-wise weights.

The global weight vector $\hat{w}_{G}^{(\tau)}=\{\hat{w}_{G,1}^{(\tau)},\ldots,\hat{w}_{G,J}^{(\tau)}\}\in\mathbb{R}^{J}$ is defined as the column-wise average of $W^{(\tau)}$, such that
\begin{equation*}
    \hat{w}_{G,j}^{(\tau)}=\frac{1}{n}\sum_{i=1}^{n} \hat{w}_{i,j}^{(\tau)}, \quad j=1,\ldots,J.
\end{equation*}

\begin{proposition}
$\hat{w}_{G}^{(\tau)}$ is a probability vector: $\hat{w}_{G,j}^{(\tau)} \in [0,1]$ for all $j=1,\ldots,J$ and $\sum_{j=1}^{J} \hat{w}_{G,j}^{(\tau)} = 1$.
\end{proposition}

\begin{proof}
Since each element of $W^{(\tau)}$, $\hat{w}_{i,j}^{(\tau)} \in [0,1]$, their arithmetic mean satisfies $0 \le \hat{w}_{G,j}^{(\tau)} \le 1$ for all $j$. Moreover,
\begin{equation*}
    \sum_{j=1}^{J} \hat{w}_{G,j}^{(\tau)} 
    = \sum_{j=1}^{J} \frac{1}{n}\sum_{i=1}^{n} \hat{w}_{i,j}^{(\tau)}
    = \frac{1}{n} \sum_{i=1}^{n} \underbrace{\sum_{j=1}^{J} \hat{w}_{i,j}^{(\tau)}}_{=1} 
    = \frac{1}{n} \cdot n = 1.
\end{equation*}
\end{proof}
Thus, ``global'' averaging preserves the stochastic property, producing a blended global set of weights. The consensus weight vector $\hat{w}_{C}^{(\tau)}=\{\hat{w}_{C,1}^{(\tau)},\ldots,\hat{w}_{C,J}^{(\tau)}\} \in \mathbb{R}^J$ is obtained by first identifying, for each observation $i$, the model index with the largest weight (i.e., the ``preferred'' model for that observation), and then normalizing the counts of such preferences. 

Formally, let $j_{i}^{(\tau)} = \underset{1 \le j \le J}{\arg\max} \;\hat{w}_{i,j}^{(\tau)}$, with a deterministic tie-breaking rule (e.g., the smallest index). Define the count for model $j$ as  $c_{j}^{(\tau)} = \Bigl|\{ i : j_{i}^{(\tau)} = j \}\Bigr|$, and set  
\begin{equation*}
    \hat{w}_{C,j}^{(\tau)} = \frac{c_{j}^{(\tau)}}{n}, \qquad j = 1,\dots,J.
\end{equation*}

\begin{proposition}
$\hat{w}_{C}^{(\tau)}$ is a probability vector: $\hat{w}_{C,j}^{(\tau)} \in [0,1]$ for all $j$ and $\sum_{j=1}^{J} \hat{w}_{C,j}^{(\tau)} = 1$.
\end{proposition}

\begin{proof}
Each observation contributes to exactly one count, so $\sum_{j=1}^{J} c_{j}^{(\tau)} = n$. Hence, %$0 \le \hat{w}_{C,j}^{(\tau)} = \frac{c_{j}}{n} \le 1$ and  
\begin{equation*}
 0 \le \hat{w}_{C,j}^{(\tau)} = \frac{c_{j}^{(\tau)}}{n} \le 1 \quad \text{and}\quad \sum_{j=1}^{J} \hat{w}_{C,j}^{(\tau)} = \frac{1}{n} \sum_{j=1}^{J} c_{j}^{(\tau)} = \frac{n}{n} = 1.
\end{equation*}
\end{proof}

This method yields a global distribution that reflects the plurality of observation-level preferences. Both aggregation schemes therefore ensure that the resulting vector is a proper probability distribution over the $J$ models, making either suitable for subsequent use as weights in ensemble or decision-theoretic applications. The choice between them depends on whether a smooth blending of all weights or a majority-rule consensus is preferred.

% ------------------------------------------------
\section{Computational details and algorithms}\label{sec:appendixB_algs}

This Section collects and complements the algorithmic details underlying the predictive procedures discussed in Section~\ref{sec:computational}. In particular, Algorithms~\ref{alg:temporal_forecasting} and~\ref{alg:spatial_prediction} formalize the implementation of dynamic Bayesian predictive stacking for temporal forecasting and spatial interpolation, respectively. 

%%%%%%%%%%%%%%%%%%%%%%%%%%%%%%%%%%%%%%%%%%%%%%%%%%%
\begin{algorithm}[t!] 
\caption{\textsc{dynbps} - temporal forecasting}\label{alg:temporal_forecasting}
\begin{algorithmic}[1]
\Require Forecast horizon $K$, stacking weights $\left\{\hat{w}^{(t)}\right\}_{t=0}^{T}$, filtering parameters and inputs from Algorithm~\ref{alg:forward_filter}
\Ensure R posterior predictive samples from the marginal predictive $\hat{p}(\Theta_{T+k},Y_{T+k}\mid \mathscr{D}_{T})$ for $k=1,\dots,K$
\MyFunction{dynbps-forecast}{\Big($\{\hat{w}^{(t)}, \{m_{t}^{(j)},C_{t}^{(j)},\Psi_{t}^{(j)},\nu_{t}^{(j)},\alpha^{(j)},\phi^{(j)}\}_{j=1}^{J}\}_{t=0}^{T}, R,K$\Big)}
\For{$r=1,\dots,R$}
\State  \textit{\# Draw model $\mathscr{M}_{j}$ such that} $j \sim\;\operatorname{Multinom}\left(1,J,\left\{\hat{w}_{1}^{(T)},\dots,\hat{w}_{J}^{(T)}\right\}\right)$
\State \textit{\# Compute joint posterior predictive parameters}
\State $A_{T}(0)\gets\;m^{(j)}_{T}, R_{T}(0)\gets\;C^{(j)}_{T}$
\For{$k=1,\dots,K$}
\State \textit{\# Compute observation and state row-covariance matrices}
\State $V_{T+k}^{(j)} \gets\; \left(\frac{1-\alpha^{(j)}}{\alpha^{(j)}}\right)\mathbb{I}_{n}$, $W_{T+k}^{(j)} \gets\;  \begin{bsmallmatrix} \mathbb{I}_{p} & 0_{p\times n} \\ 0_{n\times p} & \mathcal{K}_{T+k}(\mathcal{S},\mathcal{S};\phi^{(j)}) \end{bsmallmatrix}$
\State \textit{\# Compute posterior predictive parameters}
\State $A_{T}(k)\gets\;G_{T+k}A_{T}(k-1)$, $q_{T}(k)\gets\; F_{T+k}A_{T}(k)$
\State $R_{T}(k) \gets\; G_{T+k}R_{T}(k-1)G_{T+k}^{\T}+W_{T+k}$, $Q_{T}(k)\gets\; F_{T+k}R_{T}(k)F_{T+k}^{\T}+V_{T+k}$
\State \textit{\# Define joint posterior predictive parameters}
\State $A^{\star}(k)\gets\;\begin{bsmallmatrix} A_{T}(k) \\ q_{T}(k) \end{bsmallmatrix}$, $R^{\star}(k) \gets\; \begin{bsmallmatrix} R_{T}(k) & R_{T}(k)F_{T+k}^{\T} \\ F_{T+k}R_{T}(k) & Q_{T}(k) \end{bsmallmatrix}$
\State \textit{\# Draw $\{\Theta_{T+k}^{(r)},Y_{T+k}^{(r)}\}$ from $\mT_{p+n+n,q}\left(\nu_{T}^{(j)},A^{\star}(k), R^{\star}(k), \Psi_{T}^{(j)}\right)$}
\State \textit{\# Store $\left\{A^{\star}(k), R^{\star}(k)\right\}$ for the next iteration}
\EndFor
\EndFor
\State \Return R posterior predictive samples of $\left\{\Theta_{T+k},Y_{T+k}\right\}$ for any $k=1,\dots,K$
\EndMyFunction
\end{algorithmic}
\end{algorithm}

%%%%%%%%%%%%%%%%%%%%%%%%%%%%%%%%%%%%%%%%%%%%%%%%%%%
\begin{algorithm}[t!] 
\caption{\textsc{dynbps} - spatial prediction at time $t$}\label{alg:spatial_prediction}
\begin{algorithmic}[1]
\Require Stacking weights $\hat{w}^{(t)}$, unobserved locations $\mathcal{U}$ and their design matrix $\Tilde{X}_{t}$, filtering parameters and inputs from Algorithm~\ref{alg:forward_filter}
\Ensure R posterior predictive samples from $\hat{p}(\tilde{Y}_{t},\tilde{\Omega}_{t}\mid \mathscr{D}_{t})$
\MyFunction{dynbps-spatpredict}{\Big($\hat{w}^{(t)}, \{m_{t}^{(j)},C_{t}^{(j)},\Psi_{t}^{(j)},\nu_{t}^{(j)},\alpha^{(j)},\phi^{(j)}\}_{j=1}^{J}, R$\Big)}
\For{$r=1,\dots,R$}
\State  \textit{\# Draw model $\mathscr{M}_{j}$ such that} $j \sim\;\operatorname{Multinom}\left(1,J,\left\{\hat{w}_{1}^{(t)},\dots,\hat{w}_{J}^{(t)}\right\}\right)$
\State \textit{\# Compute observation row-covariance matrix for new locations}
\State $\Tilde{V}_{t}(\alpha^{(j)})\gets\;(\frac{1-\alpha^{(j)}}{\alpha^{(j)}})\mathbb{I}_{m}$
\State \textit{\# Compute Schur complement parameters}
\State $\Tilde{M}_{t}\gets\;\mathcal{K}_{t}(\mathcal{U},\mathcal{S};\phi^{(j)})\mathcal{K}^{-1}_{t}(\mathcal{S},\mathcal{S};\phi^{(j)})$, $\Tilde{W}_{t}(\phi^{(j)})\gets\;t\big[\mathcal{K}_{t}(\mathcal{U},\mathcal{U};\phi^{(j)})-\Tilde{M}_{t}\mathcal{K}_{t}(\mathcal{S},\mathcal{U};\phi^{(j)})\big]$
\State \textit{\# Compute joint posterior predictive parameters}
\State $\chi_{t}\gets\;\begin{bsmallmatrix} \Tilde{X}_{t} & \Tilde{M}_{t} \\ 0 & \Tilde{M}_{t} \end{bsmallmatrix}$, $N_{t}\gets\;\begin{bsmallmatrix} \Tilde{V}_{t}(\alpha^{(j)})+\Tilde{W}_{t}(\phi^{(j)}) & \Tilde{W}_{t}(\phi^{(j)}) \\ \Tilde{W}_{t}(\phi^{(j)}) & \Tilde{W}_{t}(\phi^{(j)}) \end{bsmallmatrix}$
\State $\mu_{t}\gets\;\chi_{t}m_{t}^{(j)}$, $E_{t}\gets\;\chi_{t}C_{t}^{(j)}\chi_{t}^{\T}+N_{t}$
\State \textit{\# Draw $\{\Tilde{Y}_{t}^{(r)},\Tilde{\Omega}_{t}^{(r)}\}$ from $\mT_{2m,q}(\nu_{t}^{(j)}, \mu_{t}, E_{t}, \Psi_{t}^{(j)})$}
\EndFor
\State \Return R posterior predictive samples of $\left\{\tilde{Y}_{t},\tilde{\Omega}_{t}\right\}$
\EndMyFunction
\end{algorithmic}
\end{algorithm}

The former describes a recursive procedure for generating one-step and multi-step-ahead posterior predictive samples for temporal forecasting using spatiotemporal models in Equation~\eqref{eq:DLM_spatiotemporal}, by sequentially propagating uncertainty through time. The latter outlines the mechanism for spatial interpolation at a generic time point $t$, whereby posterior predictive samples of the multivariate latent spatial process and responses are drawn at unobserved locations using the results presented in Equation~\eqref{eq:Predictive_posterior}. Together with the methods presented in Section~\ref{sec:computational}, these procedures are implemented in the \if1\blind{\href{https://github.com/lucapresicce/spFFBS}{\texttt{spFFBS }}}\fi\if0\blind{\texttt{REDACTED }}\fi R package to provide a framework for spatiotemporal modeling with uncertainty quantification via \textsc{dynbps}.

% ------------------------------------------------
\section{Additional simulation experiment}\label{sec:appendixC_sim}

% %%%%%%%%%%%%%%%%%%%%%%%%%%%%%%%%%%%%%%%%%%%%%%%%
\subsection{Computational comparison against \texorpdfstring{\textsc{mcmc}}{MCMC}}\label{sec:sim_mcmc}

We compare \textsc{dynbps} directly with standard \textsc{mcmc} in terms of computational cost and predictive accuracy for the multivariate spatial response $Y$ and the spatial process $\Omega$. This experiment aims to quantify the computational advantage delivered by \textsc{dynbps} and whether it comes at the expense of predictive performance.

We generate $10$ replicate datasets $Y$ from \eqref{eq:DLM_spatiotemporal}, for $q=2$ correlated outcomes, over $T=10$ training time points and one held-out time point, $n=250$ spatial locations that remain fixed across the datasets, and a design matrix $X$, fixed over time, with $p=2$ comprising an intercept and a single predictor whose values were sampled independently from a uniform distribution over $[-1,1]$. The multivariate spatial process $\Omega$ and the regression coefficients $B$ evolve according to the state equation, with the starting value $\Theta_{0}=[B_{0}^{\T}:\Omega_{0}^{\T}]^{\T}$ drawn from its prior distribution; see Section~\ref{sec:dlm}. We fix the column covariance matrix as $\Sigma = \begin{bsmallmatrix} 1 & -0.3 \\ -0.3 & 1 \end{bsmallmatrix}$, and set $\alpha = 0.8$ and $\phi = 4$. Locations lie in the unit square, the spatial correlation function is exponential, and the prior hyperparameters are specified as in Section~\ref{sec:sim_Msettings}.

For each of the $10$ replicates, we fit \textsc{dynbps} with $\alpha \in \{0.7, 0.8, 0.9\}$ and $\phi \in \{2, 4, 6\}$, characterizing $J=9$ parallel learning flows. The grid contains the data-generating values, so this is an $\mathscr{M}$-closed comparison. On the same data, we run a standard batch Markov chain Monte Carlo sampler targeting the full \textsc{dlm} posterior over the entire training set. The \textsc{mcmc} chain runs for $K_{\mathrm{iter}}=2000$ iterations, discarding the first $1000$ as burn-in. The chain explores $\{\alpha,\phi\}$, with $(\Theta,\Sigma)$ available in closed form conditionally. The effective sample sizes show no evidence of convergence issues. From each fitted model, we draw $R=200$ samples from the one-step-ahead posterior predictive distribution of $\{Y_{T+1},\Omega_{T+1}\}$. % and compute the $2.5$, $50$, and $97.5$ predictive quantiles. 
Performance is assessed using root mean squared prediction error (\textsc{rmspe}), empirical coverage at the $95\%$ nominal level, and interval score (\textsc{is}). The computational cost is measured as the average wall-clock time required to fit each dataset across the replications. Results are reported in Table~\ref{tab:mcmc_comparison}.

\textsc{dynbps} fits each dataset in $1.8$ seconds on average, compared to $1344.6$ seconds for \textsc{mcmc}, making it approximately $750$ times faster. 
Timings refer to the hardware described in Section~\ref{sec:simulations}. Both approaches share the same $O(n^{3})$ cost per evaluation, but \textsc{dynbps} requires $J$ such evaluations per time step against $K_{\mathrm{iter}}$ for \textsc{mcmc}, so the operation count ratio is governed by $K_{\mathrm{iter}}/J$ rather than by $n$. In the Copernicus analysis of Section~\ref{sec:dataappl}, with $J=4$, the speedup exceeds $6400\times$.

The speedup does not come at the expense of predictive accuracy: the two methods attain nearly identical \textsc{rmspe} for $Y$ ($1.449$ versus $1.443$), and \textsc{dynbps} achieves a slightly lower \textsc{rmspe} for $\Omega$ ($1.190$ versus $1.237$). 
Both methods are conservative, particularly for $\Omega$, with \textsc{dynbps} closer to the nominal level for $Y$ ($0.978$ versus $0.987$) and essentially identical coverage for $\Omega$ ($0.995$ versus $0.996$). The interval score of \textsc{dynbps} is lower for both $Y$ and $\Omega$, indicating sharper predictive intervals at no cost in calibration. 
Overall, \textsc{dynbps} delivers predictive accuracy comparable to \textsc{mcmc} at a fraction of its computational cost, a result that is confirmed on real data in Section~\ref{sec:dataappl}, where \textsc{dynbps} is compared against \textsc{eb}, \textsc{mcmc}, \textsc{eof}, and \textsc{inla}.

\begin{table}[t!]
\centering
\setlength{\tabcolsep}{6pt}
\begin{tabular}{lccccccr}
\toprule
& \multicolumn{2}{c}{\textsc{rmspe}}
  & \multicolumn{2}{c}{Coverage}
  & \multicolumn{2}{c}{\textsc{is}}
  & \\
\cmidrule(lr){2-3}\cmidrule(lr){4-5}\cmidrule(lr){6-7}
Method
  & $Y$ & $\Omega$
  & $Y$ & $\Omega$
  & $Y$ & $\Omega$
  & Time (s) \\
\midrule
\textsc{dynbps} & 1.449 & 1.190 & 0.978 & 0.995 & 6.996 & 7.049 & 1.8 \\
\textsc{mcmc}   & 1.443 & 1.237 & 0.987 & 0.996 & 7.115 & 7.186 & 1344.6 \\
\bottomrule
\end{tabular}
\caption{One-step-ahead predictive performance and computational cost of \textsc{dynbps} and \textsc{mcmc}. Metrics are averaged over 10 replicates. Time reports the average wall-clock time per dataset fit.}
\label{tab:mcmc_comparison}
\end{table}

% %%%%%%%%%%%%%%%%%%%%%%%%%%%%%%%%%%%%%%%%%%%%%%%%
\subsection{Space-time weights dynamics}\label{sec:sim_weights}

In this simulation experiment, we investigated the different behaviors of Dynamic Bayesian predictive stacking weights. In Section~\ref{sec:propagation}, we introduce the concept of ``global'' weights (\textsc{g}) and ``consensus'' weights (\textsc{c}). This stems from the need to concentrate local information on weights, which \textsc{dynbps} actually provides (see Section~\ref{sec:propagation} for motivation and further technical details).
Both global and consensus weight sets work as central tendency extrapolations from their spatial distributions, allowing \textsc{ffbs} algorithm implementation, as detailed in Section~\ref{sec:computational}. In particular, they reflect information on the average weight and the ``consensus'' among locations for each of the $J$ candidate models at any time point. 
Although they mainly serve to retain computational admissibility, there are many interesting features we can obtain by investigating local weights and their dynamics across time. Thus, we report here a simulation experiment that provides more insight into these matters, allowing us to obtain model preferences ``locally''.

We start our simulation experiment by generating a set of spatiotemporal synthetic data. We mainly follow the same simulation framework used in Section~\ref{sec:sim_Msettings}, without replications, as we focus on the clustering of weights. Then, we have values on $n\times q\times T$ outcome $Y$ generated from \eqref{eq:DLM_spatiotemporal} with $n=500$, $q=3$, $T=100$ and $p=2$, $X$ includes two predictors generated from a standard uniform distribution over $[0,1]$. The $n\times n$ spatial correlation matrix is specified using an exponential correlation function with $\phi=4$, and we set $\alpha=0.8$.
See Section~\ref{sec:sim_Msettings} for further details.

\begin{figure}[t!]
    \centering
    \includegraphics[width=0.8\linewidth]{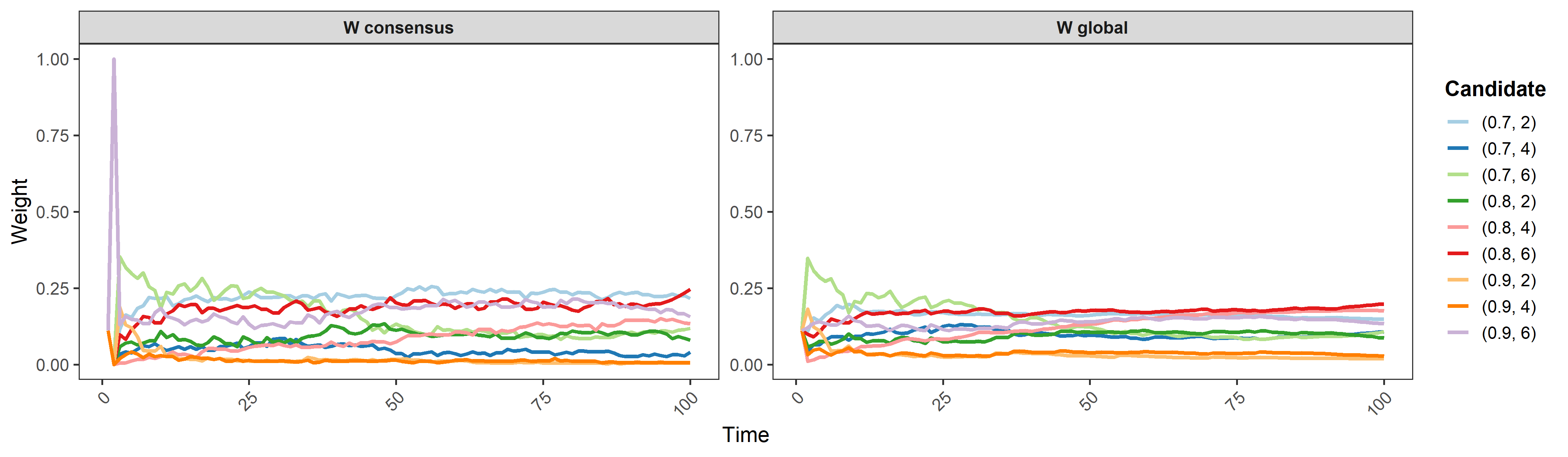}
    \caption{\textsc{dynbps} weights dynamics comparison over time: consensus weights (left panel), global weights (right panel)}
    \label{fig:weights_dynamic}
\end{figure}

We start by examining Figure~\ref{fig:weights_dynamic}. We report here two panels, from left to right: ``consensus'' weights ($\hat{w}_{C}^{(t)}$) and ``global'' weights ($\hat{w}_{G}^{(t)}$) dynamics over time-points. First, by construction of \textsc{dynbps}, as the time instant grows, more information is available to determine the weights. This leads to the possible interpretation of the panels in Figure~\ref{fig:weights_dynamic} as trajectories, i.e., reflecting the convergence of weights. That said, we notice a decreasing variability in the magnitude of weights as time passes. Both approaches completely stabilize the weights' central tendency after about $30$ time-points, even if $\hat{w}_{G}^{(t)}$ appears to have smoother and faster transitions. Moreover, we also noticed a greater separation in magnitude for $\hat{w}_{C}^{(t)}$, while spatially averaged weights ($\hat{w}_{G}^{(t)}$) tend to concentrate. Figure~\ref{fig:weights_dynamic} also shows a model-ranking coherence between the two approaches, which means they give the same predictive ``relevance'' to the same model. Therefore, we can state that the two methods yield comparable results. 

\begin{figure}[t!]
    \centering
    \includegraphics[width=0.8\linewidth]{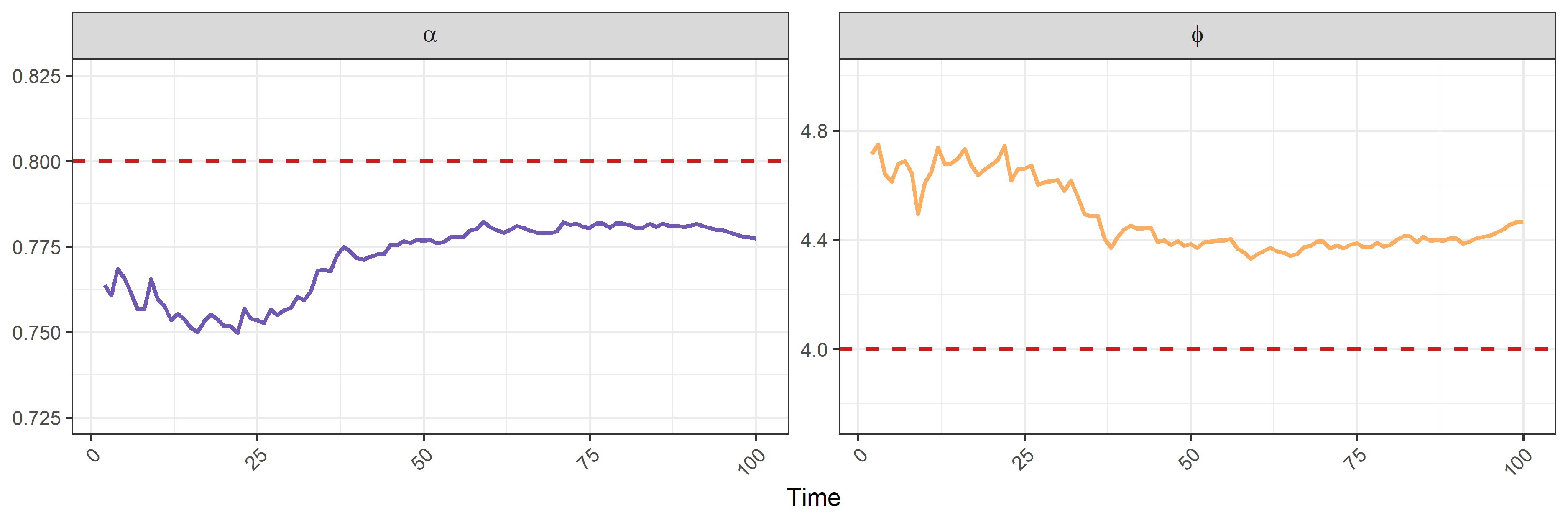}
    \caption{Parameter estimates dynamics against true value (red dashed line) over time: $\alpha$ point-estimates (left panel), $\phi$ point-estimates (right panel)}
    \label{fig:weights_par_dynamic}
\end{figure}

Hereafter, we will consider global weights $\hat{w}_{G}^{(t)}$. Figure~\ref{fig:weights_par_dynamic} represents the empirical estimates for $\{\alpha,\phi\}$ at different time points. As shown in Figure~\ref{fig:weights_dynamic}, we could interpret these as trajectories. Let us note that the estimates are given as follows: $\hat{\alpha}_{t}=\sum_{j=1}^{J}\hat{w}_{G,j}^{(t)} \alpha^{(j)}$ and $\hat{\phi}_{t}=\sum_{j=1}^{J}\hat{w}_{G,j}^{(t)} \phi^{(j)}$. In Figure~\ref{fig:weights_par_dynamic}, the leftmost panel shows the trajectory for $\alpha$, while the rightmost panel shows the trajectory for $\phi$. Both are, in some sense, attracted to the true values (dashed lines), but for $\alpha$, there is an underestimation, in contrast to $\phi$, which is overestimated. 
The latter corresponds to ``undersmoothing'', i.e., a mild preference for rougher latent surfaces, in contrast with the oversmoothing often reported for distributed learning approaches \citep[see e.g.,][]{guhaniyogi_meta-kriging_2018, presicce_bayesian_2024}. 
The concurrent underestimation of $\alpha$ is consistent with the weak identifiability of the exponential covariance parameters under fixed-domain asymptotics \citep{zhang_inconsistent_2004,tang_identifiability_2021}. Only the product of the spatial variance and the decay $\phi$, and the nugget variance, are consistently estimable.
Since the nugget is proportional to $(1-\alpha)/\alpha$ relative to the spatial variance, a larger $\phi$ is compensated by a smaller spatial variance and thus a smaller $\alpha$, leaving the predictive behavior essentially unchanged.

\begin{figure}[t!]
    \centering
    \includegraphics[width=0.8\linewidth]{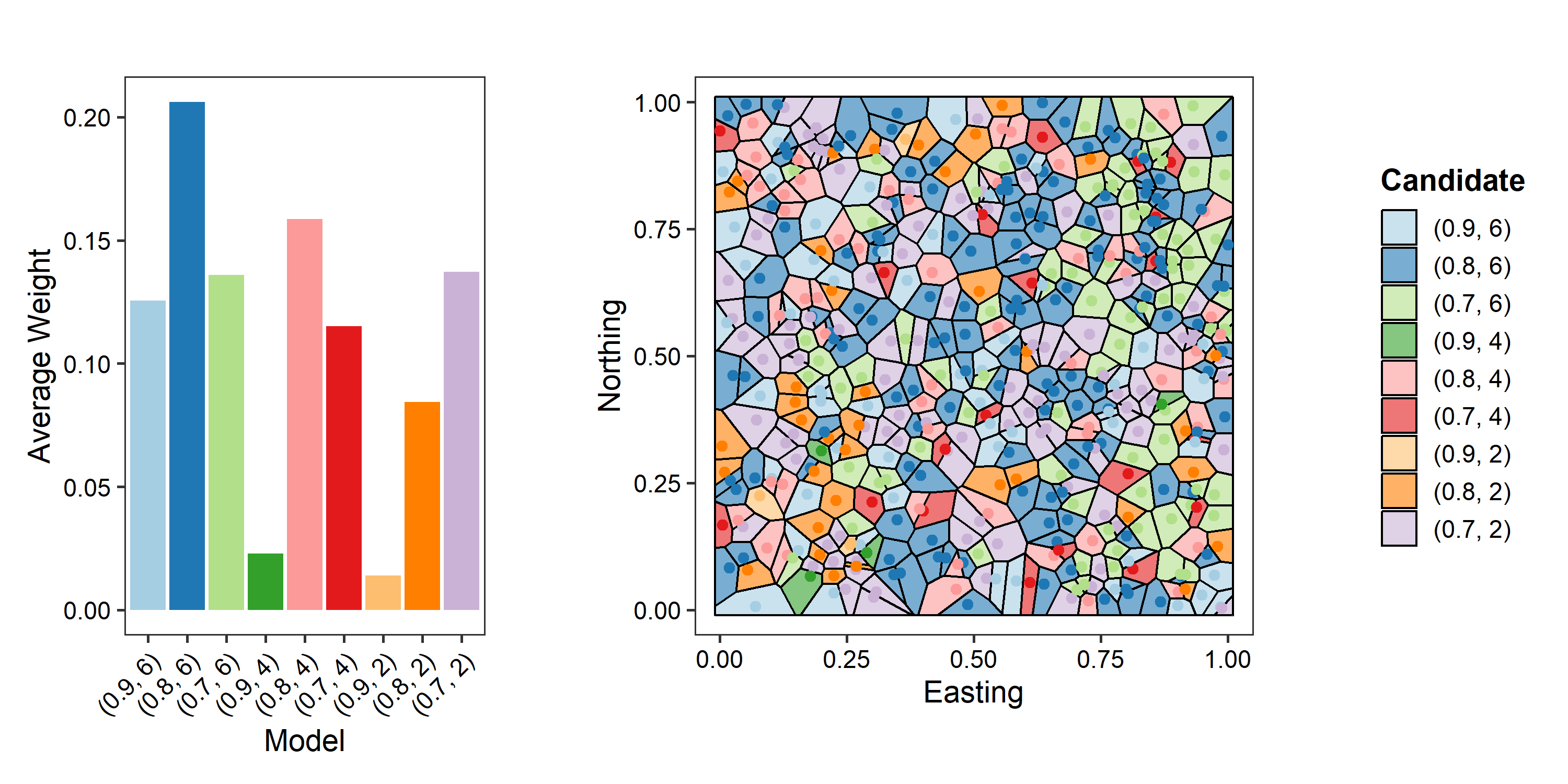}
    \caption{Location-wise weights spatial preference for each competitive model: bar plot of preference proportion (left panel), location-wise preferred model over Voronoi tessellation}
    \label{fig:weights_distrib}
\end{figure}

Lastly, we extrapolate insights from location-wise weights, which are shown in Figure~\ref{fig:weights_distrib}. This panel has two images: the share of locations preferring each model and the spatial distribution of weights colored by model. The latter makes use of the Voronoi tessellation to aid interpretation. To better understand the leftmost subplot in Figure~\ref{fig:weights_distrib}, consider each colored bar as the frequency of locations that ``prefer'' that model, i.e., assign it the greatest weight. The same reasoning applies to the right subplot: it shows for each location which model received the greatest weight, i.e., the preferred model. Figure~\ref{fig:weights_distrib} clearly indicates that the largest share of locations prefers model $\{\alpha=0.8,\phi=6\}$, but the second preferred model is the true one, i.e., $\{\alpha=0.8,\phi=4\}$. However, many locations do not prefer the true model, as in the data-generating process, the conjunction between spatial variability and non-spatial associations among variables leads to sample georeferenced points with individual characteristics, for which the posterior predictive distribution pushes towards different values. We actually average out locally dependent micro-characteristics, as justified by the backward sampling, which aims to identify a common set of parameters. However, the methodological contribution allows for further investigation of many other perspectives, giving users a flexible level of interpretability for the analysis.
From Figure~\ref{fig:weights_distrib}, we can conclude that every location has a personal set of weights based on the local evaluation of posterior predictive distributions. Even though it may be computationally cumbersome to start a local analysis for all the points, there is still the opportunity to individually investigate each point, which corresponds to a multivariate time series by itself. This implies that a subset of points of interest for a specific research question can be studied. For a single-location analysis, once forward filtering with weights is complete, backward sampling can be performed for the point of interest using an individual set of weights that reflects local model preferences.

% ------------------------------------------------
\section{Exploratory data analysis}\label{sec:appendix_eda}

\begin{table}[t!]
\centering
\caption{Summary statistics and visual representation of response variables.}
\label{tab:eda_summary}
\begin{tabular}{@{}lrrrrcc@{}}
\hline
Variable & \multicolumn{1}{c}{Mean} & \multicolumn{1}{c}{Std.Dev} 
         & \multicolumn{1}{c}{Min.} & \multicolumn{1}{c}{Max.} 
         & Histogram & Boxplot \\
\hline
Monthly temperature    & 13.043 & 8.444  & -17.059 & 35.032 
                       & \raisebox{-0.2\totalheight}{\includegraphics[width=0.08\textwidth]{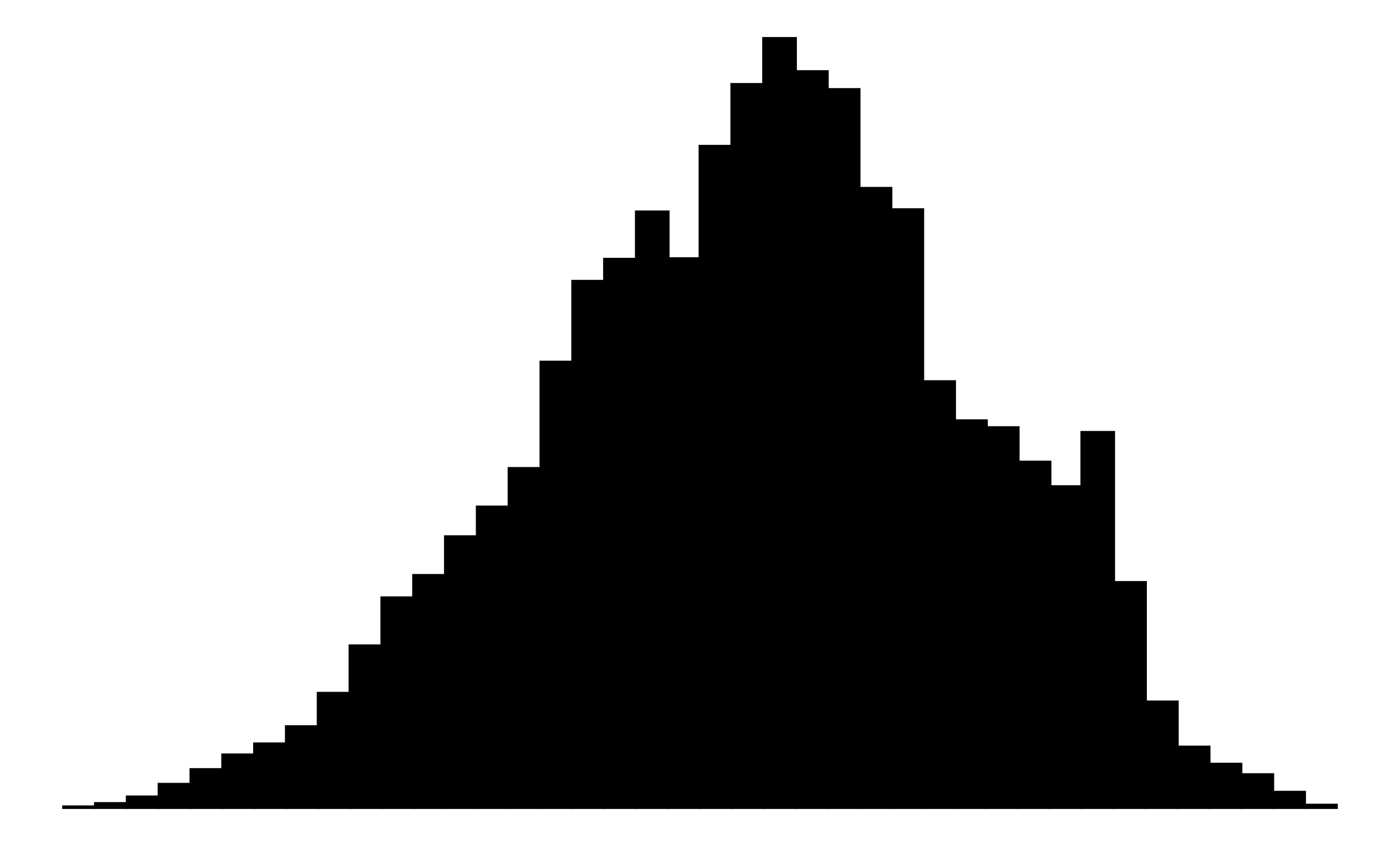}}
                       & \raisebox{-0.2\totalheight}{\includegraphics[width=0.08\textwidth]{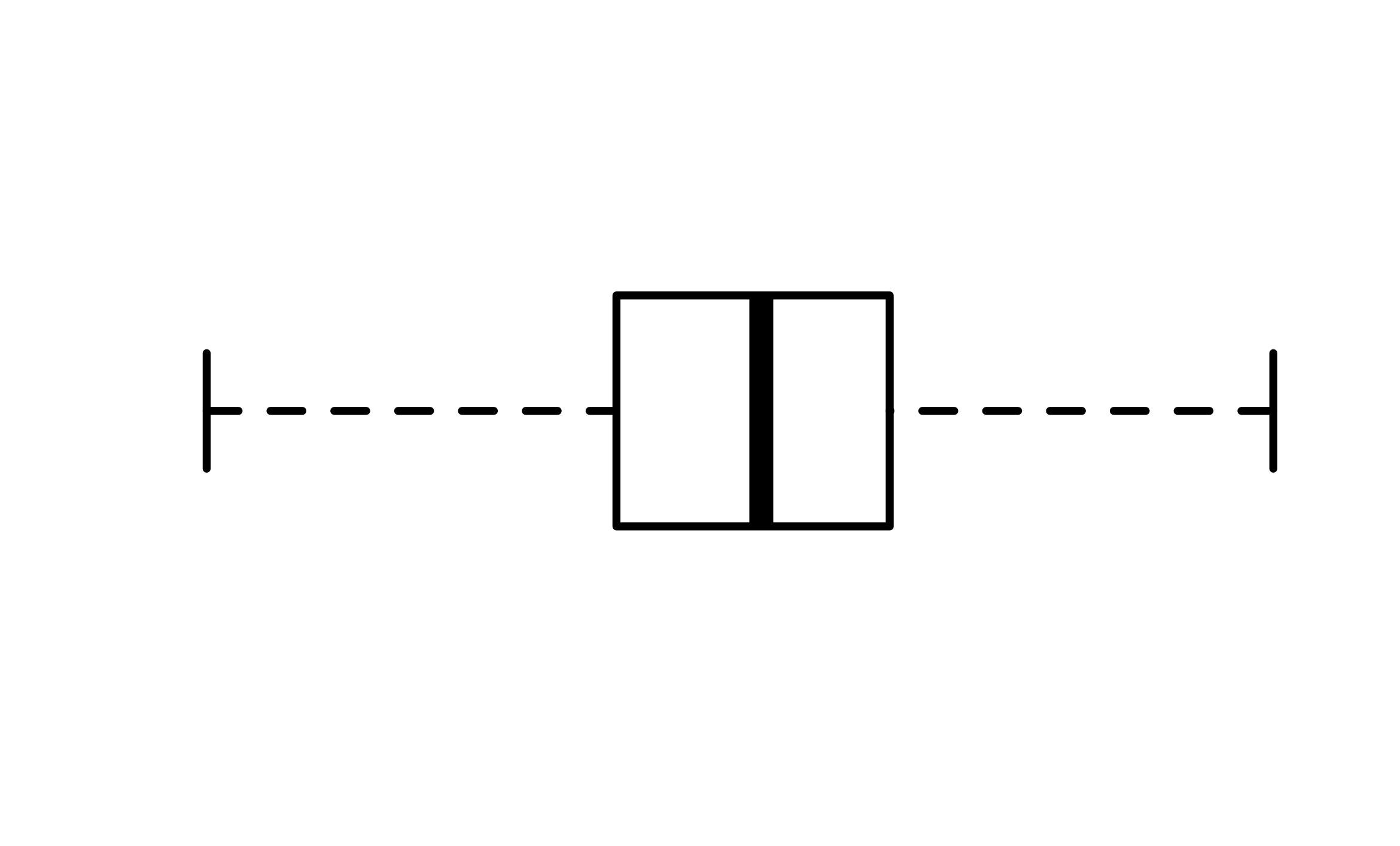}} \\

Monthly precipitation  & 1.787  & 1.082  & 0.002   & 7.163  
                       & \raisebox{-0.2\totalheight}{\includegraphics[width=0.08\textwidth]{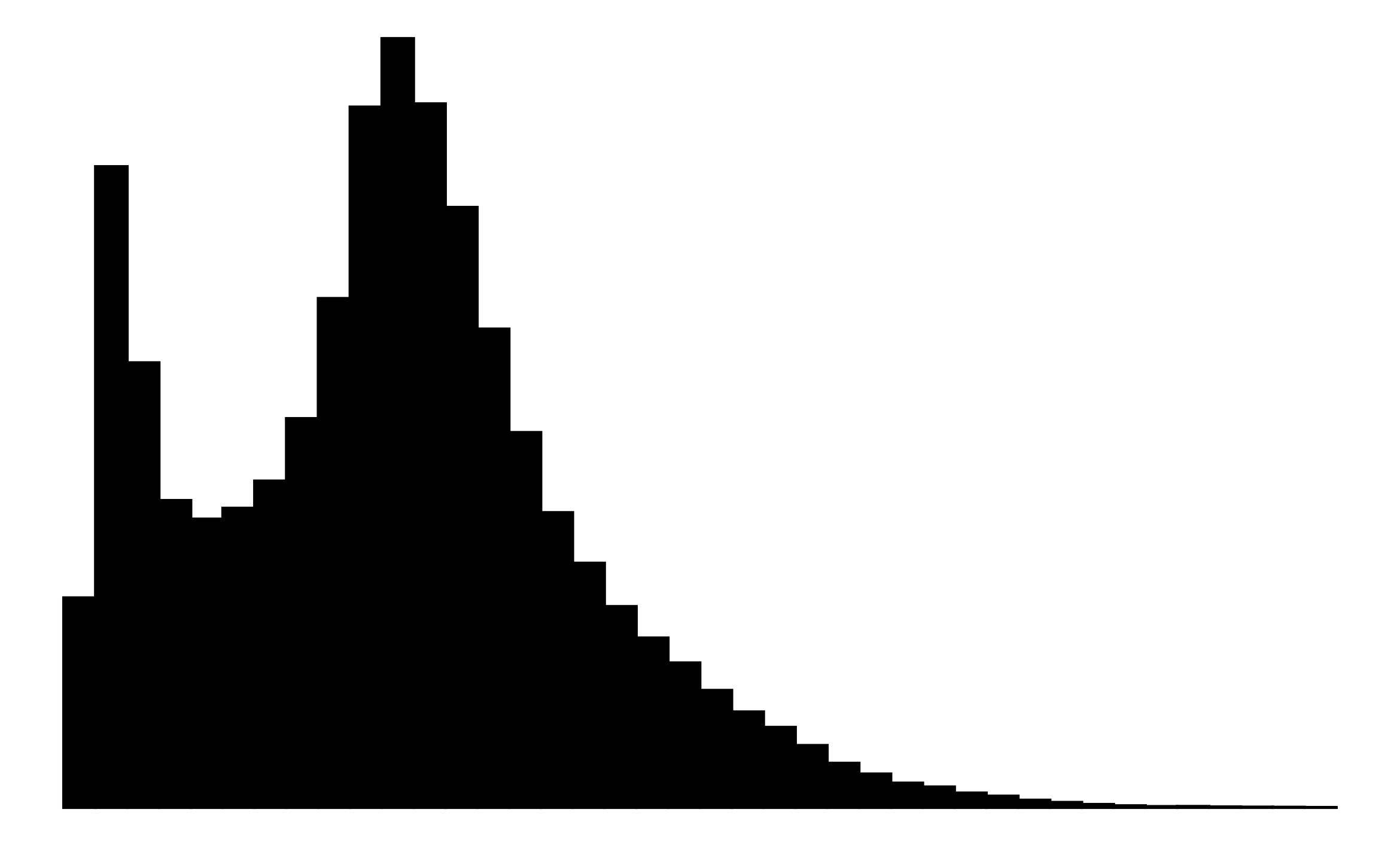}}
                       & \raisebox{-0.2\totalheight}{\includegraphics[width=0.08\textwidth]{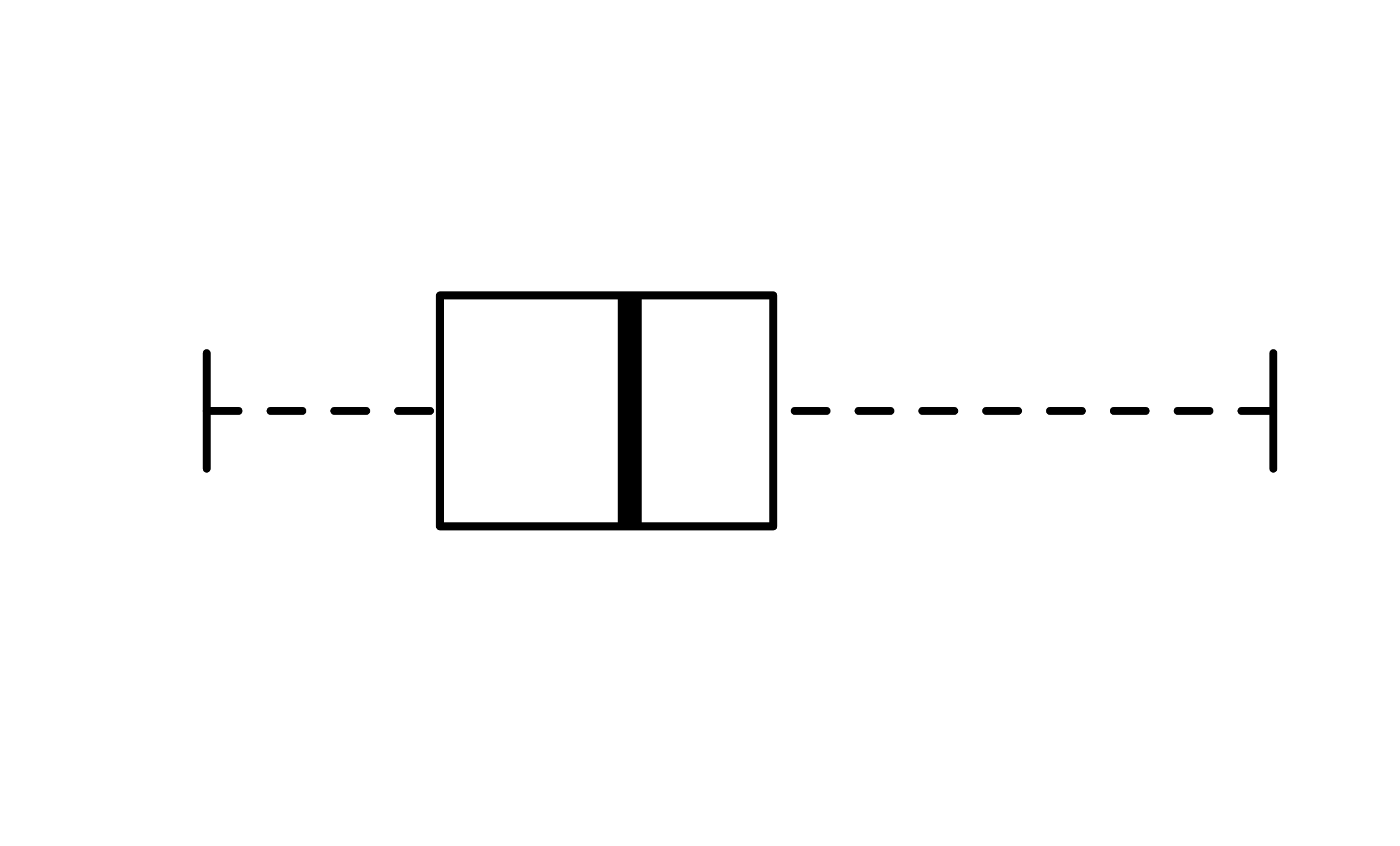}} \\

Monthly wind speed     & 4.541  & 1.685  & 1.795   & 12.067 
                       & \raisebox{-0.2\totalheight}{\includegraphics[width=0.08\textwidth]{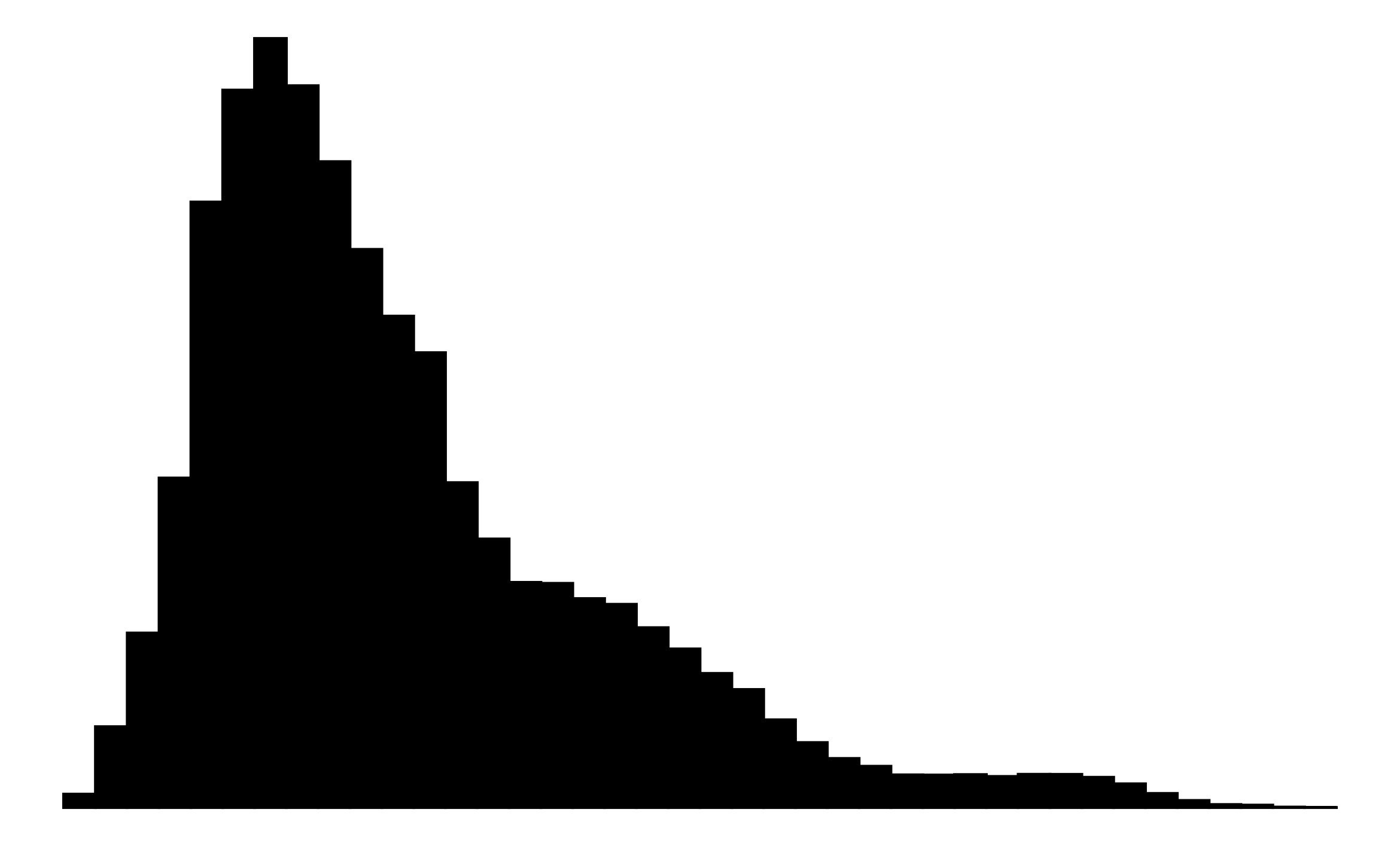}}
                       & \raisebox{-0.2\totalheight}{\includegraphics[width=0.08\textwidth]{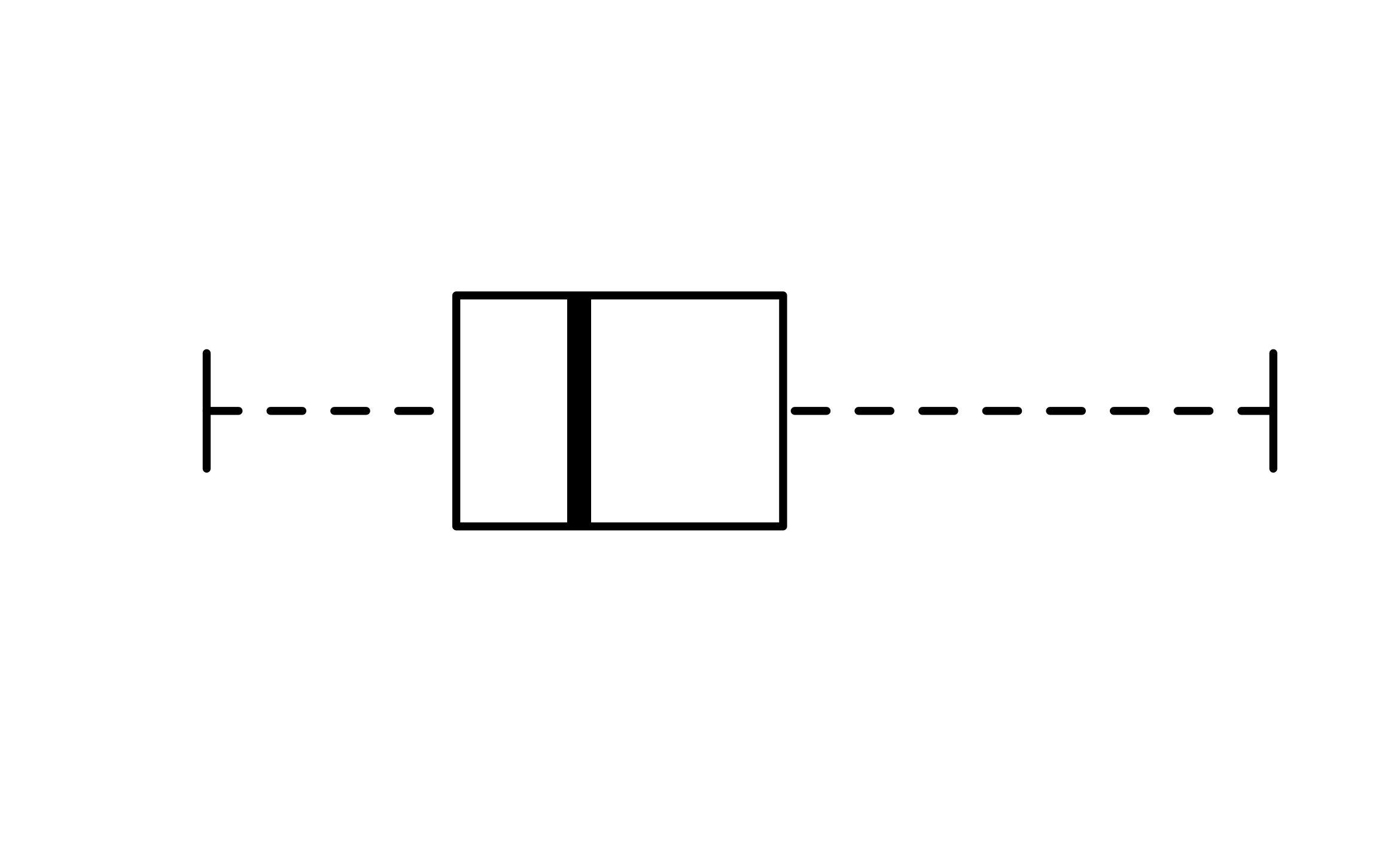}} \\

Monthly evaporation    & 1.858  & 1.165  & 0.030   & 6.384  
                       & \raisebox{-0.2\totalheight}{\includegraphics[width=0.08\textwidth]{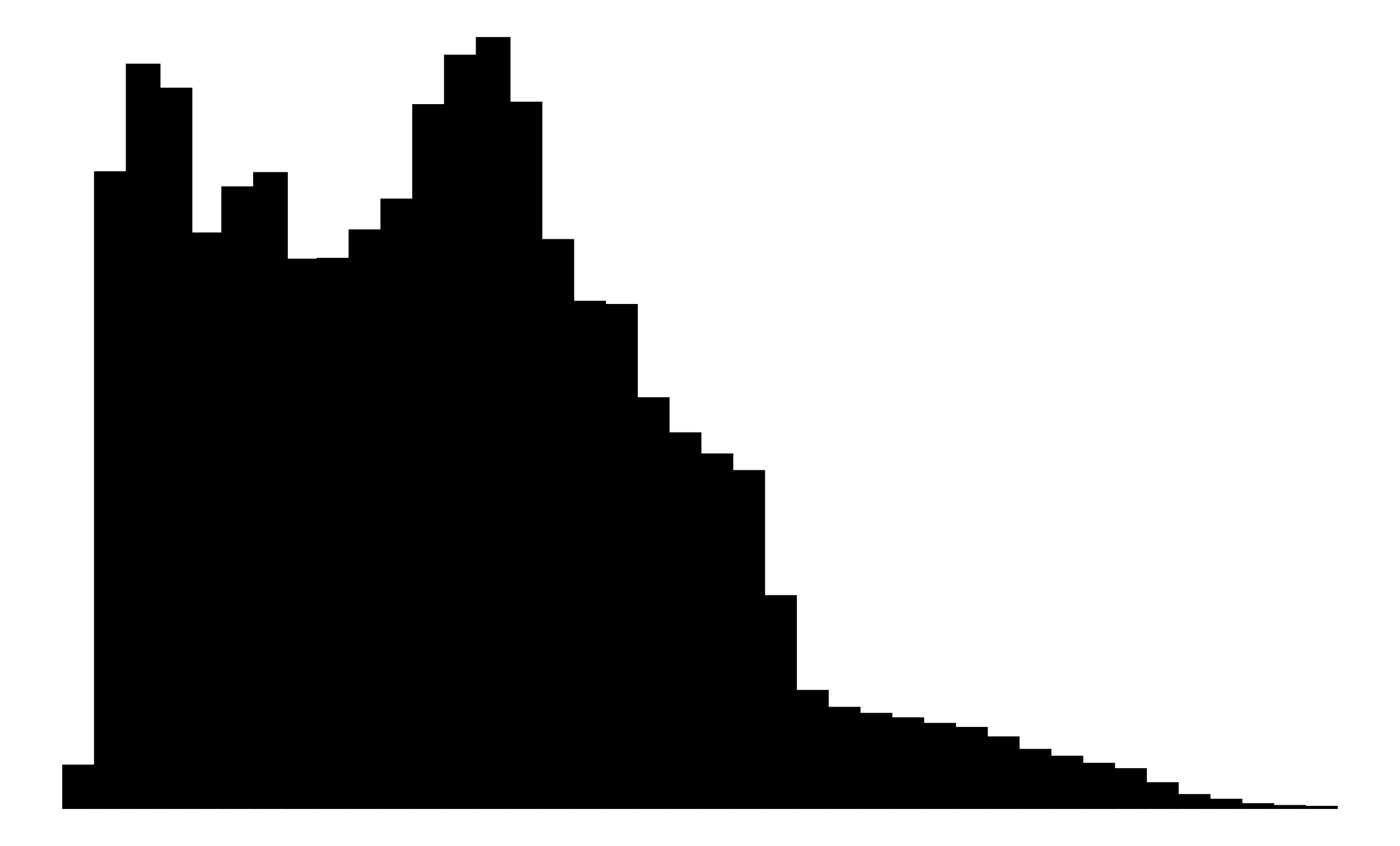}}
                       & \raisebox{-0.2\totalheight}{\includegraphics[width=0.08\textwidth]{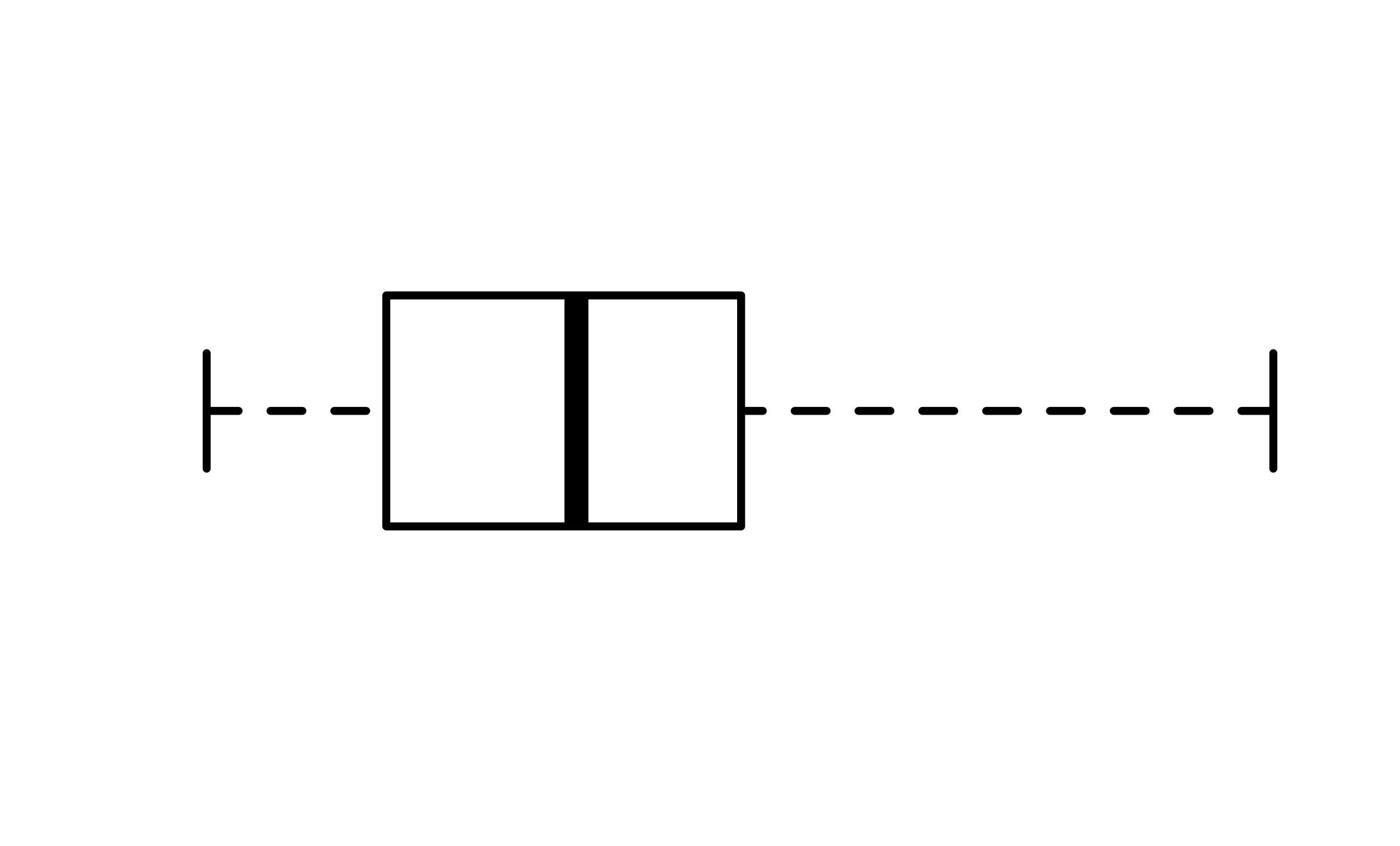}} \\
\hline
\end{tabular}
\end{table}

Our current investigation utilizes high-resolution climate data from the Copernicus Climate Change Service (\textsc{c3s}) Atlas, a multi-source dataset developed within the European Union's Copernicus Programme and accessed through the \href{https://www.copernicus.eu}{Copernicus Data Space Ecosystem (CDSE)}. The \textsc{c3s} Atlas compiles harmonized climate variables from different sources, including satellite remote-sensing and in-situ monitoring networks. These datasets are accessible through the Climate Data Store (\textsc{cds}) and conform to the \textsc{acdd} ``Netcdf4metadata'' conventions, ensuring compatibility, reproducibility across climate data analysis, and consistency for long-term climate monitoring.

Table~\ref{tab:eda_summary} presents summary statistics for the four climate variables, together with histograms and boxplots for their distributions, without considering their spatiotemporal structure.

\begin{figure}[t!]
\centering
\includegraphics[width=\linewidth]{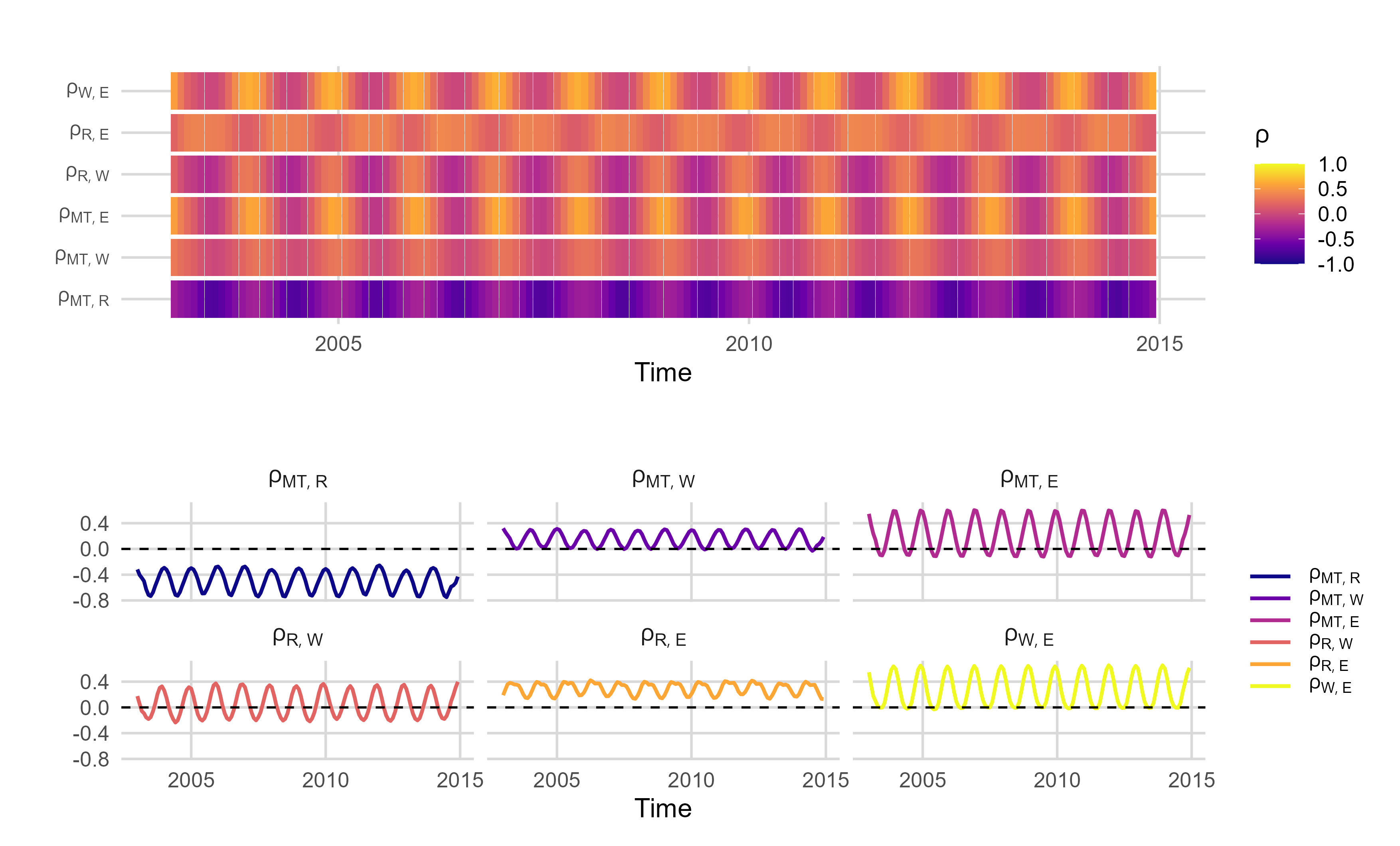} 
\caption{Correlation analysis between response variables over time: heatmap of correlation matrix elements (top panel), and time series of correlation matrix elements (bottom panel). Where MT = monthly temperature, R = monthly precipitation/rain, W = monthly wind speed, E = monthly evaporation}
\label{fig:eda_correlation}
\end{figure}

We investigate inter-variable temporal dependence by analyzing the evolution of pairwise correlations over time. Figure~\ref{fig:eda_correlation} presents two complementary views: a heatmap showing the elements of the correlation matrix at each month, and a time series of individual correlation coefficients. These diagnostics indicate pronounced seasonal modulation in all variable pairs, with coherent but magnitude-varying cycles driven by the underlying atmospheric processes.

% spatiotemporal variograms
We compute empirical spatiotemporal variograms on residuals to quantify the joint spatial and temporal dependence for each climate variable.
Due to the strong seasonal effects, the residuals were obtained by regressing each variable in $Y$ on relevant predictors and monthly seasonal dummy variables, see Section~\ref{sec:dataappl}.
We estimated the empirical spatiotemporal variograms using the \texttt{gstat} and \texttt{spacetime} in R. 

A separable parametric model was fitted for each variable. This model was chosen for its interpretability and computational efficiency, as it assumes that spatial and temporal dependence can be modeled independently while still capturing the main features of the covariance structure. As initial values, we combined an exponential spatial component with nugget $0.1$, partial sill $1$, and range $10$; and an exponential temporal component with nugget $0.1$, partial sill $1$, and range $5$ months. The variograms were computed over all the training temporal lags, from $0$ to $119$, with a $10$-month increment. 

\begin{figure}[t!]
\centering
\includegraphics[width=\linewidth]{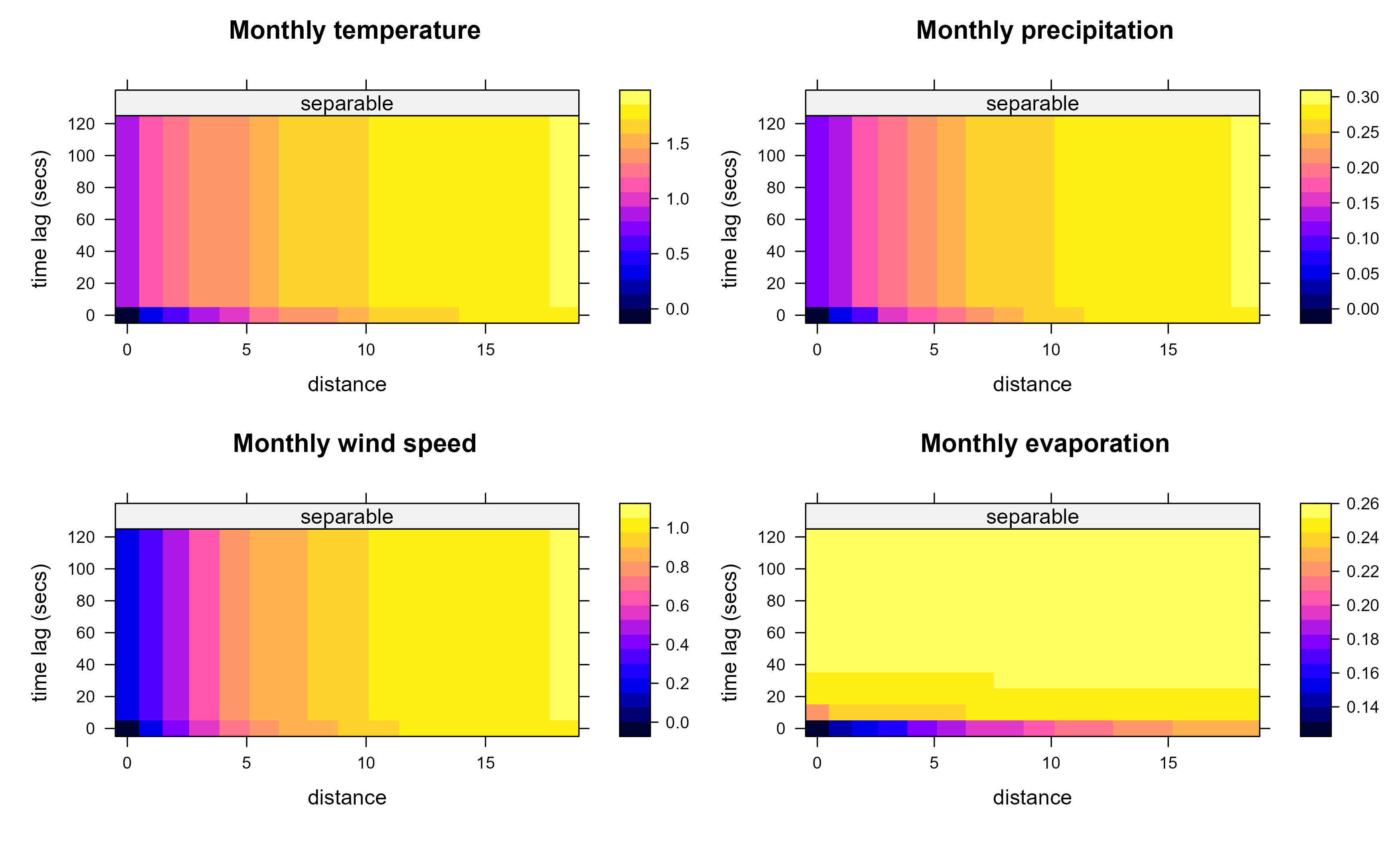} 
\caption{Spatiotemporal sample variogram computed on residuals: one panel for each response variable}
\label{fig:st_variogram}
\end{figure}

Figure~\ref{fig:st_variogram} presents the fitted spatiotemporal variogram for all four response variables. The variogram analysis reveals a clear dichotomy. Atmospheric variables, like temperature, precipitation, and wind speed, all exhibit strong and persistent temporal memory, with autocorrelation lasting for long periods; wind speed shows the most robust temporal structure.

Spatially, their patterns are more localized, with wind speed again being more scattered. In contrast, evaporation behaves uniquely with respect to the other outcomes: it has a very short temporal memory, becoming uncorrelated fairly quickly, while its spatial pattern is the most extensive but also the noisiest, being dominated by fine-scale variations. 

The fitted parameters supported the graphical interpretations, while we made use of them as starting points, composing the grid of values for $\alpha$, and $\phi$ needed to implement \textsc{dynbps}. For further details, see Section~\ref{sec:dataappl}.

% Hovmöller diagrams
To visualize large-scale spatiotemporal patterns, we constructed two Hovmöller diagrams, comparing the time evolution with respect to latitude and longitude, respectively. 

\begin{figure}[t!]
\centering
\includegraphics[width=0.9\linewidth]{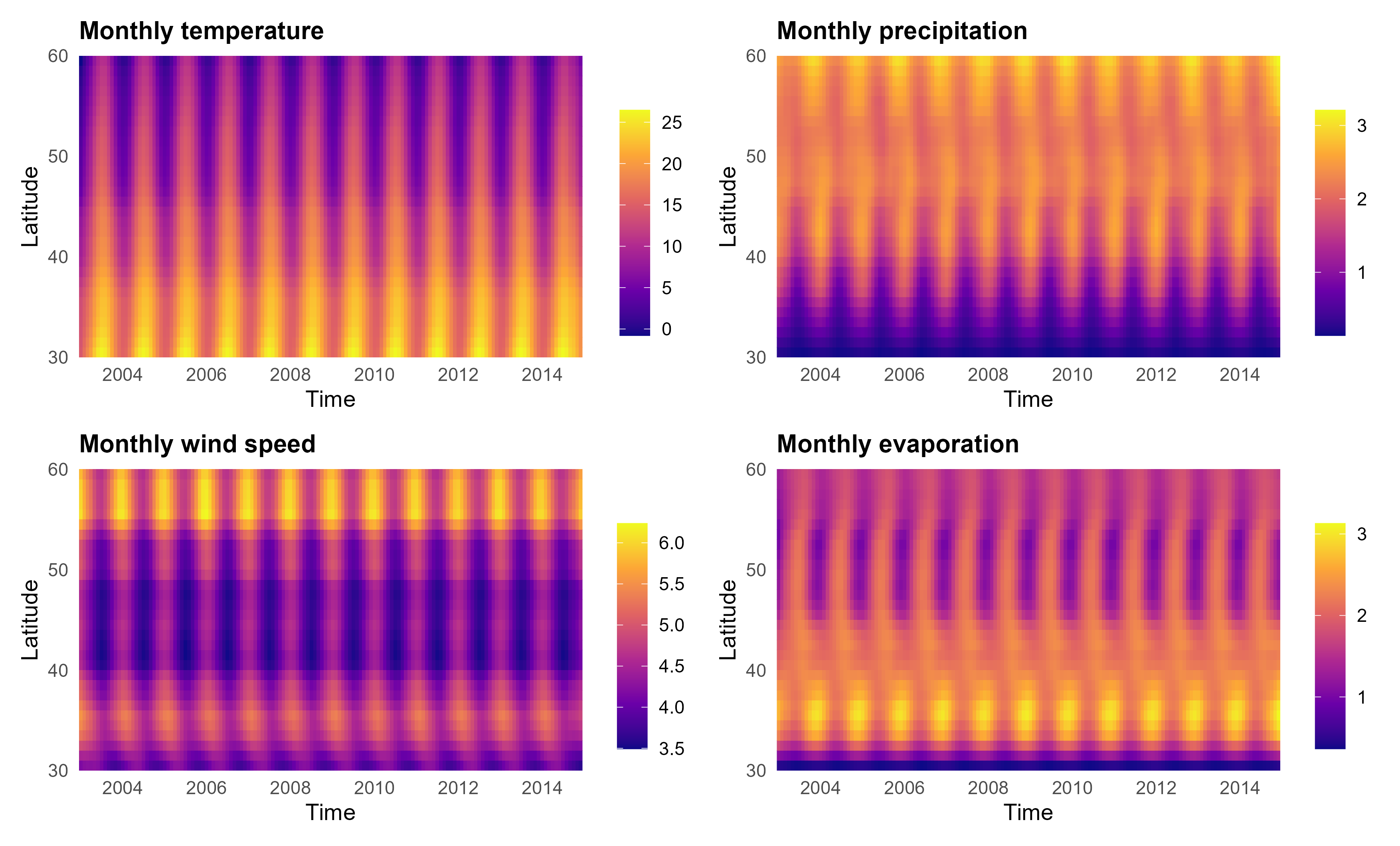} 
\caption{Hovmöller diagram by latitude for spatiotemporal atmospheric outcomes: time (x-axis) against latitude (y-axis)}
\label{fig:hovmoller_lat}
\end{figure}

Figure~\ref{fig:hovmoller_lat} shows the Latitude-wise diagram, where for each month, the variable averages were computed along longitudes to display how the variables vary with latitude over time.
Seasonal variability is clearly observed, highlighting strong and almost-linear latitudinal gradients in temperature and precipitation, while still showing smoothed but complex gradients for wind speed and evaporation. As expected, lower latitudes correspond to higher temperature and lower precipitation, with the opposite pattern moving northward. Despite the strong seasonality, all outcomes appear stationary over the whole period, with no visible long-term trend.

\begin{figure}[t!]
\centering
\includegraphics[width=0.9\linewidth]{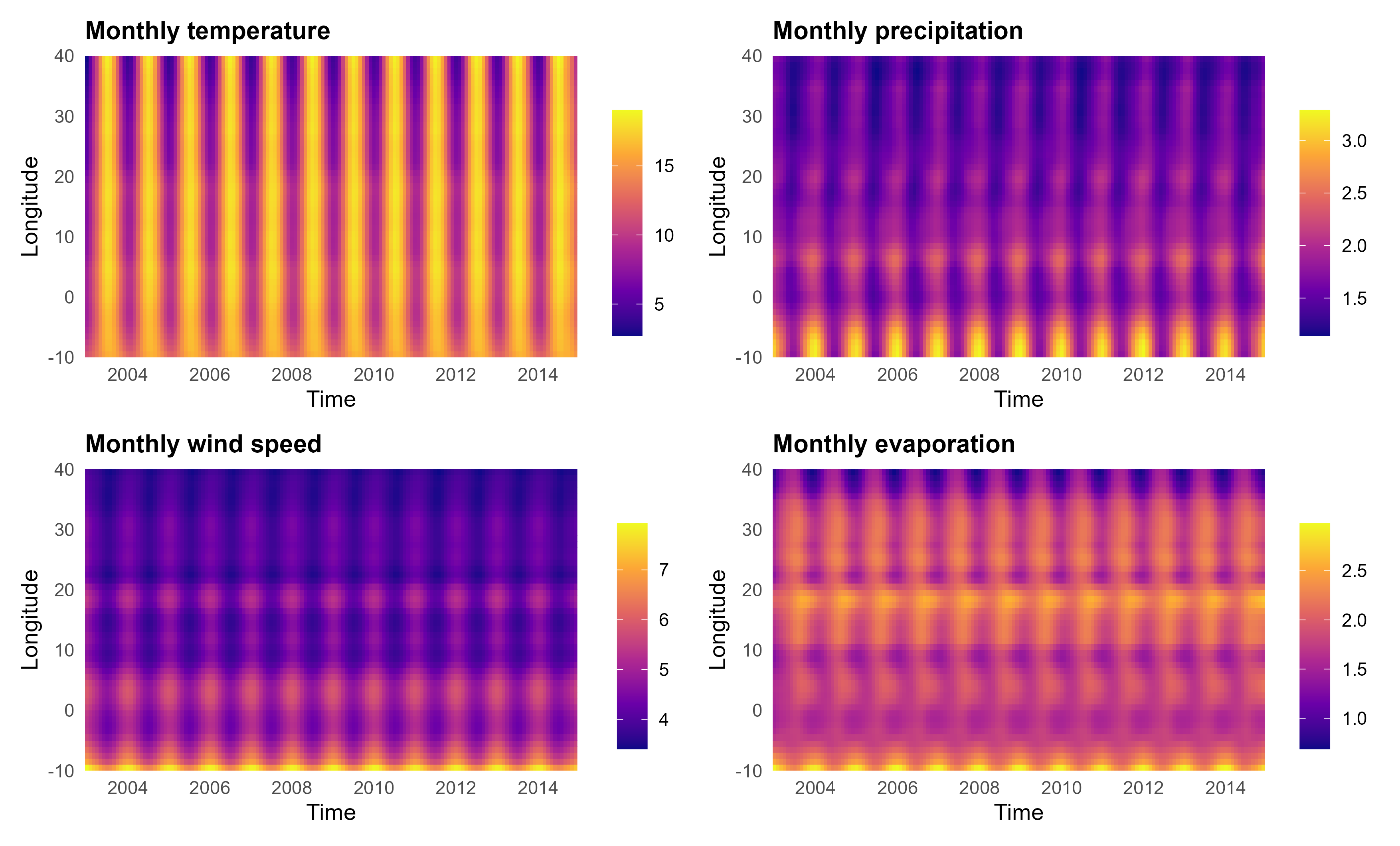} 
\caption{Hovmöller diagram by longitude for spatiotemporal atmospheric outcomes: time (x-axis) against longitude (y-axis)}
\label{fig:hovmoller_lon}
\end{figure}

Conversely, in Figure~\ref{fig:hovmoller_lon} we present Hovmöller diagrams with respect to longitude, where variable averages were computed along latitudes to illustrate zonal patterns across the study region.
Variability along longitude is lower, with smoother surfaces for all variables except evaporation, which still shows a complex structure. Strong seasonality and apparent stationarity remain along longitude.

Together, the residual-based spatiotemporal variograms and Hovmöller diagrams provide strong evidence of structured dependencies across both space and time, as well as strong seasonality and apparent stationarity. These diagnostics confirm that a multivariate spatiotemporal modeling framework is appropriate for capturing complex climate dynamics and justifies the predictive and inferential analyses presented in Section~\ref{sec:dataappl}. Such an approach exploits the evolving cross-variable dependence and enables the borrowing of information across space and time.
%We nevertheless adopt the random walk evolution in Equation~\eqref{eq:DLM_spatiotemporal}, as it preserves conjugacy and the block structure required for scalability (see Section~\ref{sec:dlm}).

% --------------------------------------------------
\section{Additional figures}\label{sec:appendixD_graphics}

\begin{figure}[H]
    \centering
    \includegraphics[width=0.90\linewidth]{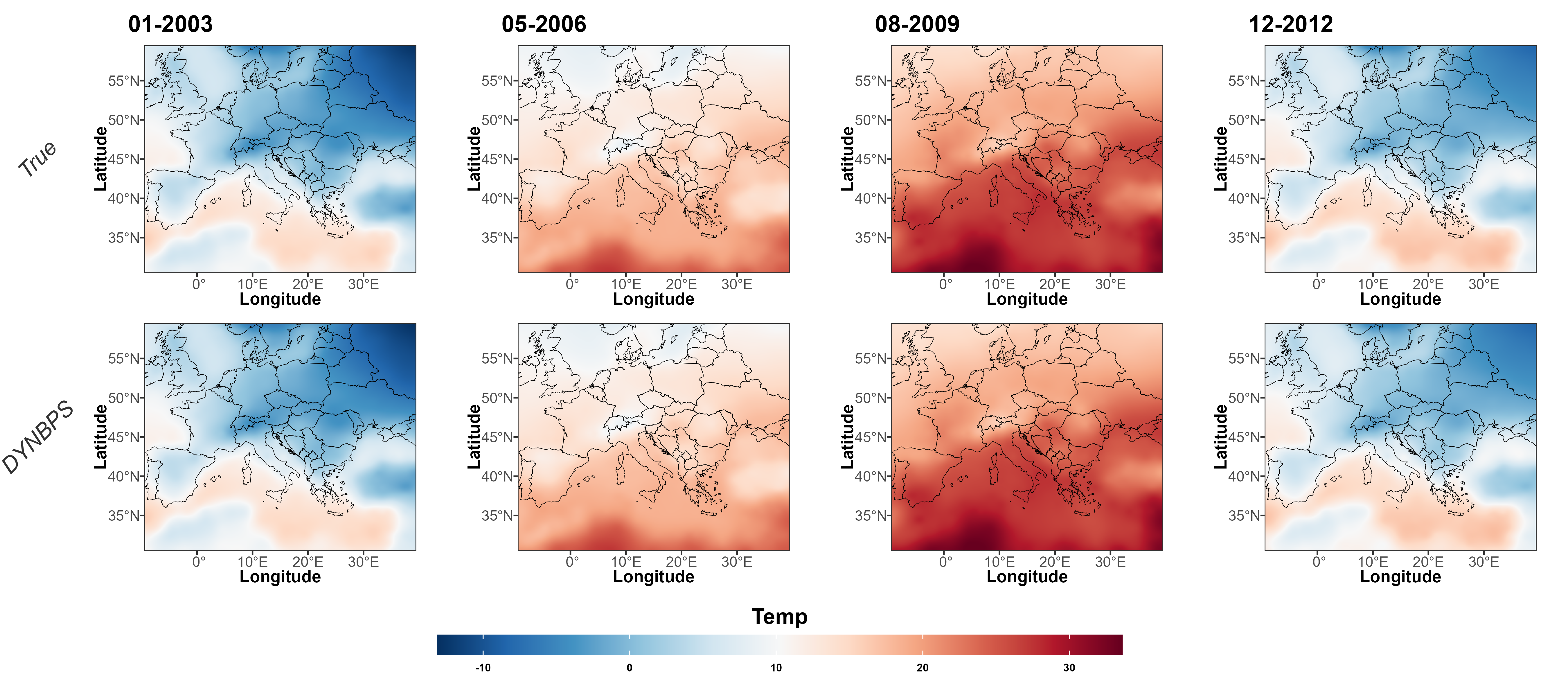}
    \caption{Monthly temperature forecast at different time points: true spatial surfaces (top row) and \textsc{dynbps} predictions (bottom row).}
    \label{fig:forecast_temp}
\end{figure}

\begin{figure}[H]
    \centering
    \includegraphics[width=0.90\linewidth]{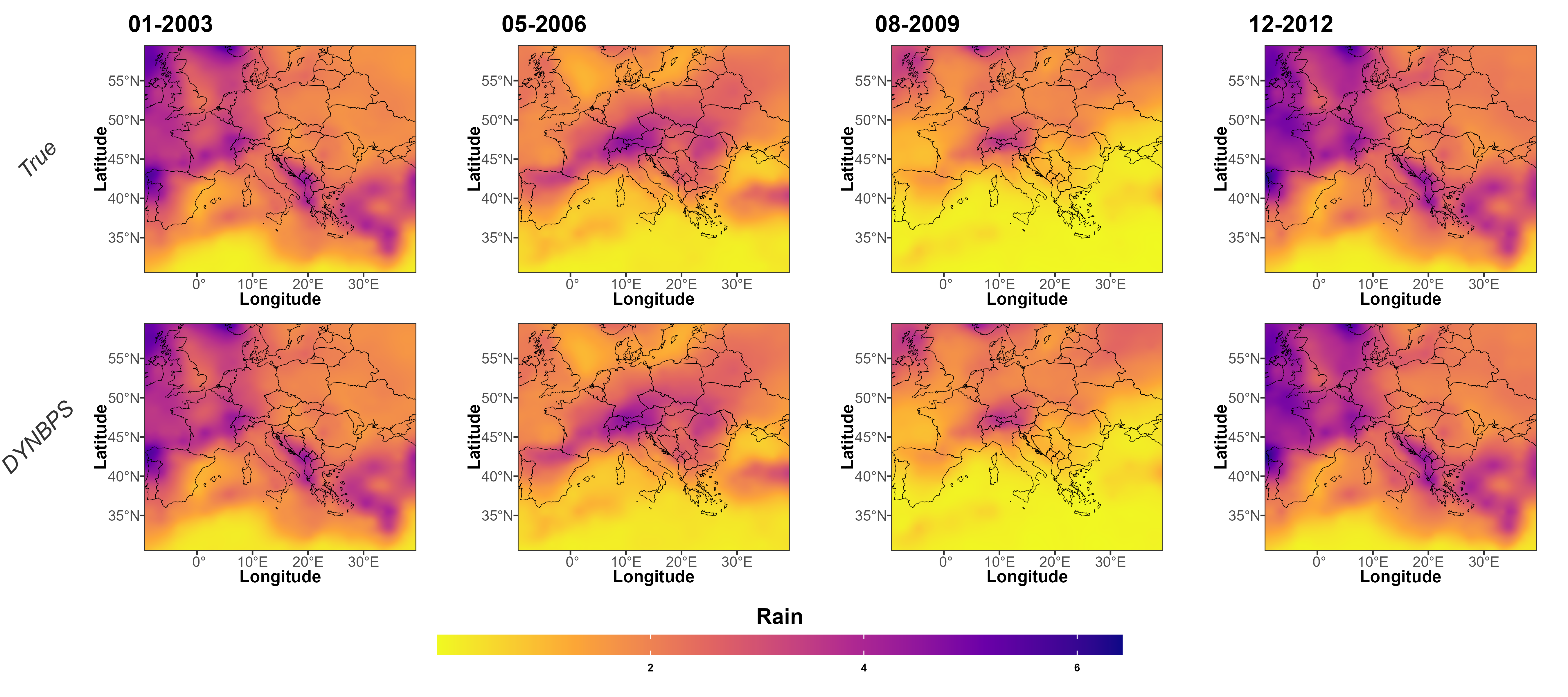}
    \caption{Monthly precipitation forecast at different time points: true spatial surfaces (top row) and \textsc{dynbps} predictions (bottom row).}
    \label{fig:forecast_rain}
\end{figure}

\begin{figure}[H]
    \centering
    \includegraphics[width=0.90\linewidth]{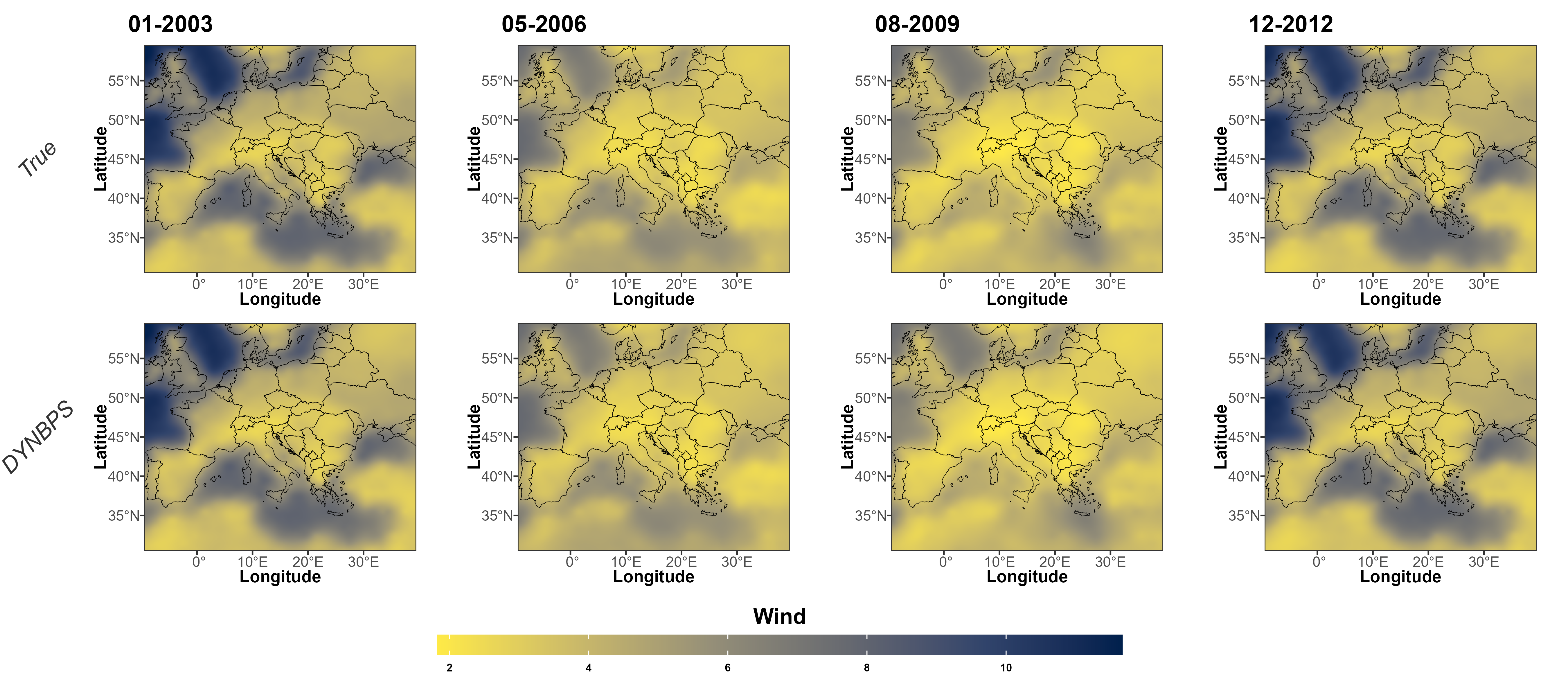}
    \caption{Monthly wind speed forecast at different time points: true spatial surfaces (top row) and \textsc{dynbps} predictions (bottom row).}
    \label{fig:forecast_wind}
\end{figure}

\begin{figure}[H]
    \centering
    \includegraphics[width=0.90\linewidth]{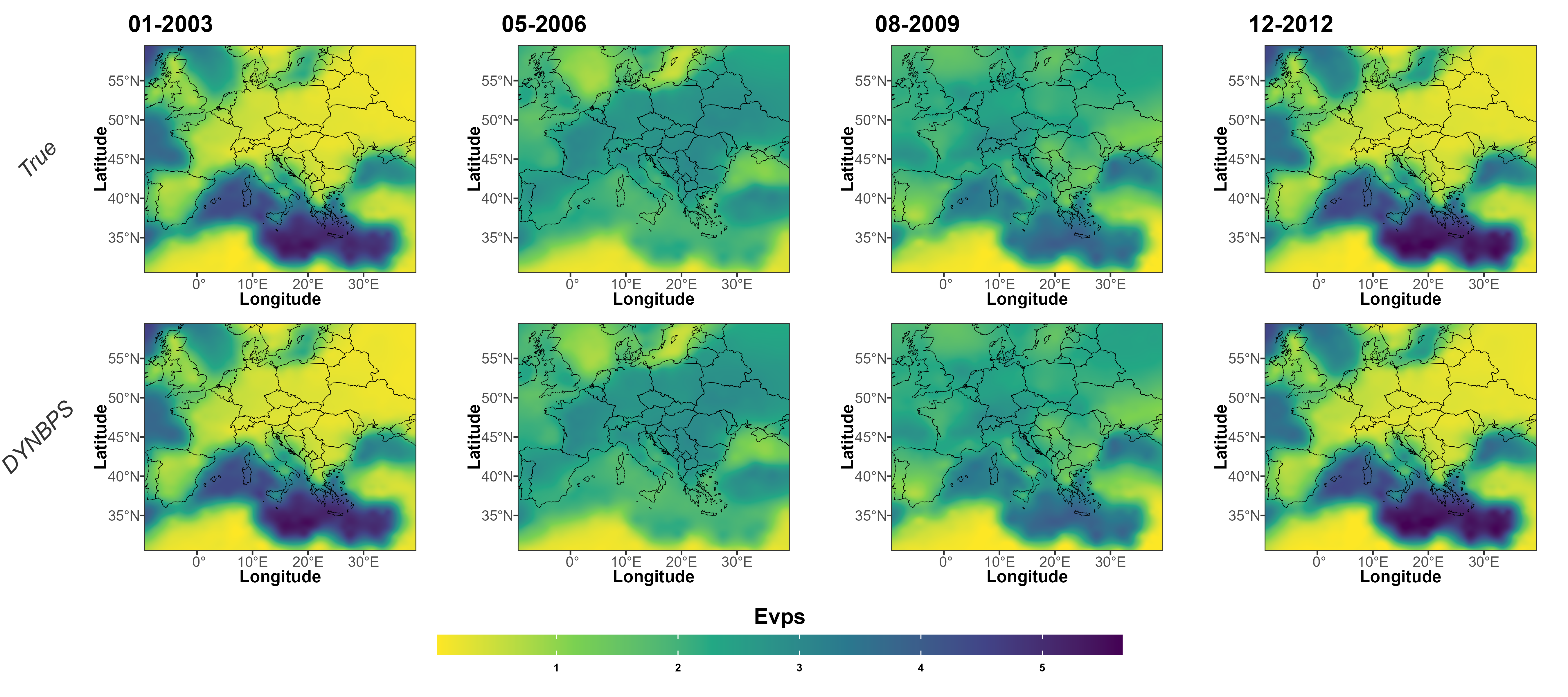}
    \caption{Monthly evaporation forecast at different time points: true spatial surfaces (top row) and \textsc{dynbps} predictions (bottom row).}
    \label{fig:forecast_evps}
\end{figure}

\begin{figure}[H]
    \centering
    \includegraphics[width=\linewidth]{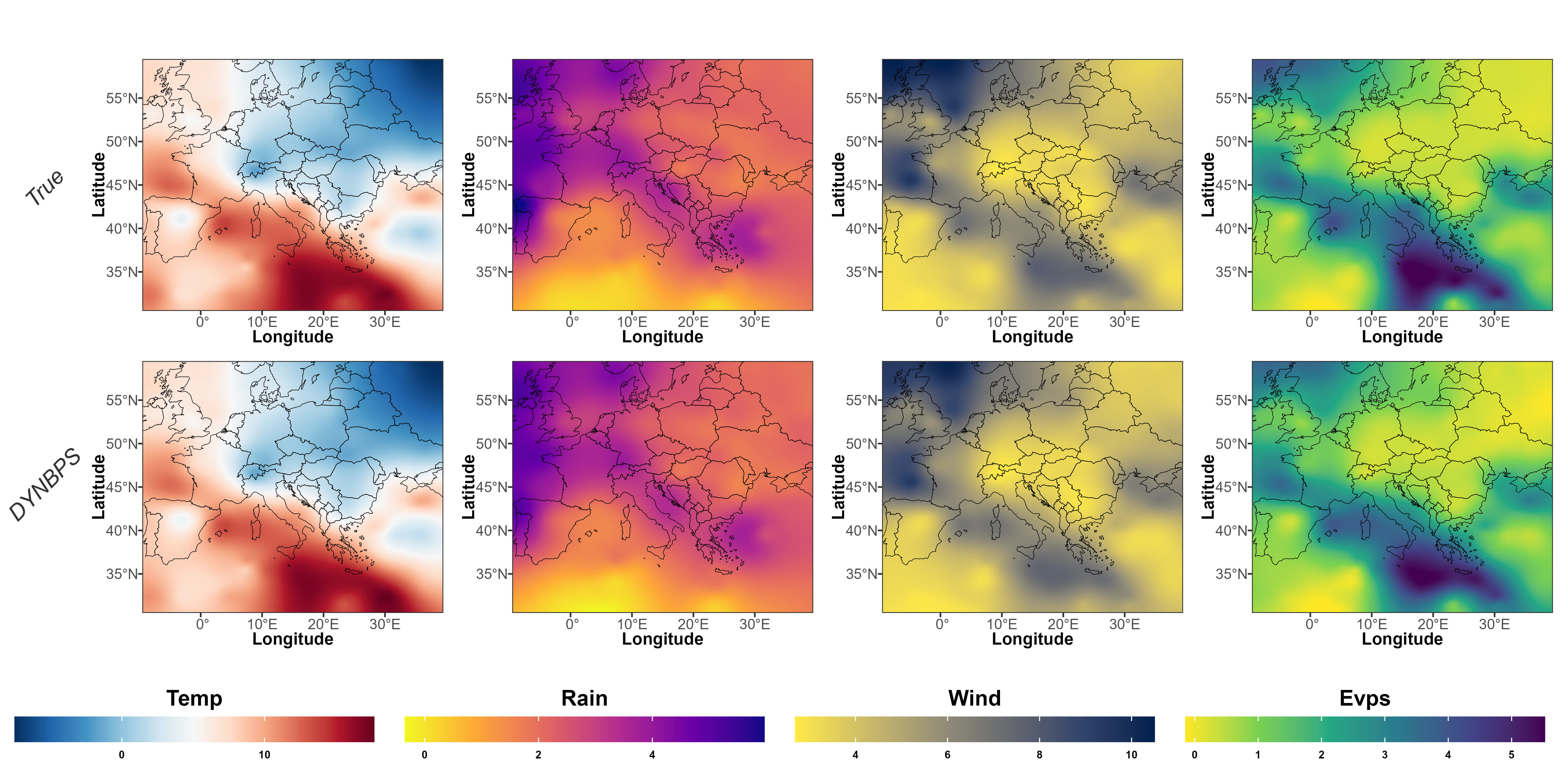}
    \caption{Spatial surface interpolation at observed time (in sample, December~2012, $t=120$): true spatial surfaces (top row) and \textsc{dynbps} predictions (bottom row) for all four climate variables.}
    \label{fig:interpolation_in}
\end{figure}

\begin{figure}[H]
    \centering
    \begin{minipage}{\textwidth}
        \centering
        \includegraphics[width=\textwidth]{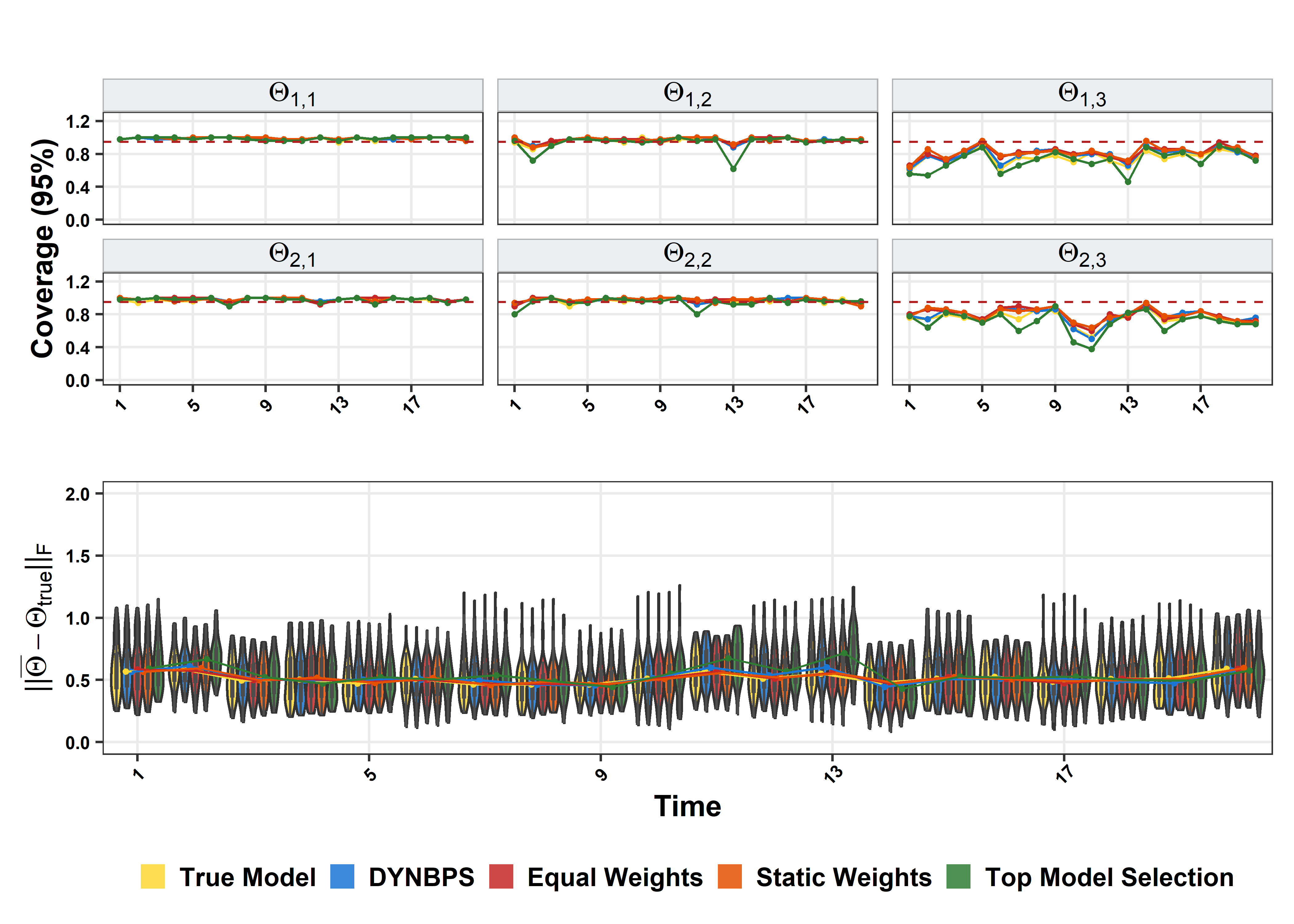}
        \small (a) $\mathscr{M}$-Closed
    \end{minipage}
    \vspace{1em}
    \begin{minipage}{\textwidth}
        \centering
        \includegraphics[width=\textwidth]{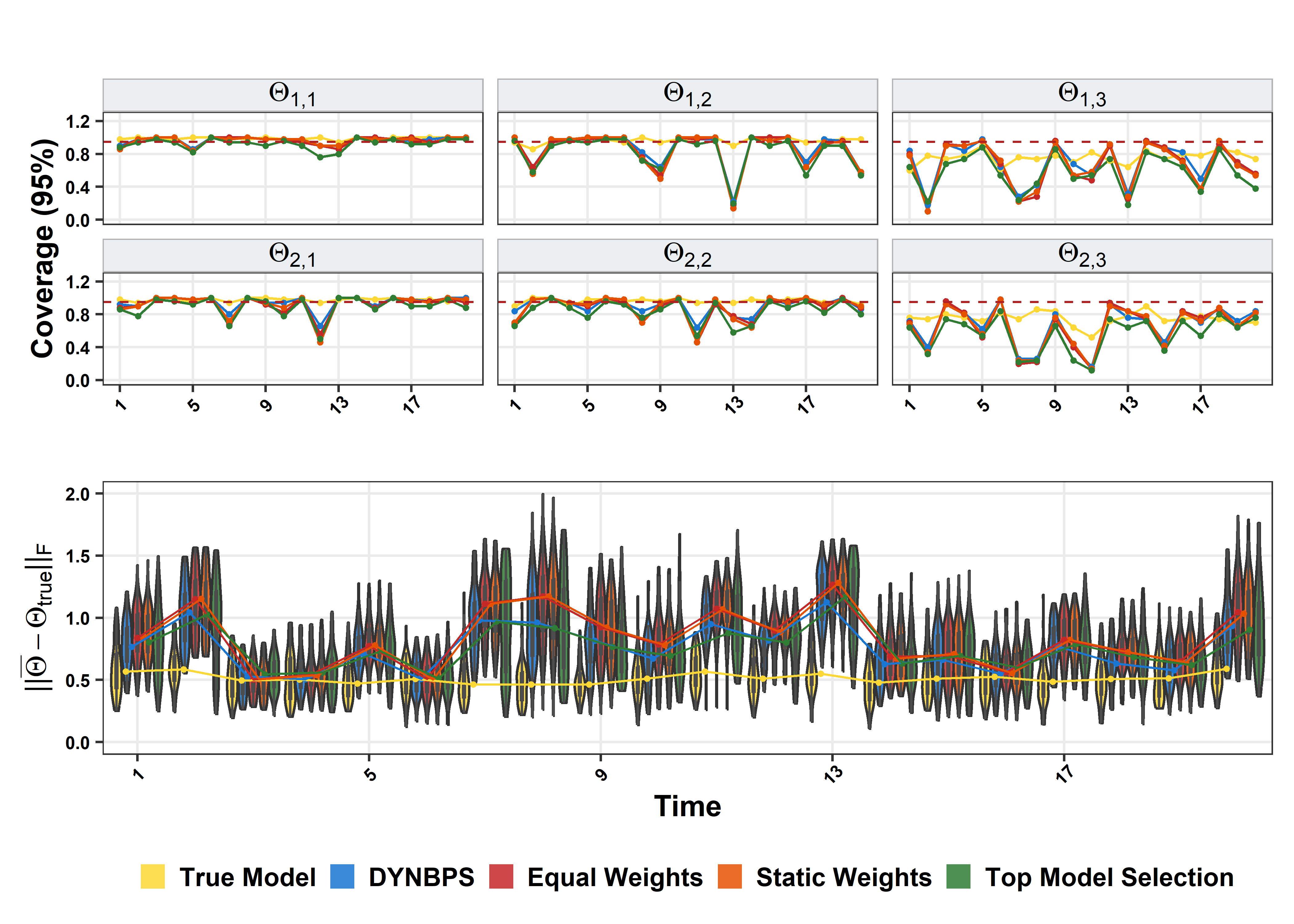}
        \small (b) $\mathscr{M}$-Open
    \end{minipage}
    \caption{Coverage ($95\%$) and Frobenius norm at each time point for regression coefficients $B$, i.e. $\Theta_{1:p,1:q}$, under the considered model configurations and 50 replications in (a) $\mathscr{M}$-closed and (b) $\mathscr{M}$-open settings.}
    \label{fig:Msettings_theta}
\end{figure}

\begin{figure}[H]
    \centering
    \begin{minipage}{\textwidth}
        \centering
        \includegraphics[width=\textwidth]{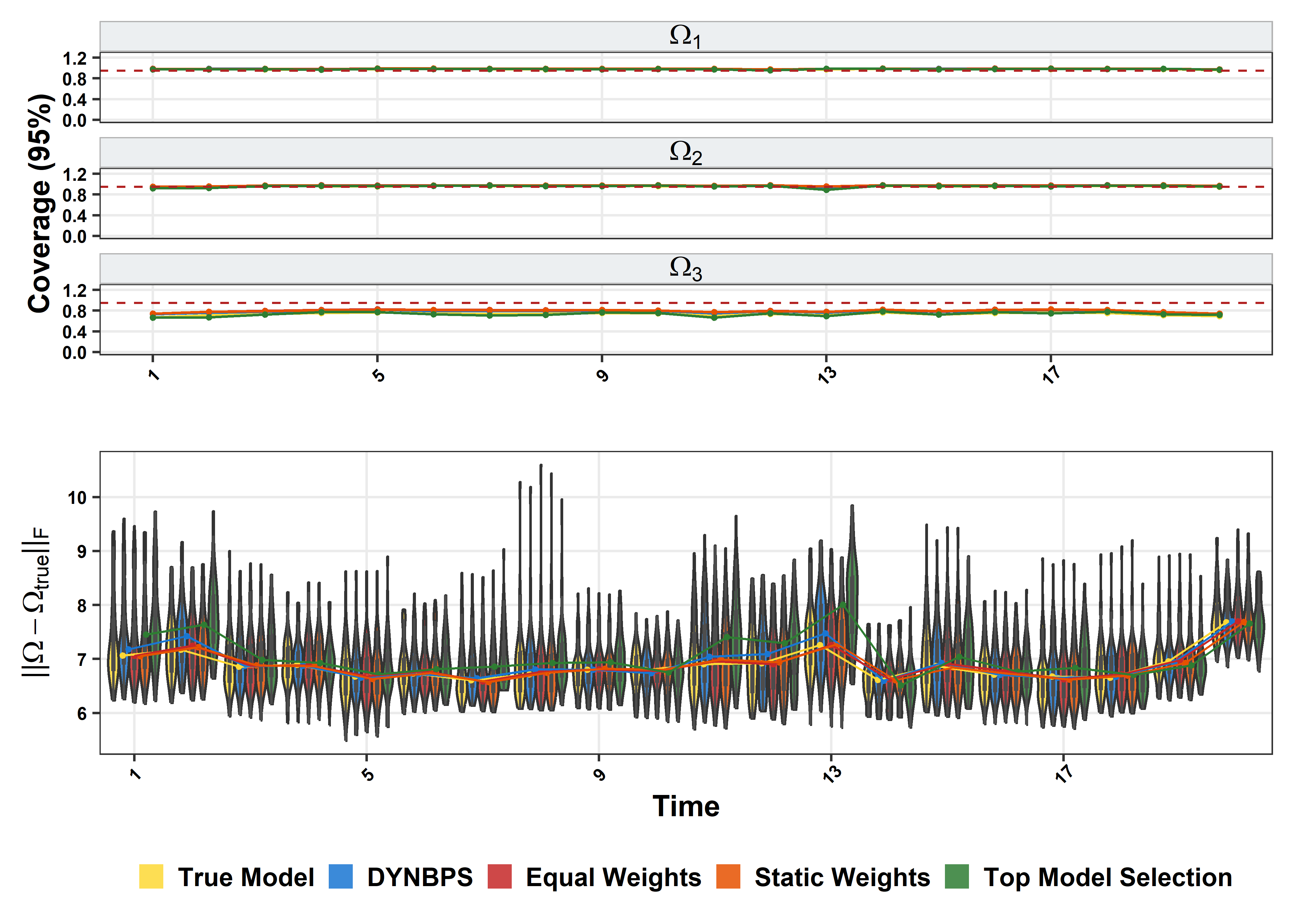}
        \small (a) $\mathscr{M}$-Closed
    \end{minipage}
    \vspace{1em}
    \begin{minipage}{\textwidth}
        \centering
        \includegraphics[width=\textwidth]{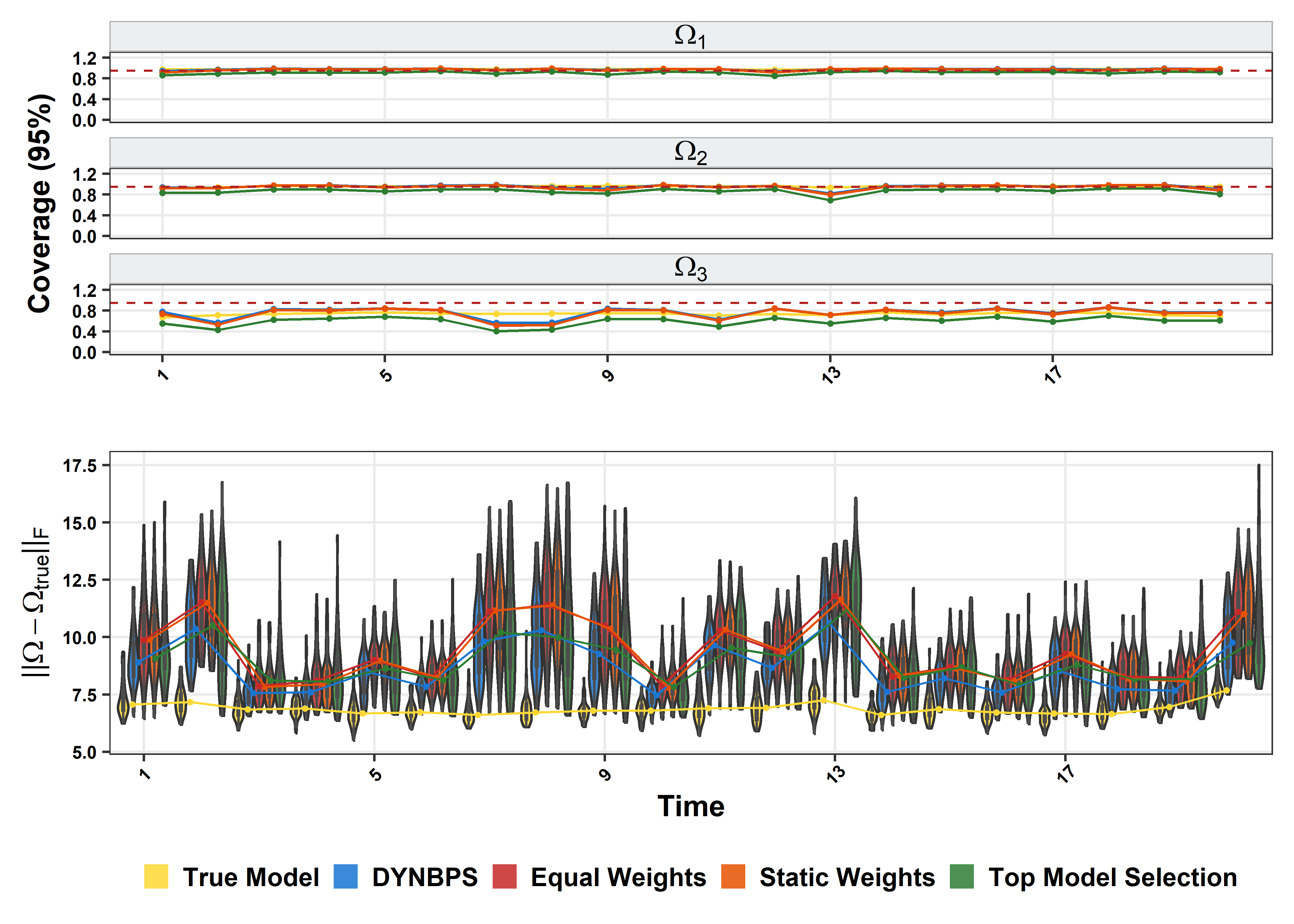}
        \small (b) $\mathscr{M}$-Open
    \end{minipage}
    \caption{Coverage ($95\%$) and Frobenius norm at each time point for each component of the multivariate spatial process $\Omega$, under the considered model configurations and 50 replications.}
    \label{fig:Msettings_omega}
\end{figure}

% --------------------------------------------------

% ################################################
%% ** The bibliograhy **
\bibliographystyle{chicago}
\bibliography{reference}% place <bib-data-file> 